\documentclass[11pt,a4paper]{article}
\usepackage{jheppub}
\usepackage{xcolor,bm} 
 
\newcommand{\dms}{\frac{(\mu^2 e^{\gamma_{\mathrm{E}}})^{\epsilon}}{\Gamma(1-\epsilon)}}
\newcommand{\dd}[1]{\frac{d^D {#1} }{(2\pi)^D}} 
\newcommand{\scone}{$\mathrm{SCET_I}$}
\newcommand{\sctwo}{$\mathrm{SCET_{II}}$}

\newcommand{\fms}[1]{{#1}\!\!\!/}

\newcommand{\mc}{\mathcal}

\newcommand{\eps}{\epsilon}  
\newcommand{\mums}{\mu^{2\epsilon}_{\overline{\mathrm{MS}}}}
 
\begin{document} 
\title{$\bm{N}$-jettiness in electroweak high-energy processes}
\author{Junegone Chay,} 
\author{Taewook Ha,}
\author{and Taehyun Kwon}
\def\KU{Department of Physics, Korea University, Seoul 02841, Korea}
\affiliation{\KU}
\emailAdd{chay@korea.ac.kr} 
\emailAdd{hahah@korea.ac.kr}
\emailAdd{aieamfirst@korea.ac.kr}

\abstract{ 
We study $N$-jettiness in electroweak processes at extreme high energies, in which the mass 
of the weak gauge bosons can be regarded as small. The description of the scattering process such as
$e^- e^+ \rightarrow \mu^- \mu^+ +X$ is similar to QCD. The incoming leptons emit
initial-state radiation and the resultant particles, highly off-shell, participate in the
hard scattering, which are expressed by the beam functions. After the hard scattering, the
final-state leptons or leptonic jets are observed, described by the
fragmenting jet functions or the jet functions respectively. At present, electroweak
processes are prevailed by the processes induced by the strong interaction, but they will be
relevant at future $e^- e^+$ colliders at high energy. The main difference between QCD
and electroweak processes is that the initial- and final-state particles should appear in
the form of hadrons, that is, color singlets in QCD, while there can be weak nonsinglets as
well in electroweak interactions. We analyze the factorization theorems for the
$N$-jettiness in $e^- e^+ \rightarrow \mu^- \mu^+ +X$, and compute the factorized parts to
next-to-leading logarithmic accuracy. To simplify
the comparison with QCD, we only consider the $SU(2)_W$ gauge interaction, and the
extension to the Standard Model is straightforward. Put it in a different way, it
corresponds to an imaginary world in which colored particles can be observed in QCD, and
the richer structure of effective theories is probed. Various nonzero nonsinglet matrix elements are 
interwoven to produce the factorized results, in contrast to QCD in which there are only 
contributions from the singlets. Another distinct feature is that
the rapidity divergence is prevalent in the contributions from weak nonsinglets due to the
different group theory factors between the real and virtual corrections. We verify that
the rapidity divergence cancels in all the contributions with a different number of
nonsinglet channels. We also consider the renormalization group evolution of each
factorized part to resum large logarithms, which are distinct from QCD.}


\maketitle
  
\section{Introduction\label{intro}} 
The understanding of high-energy scattering has reached a state of the art with the advent of 
the effective theories such as soft-collinear effective theory 
(SCET)~\cite{Bauer:2000ew,Bauer:2000yr,Bauer:2001ct,Bauer:2001yt}.  
The basic picture to this understanding lies in the following procedure. The partons from 
the incoming protons, as in Large Hadron Collider, participate in the hard scattering 
and produce a plethora of final-state particles including hadrons, electroweak gauge 
bosons, Higgs particles and possibly heavy particles beyond the Standard Model. 
Quantum chromodynamics (QCD) plays a major role in comprehending collider physics 
phenomenology because the strong interaction acts in every stage of the scattering 
process with disparate energy scales.

The essence of disentangling the strong interaction is to construct factorization theorems
which decompose the high-energy processes into hard, collinear and soft parts. SCET is
the appropriate effective theory of QCD for high-energy processes, in which energetic,
collinear particles form jets immersed in the sea of soft particles. In SCET we select the
relevant collinear and soft modes and integrate out all the other degrees of freedom.
Factorization of high-energy processes can be  accomplished in SCET by decoupling the soft
interaction from the collinear particles. Subsequently the collinear sectors in different directions 
do not interact with each other. Depending on the observables of interest, the
phase space is divided by the definite properties of the specific modes and the factorization
theorem is constructed according to how the modes or the phase spaces are organized. 
Each mode has its own characteristic scale, and typically there is a
hierarchy of such scales. The scattering cross sections or event shape observables are expressed in terms of
the logarithms of the ratios of the different energy scales, which necessitates the
resummation of the large logarithms. Because each factorized part is governed by a single
scale in each phase space, the large logarithms in each sector can
be resummed to all orders using the renormalization group (RG) equations.

So far, we have described the factorization of high-energy processes in QCD. 
Here we change gears to employ SCET in extremely high-energy electroweak processes, in
which all the masses of the particles including the weak gauge bosons can be regarded as small. 
By way of illustration, 
we consider the process $e^- e^+ \rightarrow \mu^- \ \mathrm{jet}, \mu^+ \ \mathrm{jet} +X$, where
$X$ denotes arbitrary final states, and the $\mu$ jet denotes the jet which includes the muon in the final
state.\footnote{From now on,
we will write $e^- e^+\rightarrow \mu^- \mu^+ +X$ for the jets including muons. 
If we refer to the muons instead of the muon jets, we will explicitly state it.}
In order to simplify the situation, we consider only the weak $SU(2)_W$ gauge interaction 
by turning off all the other gauge
interactions. Therefore we imagine a world with the $SU(2)_W$ weak gauge interaction out
of the full $SU(3)_C \times SU(2)_W\times U(1)_B$ gauge interactions of the Standard
Model. The extension to the Standard Model is complicated, but straightforward. 
In the high-energy limit, the incoming electron can be regarded as a collection of
the ``partons'' which consist of leptons, weak gauge bosons. These partons undergo a hard
collision and produce final-state energetic particles, along with soft particles.

This scenario may sound rather dull because everything mimics the processes in QCD, 
and one may wonder what, if any, can be learned from this
imaginary world. The most interesting issue in this context is that the weak gauge
interaction is not confining. It means that the incoming particles or the outgoing
particles do not have to be gauge singlets. This is in contrast to QCD, where all the
strongly-interacting particles are produced as color singlets, that is, hadrons. Due to this constraint,
various matrix elements of the operators in QCD in, say, the parton distribution functions (PDF) or
the jet functions are evaluated only for the color singlet configurations. However, 
the gauge singlets, as well as the gauge nonsinglets participate in
weak high-energy processes. It makes the procedure of the factorization 
more sophisticated, and requires more care in analyzing the interwoven structure.
Put it in a different way, it corresponds to
imagining QCD without confinement and ask how the factorization works if there are free
quarks and gluons. The underlying hard scattering processes can be traced directly and the
measurement of the jets and the properties can be reconstructed explicitly by the constituents without
worrying about hadronization.  
 
We sketch electroweak high-energy processes by analogy with QCD. The incoming on-shell
``partons'' (electrons in our case) possess certain fractions of the energy from the initial particles
and they radiate away gauge bosons to be far off-shell. This process is described by the electroweak beam
functions. Then the energetic partons undergo a hard scattering and the final-state
particles can be observed in terms of individual particles, described by the fragmentation functions,
or jets  by the jet functions or the fragmenting jet functions (FJF). And the soft interaction
is interspersed between the collinear parts. In electroweak high-energy processes, gauge
singlets and nonsinglets are involved in all these factorized components, which makes the analysis
more intriguing. This study will be relevant in high-energy electron-positron colliders such as 
CEPC~\cite{dEnterria:2016sca},  ILC~\cite{Djouadi:2007ik}, FCC-ee~\cite{Abada:2019zxq}, and
CLIC~\cite{Charles:2018vfv}. Our analysis can be extended to include the production of the Higgs boson,
weak gauge bosons and top quarks.

There appear many different energy scales, the energy of the hard
scattering $Q$, the invariant masses of the initial- and final-states as a
typical collinear scale, and the soft scale, and possibly more if we are interested in
more differential processes describing event shapes. In addition, the mass of the weak gauge bosons $M$
also enters into the picture as a physical mass. In radiative corrections, there are
logarithms of the ratios of these scales and we need to resum the large logarithms to all
orders. In SCET, the factorization is achieved by dissecting the phase space and devising the
modes in that phase space such that the radiative corrections depend on a single scale in each
phase space. Then the resummation is obtained by solving the RG equation.

Besides the conventional RG behavior, there is additional rapidity
divergence because the phase space is divided into different regions.
It is obviously absent in the full theory because there
is no separation of the phase space. Therefore it is a good consistency check for an
effective theory to see whether the rapidity divergences are cancelled when all the
factorized contributions are added. It is one of the goals in this paper to check this point
even in the presence of the nonsinglets participating in the scattering.

Though the sum of the rapidity divergences cancels, the rapidity
divergence remains in each sector and it affects the RG behavior of the factorized parts. The
rapidity divergence has been regulated using diverse methods, such as the use of the
Wilson lines off the lightcone~\cite{collins_2011}, the
$\delta$-regulator~\cite{Idilbi:2007ff,Idilbi:2007yi}, the analytic
regulator~\cite{Becher:2011dz}, the rapidity regulator~\cite{Chiu:2011qc,Chiu:2012ir}, the
exponential regulator~\cite{Li:2016axz}, and the pure rapidity
regulator~\cite{Ebert:2018gsn}, etc.. Recently one of us has proposed a consistent scheme
of applying rapidity regulators to the soft and collinear sectors~\cite{Chay:2020jzn}. It
correctly produces the directional dependence in the soft function, which is essential in
our case because there are four different collinear directions involved. The independence
of  the rapidity scale in the cross section critically depends on
the interplay between the collinear and the soft functions.
 
The dependence on the rapidity scale in each sector becomes highly nontrivial and 
more interesting when gauge 
nonsinglets as well as singlets participate in high-energy scattering. First of all, let us 
describe the rapidity divergence for singlets, which is well understood in QCD. 
The rapidity divergence appears distinctively in $\mathrm{SCET_I}$  and
$\mathrm{SCET}_{\mathrm{II}}$.  In \scone, since collinear 
and soft particles have different offshellness,  there is no overlap in rapidity between these modes, hence no
rapidity divergence arises. In other words, the rapidity divergence cancels in
each sector.  On the other hand, in $\mathrm{SCET}_{\mathrm{II}}$, where collinear and soft particles
have the same offshellness, they overlap near the rapidity boundary.  
The soft particles with small rapidity cannot
perceive the large rapidity region which belongs to collinear sector, causing the rapidity divergence  
in the soft sector. In the collinear sector, according to our regularization scheme of the rapidity divergence, 
which will be described in detail, the rapidity divergence arises from the 
zero-bin subtraction~\cite{Chiu:2011qc,Chiu:2012ir}. The zero-bin subtraction replaces the spurious rapidity 
divergence in the naive contribution. It is analogous to the pullup mechanism~\cite{Manohar:2000kr,Hoang:2001rr}
in the dimensional regularization,
in which the IR divergence is replaced by the UV divergence.
The rapidity divergence may survive in each sector, 
but their sum cancels.  However, the persistent existence of the rapidity divergence in each sector 
offers additional evolution to complete the resummation.

In weak interaction, the structure of the rapidity divergence is more intricate. For gauge singlets,
there is no rapidity divergence as in QCD. 
For gauge nonsinglets, regardless of $\mathrm{SCET_I}$ or $\mathrm{SCET_{II}}$,  the rapidity divergence is
not cancelled in each sector when the weak charges of the initial or final states are specified. 
Nor are the Sudakov logarithms. 
The non-cancellation of the electroweak logarithms 
is known as the  Block-Nordsieck violation in electroweak processes~\cite{Ciafaloni:2000df,
Ciafaloni:2001vt,Ciafaloni:2006qu,Manohar:2014vxa}. In contrast,  the Sudakov logarithms in QCD
cancel from virtual and real contributions in inclusive processes. It yields, for example, the  
Dokshitzer-Gribov-Lipatov-Altarelli-Parisi (DGLAP) evolution of the PDF. The non-cancellation in weak 
interaction affects the UV behavior as well as the rapidity behavior. Our result 
is, in some sense, a manifestation of the non-cancellation in considering the event shape through the
jettiness.

In the framework of SCET, it was pointed out in ref.~\cite{Manohar:2018kfx} that the
nonsinglet electroweak PDFs possess this property. It is
also true for the beam function in the initial state, and for the jet functions, or the FJFs in the final state,
and for the soft function. The purposes
of this paper are to analyze the structure of the factorization in the weak processes, and to
resum large logarithms by probing the structure of the divergences
including the rapidity divergence from the nonsinglet contributions. 

As a concrete example, we consider an event shape, especially
the $N$-jettiness~\cite{Stewart:2010tn, Jouttenus:2011wh} (in fact, 2-jettiness) in 
$e^-e^+ \rightarrow \mu^- \mu^+ +X$. By considering the jettiness, we can probe the hierarchy
of scales in effective theories, and the characteristics of a measurement-dependent 
factorization can be discussed. In addition, we are interested in the case in which there are 
four distinct lightlike directions. For this reason, $N$-jettiness is better suited than the 
beam thrust in the Drell-Yan process~\cite{Stewart:2010pd} or the jet thrust in $e^- e^+$ collisions.
In terms of QCD, it corresponds to the $N$-jettiness from the partonic subprocess
$q\overline{q} \rightarrow q' \overline{q}' +X$.

The rest of the paper is organized as follows. In section~\ref{factew}, we explain the $N$-jettiness,
and establish the power counting of the relevant momenta and the jettiness to choose
the appropriate effective theories. In section~\ref{jettifac}, we first construct the factorization of the 
$N$-jettiness in \scone, in which the ingredients of the factorized parts are extracted and defined. 
The beam functions,  the semi-inclusive jet functions, and the soft functions are constructed in the context
of \scone\ with the weak singlet and nonsinglet contributions. Then 
we scale down to \sctwo\  and establish the factorization by probing the relations 
between the beam function in \scone, and the PDF in \sctwo, or those between the fragmenting 
jet functions and the fragmentation
functions.  In section~\ref{rapidity}, we examine the source of the rapidity divergence, set up the 
rapidity regulators in the collinear and the soft sectors and discuss their characteristics.

In section~\ref{collinear}, we present the radiative corrections of the collinear parts at
next-to-leading order (NLO), which consist of the beam functions, the PDFs,
the jet functions, the fragmentation functions and the FJFs.  In section~\ref{hardfun}, the hard functions are 
collected for the process $\ell_e \overline{\ell}_e \rightarrow \ell_{\mu} \overline{\ell}_{\mu}$ 
with the left-handed electron and muon doublets, and the hard anomalous dimension matrix is presented. The
soft functions are analyzed in section~\ref{softs}, which are expressed in terms of the
matrices in weak-charge space. In section~\ref{rge}, we combine all the factorized parts
to show the RG evolution of the $N$-jettiness, and we conclude in section~\ref{conc}. Long
technical calculations are relegated to appendices. In appendix~\ref{laplace}, we list the
Laplace transforms of the distributions. In appendices~\ref{beamzero} and \ref{jetzero}, we present
how to obtain the collinear functions and the matching coefficients in the limit of small $M$. 
In appendix~\ref{samat}, we illustrate the color matrices for the soft functions at tree level,
and we show that there is no mixing at one loop for the soft functions with four
nonsinglets.
 
\section{SCET setup  for the jettiness in $e^- e^+ \rightarrow \mu^- \mu^+ X$  \label{factew} }
We consider the $N$-jettiness\footnote{We keep using the terminology $N$-jettiness, bearing in mind 
that we actually consider the 2-jettiness $\mathcal{T}_2$ in our case.}, which is defined 
as~\cite{Stewart:2010tn, Jouttenus:2011wh} 
\begin{equation} \label{defjet}
\mathcal{T}_N = \sum_k \mathrm{min} \ \Bigl\{ \frac{2 q_i \cdot p_k}{\omega_i} \Bigr\},
\end{equation} 
where $i$ runs over 1, 2 for the beams, and $3, \cdots, N+2$ for the final-state jets.  
Here $q_i$ are the reference momenta of the beams and the jets with the normalization 
factors $\omega_i = \overline{n}_i \cdot q_i$, and $p_k$ are the momenta of all the measured
particles in the final state. 
\begin{equation} \label{omegai}
q_{1,2}^{\mu} =\frac{1}{2} z_{1,2} E_{\mathrm{cm}} n_{1,2}^{\mu}= 
\frac{1}{2} \omega_{1,2} n_{1,2}^{\mu}, \ q_{i}^{\mu} =\frac{1}{2}\omega_i
n_i^{\mu}, (i=3,\cdots, N+2),
\end{equation} 
where $z_{1,2}$ are the momentum fractions of the beams.
Here $n_1$ and $n_2$ are lightcone vectors for
the beams, which are aligned to the $z$ direction, 
$n_1^{\mu} = (1,0,0,1),  \ n_2^{\mu} = (1, 0, 0, -1)$,
and $n_i$ are the lightcone vectors
specifying the jet directions. Typically we write the lightlike
vectors as $n_i = (1, \mathbf{n}_i)$ , $\overline{n}_i = (1, -\mathbf{n}_i)$,
where the unit vector $\mathbf{n}_i$ denotes the direction of the spatial momentum 
$\mathbf{p}_i$.  The $N$-jettiness in eq.~\eqref{defjet} can also be expressed 
in terms of some arbitrary hard scales instead of the normalization factor $\omega_i$.
 
In references~\cite{Jouttenus:2011wh, Bertolini:2017efs}, the authors have considered the 
differential distribution with respect to the individual jettiness, which is defined as
\begin{equation} \label{indijet}
\tilde{\mathcal{T}}_i = \sum_k \frac{2 q_i \cdot p_k}{\omega_i} \prod_{i\neq j} 
\theta (n_j \cdot p_k - n_i \cdot p_k).
\end{equation}
Here we consider the total $N$-jettiness $\mathcal{T}_N$,
which is given by $\mathcal{T}_N = \sum_i \tilde{\mathcal{T}}_i$.

For $\mathcal{T}_2 \ll Q$, SCET can be applied to the process 
$e^- e^+ \rightarrow \mu^- \mu^+ +X$, 
but we should determine which effective theories are to be employed, depending on the 
hierarchy of the scales in the collinear and soft momenta and the magnitude of the jettiness.
The $n$-collinear momentum scales as $p_c^{\mu} = (\overline{n}\cdot p_c,
p_{c\perp}, n\cdot p_c) =(p_c^-, p_{c\perp}, p_c^+) \sim Q(1, \lambda, \lambda^2)$, where 
$\lambda \sim p_{c\perp}/\overline{n}\cdot p_c$ 
is the small parameter in SCET. We can consider either \scone, or continue down to
\sctwo, depending on the power counting of the soft momentum. 

In $\mathrm{SCET}_{\mathrm{I}}$,
the ultrasoft (usoft) momentum scales as $p_{us}^{\mu} =(p_{us}^-, p_{us\perp}, p_{us}^+)
\sim Q(\lambda^2, \lambda^2, \lambda^2)$, while the soft momentum in
$\mathrm{SCET}_{\mathrm{II}}$ scales as $p_s^{\mu} =(p_s^-, p_{s\perp}, p_s^+) \sim Q(\lambda,
\lambda, \lambda)$.  The gauge boson mass $M$ is another mass scale which enters into the system.
When the usoft or the soft gauge bosons are on their mass shell, $p_{us}^2 \sim p_s^2 \sim M^2$,
it implies that the gauge boson mass $M$ is power counted as $Q\lambda^2$ in \scone, 
and $Q\lambda$ in \sctwo. 

The scale of the $N$-jettiness $\mathcal{T}_N$ from eq.~\eqref{defjet},
is extracted from the lightcone component $p^+$. Note, however, that the magnitudes of $p^+$ for
the collinear and the usoft momenta in \scone\ are comparable to each other, hence the contribution
to the $N$-jettiness comes both from the collinear and the usoft sectors. On the other hand,
the magnitude of $p^+$ for the collinear momentum is much smaller than that for the 
soft momentum in \sctwo. In this case, the contribution to the $N$-jettiness comes only from the soft
sector. Therefore we expect that the structure of the factorization takes different forms in \scone\ and
in \sctwo.
 
In \scone,  the hierarchy of the scales 
is given as $\mathcal{T}^2 \sim M^2 \ll p_c^2 \sim Q\mathcal{T} \ll Q^2$. The relevant modes scale as
\begin{align}
&n\mbox{-}\mathrm{collinear:} \ p_c^{\mu} \sim (Q, \sqrt{Q\mathcal{T}},\mathcal{T}) \sim (Q, \sqrt{QM}, M)
\sim (Q, p_{\perp}, p_{\perp}^2/Q), \nonumber \\
&\mathrm{usoft:} \ p_{us}^{\mu} \sim (\mathcal{T}, \mathcal{T}, \mathcal{T})\sim (M, M, M). 
\end{align}
The $N$-jettiness probes the scale of order 
$p_c^+ \sim  p_{us}^+\sim Q\lambda^2 \sim \mathcal{T}$, hence
both the collinear and the usoft parts contribute to the $N$-jettiness. 
 
If we consider the kinematical situation in which $M^2 \ll p_c^2 \sim Q\mathcal{T}$, 
the framework of \scone\ suffices to describe the $N$-jettiness. However, we would like to include 
another kinematical case, in which the jet becomes narrower, and we reach the region $M^2 \sim p_c^2 
\sim  \mathcal{T}^2$. Then the plus component $p_c^+ \sim \mathcal{T}^2/Q$ does not contribute to the
jettiness. However, there remains large logarithms associated with $Q/M$, which should be resummed. 
This can be achieved by going from \scone\ to \sctwo\ through the matching. The beam functions and the
jet functions in \scone\ have virtuality $p_c^2 \sim QM$, and we need a second stage of matching
to separate these scales.  And the ultrasoft function is scaled down to the soft function, which develops 
the rapidity divergence. These necessitate the use of \sctwo. 

In \sctwo, the hierarchy of scales is given by
$\mathcal{T}^2 \sim M^2 \sim p_c^2 \ll  Q\mathcal{T} \ll Q^2$. The hard-collinear, collinear 
and soft modes scale as
\begin{align}
&n\mbox{-}\mathrm{hard}\mbox{-}\mathrm{collinear:} \ p_{hc}^{\mu} \sim 
(Q, \sqrt{Q\mathcal{T}},\mathcal{T}) \sim (Q, \sqrt{QM}, M),\nonumber \\
&n\mbox{-}\mathrm{collinear:} \ p_c^{\mu} \sim (Q,  \mathcal{T},\mathcal{T}^2/Q) 
\sim (Q, M, M^2/Q)
\sim (Q, p_{\perp}, p_{\perp}^2/Q), \nonumber \\
 &\mathrm{soft:} \ p_{s}^{\mu} \sim (\mathcal{T}, \mathcal{T}, \mathcal{T}) \sim (M, M, M).
\end{align}
In fact, the hard-collinear modes are not the dynamical degrees of freedom in \sctwo, but
these are the modes from \scone\ to be integrated out to obtain \sctwo.
The $N$-jettiness probes the scale of order $p_s^+
\sim Q\lambda \sim M$, and the collinear contribution with $p_c^+ 
\sim Q\lambda^2 \sim M^2/Q$ does not
contribute to the jettiness. This is also recognized in ref.~\cite{Lustermans:2019plv} in a different context of 
measuring the transverse momentum $q_T$ and the 0-jettiness $\mathcal{T}_0$.

 We first describe $\mathrm{SCET_I}$ 
in order to set up the elements of the factorization. In order to obtain $\mathrm{SCET_{II}}$,  the hard-collinear
modes, which are previously collinear modes in $\mathrm{SCET_I}$, are integrated out to reach the collinear 
modes in $\mathrm{SCET_{II}}$. Note that the scaling of the usoft momentum in \scone, and the soft 
momentum in \sctwo\ remain the same, but the small parameter $\lambda$, responsible for the
power counting, is rescaled from $\sqrt{\mathcal{T}/Q}$ to $\mathcal{T}/Q$.
We present the results in both cases. 
 
\section{Factorization for the  $N$-jettiness\label{jettifac}}

\subsection{$\mathrm{SCET_I}$: $\mathcal{T}^2 \sim M^2 \ll p_c^2 \sim Q \mathcal{T} \ll Q^2$}

The procedure of obtaining the effective 
operators by integrating out the degrees of freedom of order $Q$ with their
Wilson coefficients was previously studied extensively in constructing the weak effective Hamiltonian 
with the QCD radiative corrections~\cite{Buchalla:1995vs}.  We follow the same technique, and 
the Wilson coefficients $D_I$ for the operators $O_I$ are obtained by matching the full theory onto 
SCET at any fixed order. 
The four-lepton operators for the process $e^- e^+ \rightarrow \mu^- \mu^+ X$ in SCET are given 
as\footnote{In ref.~\cite{Manohar:2018kfx}, other four-fermion operators are listed for neutrino scattering 
$\nu p\rightarrow \ell X$.}
\begin{equation}  \label{oi}
O_I  (x)= \overline{\ell}_{L2} (x) T_I \gamma^{\mu} \ell_{L1} (x)
\cdot \overline{\ell}_{L3} (x) T_I \gamma_{\mu} \ell_{L4} (x), \ (I = 1, 2),
\end{equation}
at leading order in $M/Q$.  We  label the incoming leptons as 1 and 2, and the outgoing leptons as 3 and 4. 
The index $I$ refers to the weak charge ($T_1 =t^a$ for the nonsinglet and $T_2=1$ for the singlet).
At tree level, $O_1$ is obtained by the exchange of a gauge boson, which leads to the matching coefficient
\begin{equation}
D_1^{(0)} = \frac{ig^2}{2p_1 \cdot p_2},
\end{equation}
where the incoming momenta are $p_1$ and $p_2$ 
for the weak doublets $\ell_{L1}$ and $\ell_{L2}$ with $2p_1 \cdot p_2 \sim Q^2$. 
The matching coefficient $D_2$ for $O_2$ begins at order $g^4$. 
 Here we can utilize the results of the corresponding four-quark operators in QCD 
for $q\overline{q} \rightarrow q' \overline{q}'$ at NLO,  and the Wilson coefficients  can be read off from 
those in QCD  in ref.~\cite{Kelley:2010fn} by adjusting the group theory factors for $SU(2)_W$.  In constructing the Wilson
coefficients for our process, note that  only the left-handed doublets 
participate in the scattering.
The effective Lagrangian for the leptonic high-energy scattering can be written as
\begin{equation}
\mathcal{L}_{\mathrm{eff}} = -i\sum_I  D_I  O_I  + \mathrm{hermitian \ conjugate}.
\end{equation}
 
The fields in the operators $O_I$ are now expressed in terms of the collinear fields in SCET. 
We refer to refs.~\cite{Bauer:2000ew,Bauer:2000yr,Bauer:2001ct,Bauer:2001yt} for 
the detailed formulation of SCET, 
and here we collect the necessary ingredients to express the operators in SCET. 
We introduce a lightcone vector
$n^{\mu}$,  and its conjugate light-cone vector $\overline{n}^{\mu}$
such that $n^2 = \overline{n}^2 =0$ and $n\cdot \overline{n}=2$. A four-vector $p^{\mu}$ can be
decomposed as $p^{\mu} = (p^-, p_{\perp}, p^+)$, and 
the $n$-collinear momentum $p_c^{\mu}$ scales as
$p_c^{\mu} = (p_c^-, p_{c\perp}, p_c^+) \sim p_c^- (1, \lambda, \lambda^2)$, where 
$\lambda$ is a small parameter in SCET. The usoft momentum scales as 
$p_{us}^{\mu} =  (p_{us}^-, p_{us\perp}, p_{us}^+) \sim p_c^- (\lambda^2, \lambda^2, \lambda^2)$.

The collinear operators, which are invariant under collinear gauge transformations, are constructed in terms
of the product of the fields and the Wilson lines. The basic building blocks for the lepton and the gauge
bosons are defined as
\begin{equation}
\ell_n (x) = W_n^{\dagger} (x) \xi_n (x),  \ \ \mathcal{B}_{n\perp}^{\mu} (x) =\frac{1}{g}
[ W_n^{\dagger} (x) iD_{n\perp}^{\mu} W_n (x)], 
\end{equation}
where $iD_{n\perp}^{\mu} = \mathcal{P}_{n\perp}^{\mu} + gA_{n\perp}^{\mu}$ is the covariant 
derivative.  The collinear Wilson line is given as
\begin{equation}
W_n (x)= \mathrm{P} \exp \Bigl( ig \int_{-\infty}^0 ds \overline{n} \cdot A_{n } (x+s\overline{n} )
\Bigr)  
=   \sum_{\mathrm{perm.}} \exp\Bigl[ -g \frac{\overline{n} \cdot 
A_{n }(x)}{\overline{n} \cdot \mathcal{P}} \Bigr],
\end{equation}
where P denotes the path ordering along the integration path.

At leading order in SCET, the interactions of soft gauge bosons with collinear fields exponentiate to 
form eikonal soft Wilson lines\footnote{When no confusion arises, we refer to the usoft momentum as
the soft momentum.}. The soft gauge bosons are decoupled by the field 
redefinition~\cite{Bauer:2001yt}
\begin{equation} \label{redef}
\ell_{n }^{(0)} (x) = Y_n^{\dagger} (x) \ell_n (x), \ \  
\mathcal{B}_{n\perp}^{\mu (0)} (x) = Y_n^{\dagger} (x)  \mathcal{B}_{n\perp}^{\mu} (x) Y_n (x).
\end{equation}
In this paper we use the fields after the decoupling and we drop the superscript (0) for simplicity. Here
$Y_n (x)$ is the soft Wilson line in the fundamental representation
\begin{equation}
Y_n (x)= \mathrm{P} \exp \Bigl( ig \int_{-\infty}^0 ds n \cdot A_{us} (x+sn )
\Bigr)  
=  \sum_{\mathrm{perm.}} \exp \Bigl[ -g \frac{n \cdot A_{us} (x)}{n\cdot \mathcal{P}}   \Bigr].
\end{equation}

Employing SCET, the operators $O_I$ in eq.~\eqref{oi} are written as 
\begin{equation}  \label{weakop}
O_I =  \overline{\ell}_{L2} Y_2^{\dagger} \gamma^{\mu} T_I Y_1 \ell_{L1}\cdot 
 \overline{\ell}_{L3} Y_3^{\dagger} \gamma_{\mu} T_I Y_4 \ell_{L4}.
\end{equation} 
The differential cross section for the 2-jettiness $\mathcal{T}_2$ is given by~\cite{Bauer:2008jx}
\begin{equation} \label{observ}
\frac{d\sigma}{d\mathcal{T}_2} = \frac{1}{2s} \int d^4 x \sum_X \langle I| 
\mathcal{L}_{\mathrm{eff}} (x)|
 X \rangle \langle X|\mathcal{L}_{\mathrm{eff}} (0)|I\rangle  
 \delta \Bigl( \mathcal{T}_2 - g (X)\Bigr),
\end{equation}
where $|I\rangle$ represents the initial state,  $|X\rangle$ denotes
the final state, and the  sum over $X$ includes a sum over states with the
appropriate phase space.  The
function $g  (X)$ computes the value of the jettiness for the final state $X$.  
In SCET, the final states $|X\rangle$ consist of the collinear states $|X_i\rangle$ in 
the $n_i$ directions ($i=1, 2, 3, 4$) and the soft states $|X_s\rangle$.  Since the $n_i$-collinear particles
do not interact with each other, and the soft particles are decoupled from the collinear sectors by the
redefinition in eq.~\eqref{redef}, the final states $|X\rangle$ in the Hilbert space can be expressed 
in terms of the tensor product of the collinear states $|X_i\rangle$ and the soft states $|X_s\rangle$ as
\begin{equation}
|X\rangle = |X_1\rangle \otimes |X_2\rangle \otimes |X_3\rangle \otimes |X_4\rangle \otimes 
|X_s\rangle.
\end{equation}

The factorization in SCET is established in the scattering cross section, that is, 
at the amplitude-squared level, instead of at the amplitude level, as shown in eq.~\eqref{observ}. 
Therefore we need to factorize 
the product of the operators $O_I^{\dagger}(x) O_J (0)$, which will be reorganized in terms of the
collinear operators in each collinear direction along with the soft Wilson lines, and it will be implemented
in eq.~\eqref{observ} to express the factorized result for the $N$-jettiness. The 
product of the operators $O_I^{\dagger}(x) O_J (0)$ is given as 
\begin{align} \label{oprod} 
O_I^{\dagger} (x) O_J (0) &= \Bigl( \overline{\ell}_{L1}
Y_1^{\dagger} \gamma^{\nu} T_I Y_2 \ell_{L2}\cdot \overline{\ell}_{L4}
Y_4^{\dagger} \gamma_{\nu} T_I Y_3 \ell_{L3}\Bigr) (x)  \nonumber \\
&\times \Bigl(
\overline{\ell}_{L2} Y_2^{\dagger} \gamma^{\mu}T_J Y_1 \ell_{L1}\cdot
\overline{\ell}_{L3} Y_3^{\dagger} \gamma_{\mu} T_J Y_4 \ell_{L4}\Bigr) (0).
\end{align} 
We rewrite the product by grouping the fields in respective collinear
directions, and use the relation~\cite{Manohar:2018kfx} 
\begin{align} \label{decomp}
(\overline{\ell}_n)^i_{\alpha} (\ell_n)^j_{\beta} &= \frac{1}{2N} \delta^{ij} (\mathcal{P}_L
\fms{n})_{\beta\alpha} \overline{\ell}_n \frac{\fms{\overline{n}}}{2} \ell_n +(t^c)^{ji}
(\mathcal{P}_L\fms{n})_{\beta\alpha} \overline{\ell}_n \frac{\fms{\overline{n}}}{2} t^c \ell_n 
\nonumber \\
&\equiv \delta^{ij} (\mathcal{P}_L\fms{n})_{\beta\alpha} C_{\ell}^0 +(t^c)^{ji}
(\mathcal{P}_L\fms{n})_{\beta\alpha} C_{\ell}^c , 
\end{align} 
where $i$, $j$ are the gauge indices, $\alpha$, $\beta$ are the Dirac indices, and 
$\mathcal{P}_L = (1-\gamma_5)/2$. 
To make the notation concise, we have extended the index to 0 such that 
\begin{equation} 
C_{\ell}^0 (x, y) = \frac{1}{2N}\overline{\ell}_{Ln} (x)
\frac{\fms{\overline{n}}}{2} T^0 \ell_{Ln} (y) , \ 
C_{\ell}^a (x, y) = \overline{\ell}_{Ln} (x)
\frac{\fms{\overline{n}}}{2} T^a \ell_{Ln} (y), 
\end{equation} 
with $T^0 = 1$ and $T^a =t^a$ are the gauge generators for $a = 1,\cdots, N^2 -1$.
We keep $N$ as it is for the $SU(N)$ gauge interaction, and the Casimir invariants are denoted by
$C_F = (N^2 -1)/(2N)$ and $C_A =N$. For the $SU(2)_W$ gauge group, we put $N=2$.

After some manipulation, eq.~\eqref{oprod} can be written as 
\begin{align} \label{o1o2} 
&16 n_1 \cdot n_4 n_2\cdot n_3 \Bigl( Y_2^{\dagger} T_J Y_1 (0) 
\Bigr)^{i_1 k_1} \Bigl( Y_1^{\dagger} T_I  Y_2 (x)\Bigr)^{i_2 k_2} \Bigl( Y_3^{\dagger}
T_J y_4 (0)\Bigr)^{i_3 k_3} \Bigl( Y_4^{\dagger} T_I Y_3 (x)\Bigr)^{i_4 k_4} \nonumber \\
&\times \Bigl[ (T^c)^{k_2 i_1} C^c_{\ell_2} (0,x) \Bigr] \Bigl[ (T^d)^{k_1 i_2}
C^d_{\ell_1} (x,0) \Bigr] \Bigl[ (T^e)^{k_4 i_3} C^e_{\ell_3} (0,x) \Bigr] \Bigl[
(T^f)^{k_3 i_4} C^f_{\ell_4} (x,0) \Bigr] \nonumber \\
&= 16 n_1 \cdot n_4 n_2\cdot n_3  C^c_{\ell_2} (0,x)  C^d_{\ell_1} (x,0) C^e_{\ell_3} (0,x)  C^f_{\ell_4} (x,0)
\nonumber \\
&\times \mathrm{Tr}\ \Bigl[T^c Y_2^{\dagger} T_J Y_1 (0)   T^d   Y_1^{\dagger} T_I  Y_2 (x) \Bigr]
\cdot  \mathrm{Tr} \ \Bigl[ T^e Y_3^{\dagger} T_J Y_4 (0) T^f   Y_4^{\dagger} T_I Y_3 (x) \Bigr].
\end{align} 
Here the coefficient $n_1 \cdot n_4 n_2\cdot n_3$ will be absorbed into the hard function.
In order to construct the expression for the jettiness, all the collinear  and the soft parts should 
be organized in such a way that the contribution to the jettiness from each part becomes manifest. 

The $N$-jettiness (the 2-jettiness here) can be expressed in terms of the matrix elements for
each collinear part and the soft part, which is schematically expressed as
\begin{align} \label{schemefac}
\frac{d\sigma}{d\mathcal{T}_2} &= \frac{1}{2s}  \sum_{IJ} H_{JI}   
\langle  e^+ |C^c_{\ell_2} (0,x)  |e^+\rangle  \langle e^-| C^d_{\ell_1} (x,0) |e^-\rangle 
\langle 0| C^e_{\ell_3} (0,x) |0\rangle \langle 0|C^f_{\ell_4} (x,0) |0\rangle  \nonumber \\
&\times \langle 0| \mathrm{Tr}\ \Bigl[T^c Y_2^{\dagger} T_J Y_1 (0)   T^d   Y_1^{\dagger} T_I  Y_2 (x) \Bigr]
\cdot  \mathrm{Tr} \ \Bigl[ T^e Y_3^{\dagger} T_J Y_4 (0) T^f   Y_4^{\dagger} T_I Y_3 (x) \Bigr] |0\rangle
\nonumber \\
&\times \delta \Bigl( \mathcal{T}_2 - g (X)\Bigr).
\end{align} 
Eq.~\eqref{schemefac} is unavoidably 
complicated due to the presence of the gauge indices, compared to the corresponding expression in QCD, 
in which there are only singlet contributions.
 
The matrix elements between the corresponding states in the Hilbert space can be obtained because
the collinear currents $C_{\ell}^a$ in different collinear directions and the soft part are decoupled.
For example, $\langle e^-| C^d_{\ell_1} (x,0) |e^-\rangle$ yields the electron beam function, and when 
we consider the intermediate states, the projection into the $n_1$-collinear states 
$\sum|X_1\rangle \langle X_1|$
is inserted.  The matrix element $\langle 0| C^e_{\ell_3} (0,x) |0\rangle$ describes the final-state jet 
and the projection $\sum | X_3 \rangle \langle  X_3|$ is inserted for the intermediate states. 
The matrix element for the soft Wilson lines yields the soft function, and 
the intermediate states consist of $\sum |X_s\rangle \langle X_s|$. The treatment of the matrix 
elements is delineated below.

\subsubsection{The beam function}

The beam functions are obtained by taking the matrix elements of 
$C^c_{\ell_2} (0,x)$ and $C^d_{\ell_1} (x,0)$ in 
eq.~\eqref{o1o2} between the initial states $|e^- (P_1)\rangle$ or $|e^+ (P_2)\rangle$. 
For the matrix element of  
$C^d_{\ell_1} (x,0)$, the coordinate $x$ can
be expanded around the lightcone coordinate $n_1\cdot x$, and the subleading terms 
can be neglected. Then by writing $p^+ = n_1\cdot p$, $p^- =\overline{n}_1\cdot p$, 
the matrix element can be written as
\begin{align} \label{cl10}
&\langle e^- | C_{\ell_1}^d (x,0)|e^- \rangle = k_d \langle e^-| \overline{\ell}_{L1} \Bigl( x^+  
\frac{\overline{n}_1}{2}\Bigr)  \frac{\fms{\overline{n}}_1}{2}
T^d \ell_{L1} (0)|e^- \rangle  \nonumber \\
&= k_d  \int dq^+ dq^-  \langle e^- | e^{i\hat{p}^- x^+/2} \overline{\ell}_{L1} (0)
e^{-i\hat{p}^- x^+/2}\frac{\fms{\overline{n}}_1}{2}
T^d  \delta (q^+ +\mathcal{P}^+)  \delta(q^- + \mathcal{P}^-)   \ell_{L1} (0)|e^- \rangle  \nonumber %\\
\end{align}
\begin{align}
&=k_d \int dq^+ dq^-  e^{i(P_1^- -q^-) x^+/2}
\langle e^- | \overline{\ell}_{L1} (0)\frac{\fms{\overline{n}}_1}{2} T^d 
\delta (q^+ +\mathcal{P}^+)  \delta(q^- + \mathcal{P}^-)   
\ell_{L1} (0) |e^- \rangle,
\end{align}
where $k_0=1/(2N)$ for the singlet and $k_d =1$ ($d\neq 0$) for the nonsinglets.  In the second line,
the integration of the delta functions, which is equal to the identity, is inserted. 
The operators  $\mathcal{P}^{\pm}$ extract the corresponding momenta for the annihilated particles.
And $\overline{\ell}_{L1}(x^+ \overline{n}_1 /2)$ is translated to the origin using the
momentum operator $\hat{p}^{\mu}$.
 
The last part in the last    eq.~\eqref{cl10} can be manipulated as
\begin{align}
&\delta (q^- + \mathcal{P}^-) \ell_{L1} (0)|e^- (P_1)\rangle =  
\int \frac{dy}{2\pi} e^{i (\mathcal{P}^- + q^-) y} \ell_{L1} (0)
e^{-i \mathcal{P}^- y} e^{i \mathcal{P}^- y} |e^- (P_1^-) \rangle \nonumber \\
&=  \int \frac{dy}{2\pi} e^{-i (P_1^- -q^-)y} e^{i \mathcal{P}^- y} \ell_{L1} (0) 
e^{-i \mathcal{P}^- y}   |e^- (P_1^-)\rangle  \nonumber \\
&= \int \frac{dy}{2\pi} e^{-i (P_1^- -q^-)y} [e^{i \mathcal{P}^- y} \ell_{L1} (0) ]
 |e^- (P_1^-)\rangle  = [\delta (P_1^- -q^- - \mathcal{P}^- ) \ell_{L1} (0) ]
 |e^- (P_1^-)\rangle, 
  \end{align}
using the relation $e^{i \mathcal{P}^- y} \ell_{L1} (0)  e^{-i \mathcal{P}^- y} 
=[e^{i \mathcal{P}^- y} \ell_{L1} (0) ]$, where the operator $\mathcal{P}^-$ 
in the bracket acts only inside the bracket. 
 
Now we change the variables to $q^- =   (1-z_1) \overline{n}_1 \cdot P_1$, 
$q^+= t_1/\omega_1$ with 
$\omega_1 = z_1 \overline{n}_1 \cdot P_1$.  Then eq.~\eqref{cl10} is written as
\begin{align} \label{cl1}
&\langle e^- | C_{\ell_1}^d (x,0)|e^- \rangle = k_d  
 \int \frac{dz_1}{z_1} \int dt_1 \omega_1 e^{i \omega_1 n_1 \cdot x/2}  \nonumber \\
&\times   \langle e^- | \overline{\ell}_{L1} (0)\frac{\fms{\overline{n}}_1}{2}
T^d  \delta ( t_1 + \omega_1 n_1 \cdot \mathcal{P} )   
\Bigl[ \delta (\omega_1 -\overline{n}_1 \cdot \mathcal{P}) 
\ell_{L1} (0) \Bigr] |e^-\rangle  \nonumber \\
&= k_d  \int \frac{dz_1}{z_1} \int dt_1 \omega_1 e^{i \omega_1 n_1 \cdot x/2} B_e^d (t_1, z_1,  M, \mu),
\end{align}
where the electron beam function $B_e^d (t_1, z_1,  M, \mu)$ is defined as
\begin{equation} \label{beamdef}
B_e^d (t_1, z_1, M, \mu) = \langle e^- | \overline{\ell}_{L1} (0)\frac{\fms{\overline{n}}_1}{2}
T^d  \delta ( t_1 + \omega_1 n_1 \cdot \mathcal{P} )   
[\delta (\omega_1 -\overline{n}_1 \cdot \mathcal{P}) \ell_{L1} (0) ]|e^-\rangle.
\end{equation}
Note that the quantity $t_1/\omega_1$ is the contribution to the jettiness.
Similarly, the matrix element for $C_{\ell_2}^c (0,x)$ can be written as
\begin{equation} \label{cl2}
\langle e^+ | C_{\ell_2}^c (0,x)|e^+ \rangle  = k_c  \int \frac{dz_2}{z_2} dt_2 \omega_2 
e^{i \omega_2 n_2 \cdot x/2}  B_{\bar{e}}^c (t_2, z_2, M, \mu),
\end{equation}
where the anti-electron (positron) beam function $B_{\bar{e}}^c (t_2, z_2,  M, \mu)$ is defined as
\begin{equation} \label{beamdev}
B_{\bar{e}}^c (t_2, z_2,  M, \mu) = \mathrm{Tr} \,  \langle e^+ | \ell_{L2} (0) \delta (t_2 + \omega_2 n_2 
\cdot \mathcal{P}) [\delta (\omega_2 -\overline{n}_2 \cdot \mathcal{P}) \overline{\ell}_{L2} (0) 
\frac{\fms{\overline{n}}_2}{2} T^c] |e^+\rangle.
\end{equation}
 
\subsubsection{The jet function  and the semi-inclusive jet function}
The matrix element for $C_{\ell_3}^e (0,x)$ can be written, by inserting the identity 
$\int d^4 p_3 \delta^{(4)} (p_3 + \mathcal{P})$, and translating 
$\ell_{L3} (n_3\cdot x \overline{n}_3/2)$ to the origin, as
\begin{align}
\langle 0 | C_{\ell_3}^e (0,x) |0\rangle &= k_e\int d^4 p_3 e^{-i\omega_3 n_3\cdot x/2} 
\mathrm{Tr}\, \langle 0| 
 \ell_{L3} (0) \delta^{(4)} (p_3 +\mathcal{P}) \overline{\ell}_{L3} (0)  \frac{\fms{\overline{n}}_3}{2}
T^e   |0\rangle \nonumber \\
&= k_e \int \frac{d^4 p_3}{(2\pi)^3} e^{-i\omega_3 n_3\cdot x/2} \omega_3 J^e (p_3^2,  M, \mu), 
\end{align}  
where $\omega_3 = \overline{n}_3 \cdot p_3$, and the lepton jet function $J^e (p_3^2, M, \mu)$ is defined as
\begin{align} \label{jetdef}
J^e (p_3^2, M, \mu)  
&= \frac{2(2\pi)^3}{\omega_3} \mathrm{Tr} \, 
\langle 0| \ell_{L3} (0) \delta (n_3\cdot p_3 +n_3\cdot \mathcal{P}) 
\delta^{(2)}  (p_{3\perp} + \mathcal{P}_{\perp}) \nonumber \\
&\times [\delta (\omega_3 +\overline{n}_3\cdot \mathcal{P})
\overline{\ell}_{L3} (0) ]\frac{\fms{\overline{n}}_3}{2} T^e |0\rangle. 
\end{align}
The quantity $p_3^2/\omega_3$ contributes to the jettiness from the jet function.
Here the trace refers to the sum over the Dirac indices and the weak charge indices. 
The definition of the jet function in eq.~\eqref{jetdef} may look different from the conventional one, but
it turns out to be the same.  Eq.~\eqref{jetdef} can be written as
\begin{align}
J^e (p_3^2, M, \mu)  &= (2\pi)^2 \int \frac{d\overline{n}_3 \cdot y}{\omega_3} 
e^{i n_3 \cdot p_3 \bar{n}_3 \cdot y/2} \nonumber \\
&\times \mathrm{Tr} \, \langle 0| 
 \ell_{L3} (0)  e^{i n_3 \cdot \mathcal{P} \bar{n}_3 \cdot y/2} 
\delta (\omega_3  +\overline{n}_3\cdot \mathcal{P}) 
\delta^{(2)}  (p_{3\perp} + \mathcal{P}_{\perp})\overline{\ell}_{L3} (0) \frac{\fms{\overline{n}}_3}{2}
T^e  |0\rangle \nonumber \\
&= (2\pi)^2\int \frac{d\overline{n}_3 \cdot y}{\omega_3} e^{i n_3 \cdot p_3 \bar{n}_3 \cdot y/2} 
\nonumber \\
 &\times \mathrm{Tr}  \, \langle 0| 
 \ell_{L3} \Bigl(\overline{n}_3\cdot y \frac{n_3}{2}\Bigr)  
\delta (\omega_3 +\overline{n}_3\cdot \mathcal{P}) 
\delta^{(2)}  (p_{3\perp} + \mathcal{P}_{\perp})\overline{\ell}_{L3} (0) \frac{\fms{\overline{n}}_3}{2}
T^e |0\rangle,
\end{align}
which is the same as eq.~(2.30) in ref.~\cite{Stewart:2010qs} except the factor $1/(2N_c)$.  
Note that the QCD jet functions single out the color singlet components by taking the color and spin averages.
However, since we are dealing with the left-handed fields with a given weak charge, we do not
average over spin and color.

Note that $J^e (p_3^2,  M, \mu)$ is the inclusive jet function, and 
the nonsinglet part is zero, 
because the weak doublets with the opposite weak charges contribute with the opposite sign. It is the reason
why there is only the color singlet contribution in QCD.
Here we can extend the definition of the jet functions such that the individual nonsinglet contribution can be
extracted. And it can be probed by observing a muon and an antimuon in each final jet experimentally.
If we are interested in the jet in which, say, a lepton $l$ (a muon) is observed, we can define 
the semi-inclusive jet function as
\begin{align} \label{semijet}
J_{l}^a (p_3^2,  M, \mu)  &= \int \frac{d\overline{n}_3\cdot p_l}{\overline{n}_3 \cdot p_l}
\int \frac{d^2 \mathbf{p}_l^{\perp}}{\omega_3} \sum_X \mathrm{Tr} \langle 0| 
 \ell_{L3} (0) \delta (n_3\cdot p_3 +n_3\cdot \mathcal{P}) 
\delta^{(2)}  (p_{3\perp} + \mathcal{P}_{\perp})  |l X\rangle\nonumber \\
&\times   \langle l X|
[\delta (\omega_3 +\overline{n}_3\cdot \mathcal{P})
\overline{\ell}_{L3} (0) ] \frac{\fms{\overline{n}}_3}{2}T^a   |0\rangle,
\end{align}
where the lepton $l$ is specified in the final state, noting that the phase space for $l$ can be written as 
\begin{equation}
\int \frac{d^3 \mathbf{p}_l}{(2\pi)^3 2E_l} = \int \frac{d^4 p_l}{(2\pi)^3} \delta (n_3\cdot p_l
\overline{n}_3\cdot p_l -\mathbf{p}_l^{\perp 2}) = \frac{1}{2(2\pi)^3} \int 
\frac{d \overline{n}_3 \cdot p_l}{\overline{n}_3 \cdot p_l} \int d^2 \mathbf{p}_l^{\perp}.
\end{equation}
Then the semi-inclusive jet function at tree level is normalized as
$J_l^{a (0)} (p^2) = \delta (p^2) \mathrm{Tr} (T^a P_l)$,
where $P_l$ is the projection operator to the given lepton $l$. 
For example, the projection operator for the muon is given by $P_{\mu} = (1-t^3)/2$ 
in the $SU(2)$ weak interaction, and $P_{\nu_{\mu}} = (1+t^3)/2$
for the muon neutrino.   

The terminology `semi-inclusive jet function' was used in 
ref.~\cite{Kang:2016mcy}, but it is different from ours. Their definition corresponds to our fragmentation
function with the final jet instead of a final lepton. It is called the fragmentation function to a jet (FFJ) in 
ref.~\cite{Dai:2016hzf}. The relation among the FJF, the semi-inclusive 
jet function and the fragmentation function will be discussed in section~\ref{ffandfjf}.

In terms of the semi-inclusive lepton jet function,  $\langle 0| C_{\ell_3}^e (0,x)|0\rangle_l$,
with the lepton $l$ in the final state, can be written
as
\begin{equation} \label{cl3}
\langle 0 | C_{\ell_3}^e (0,x) |0\rangle_l  = k_e \int \frac{d^4 p_3}{(2\pi)^3} 
e^{-i\omega_3 n_3\cdot x/2} \omega_3 J_{l}^e (p_3^2,  M, \mu). 
\end{equation}  
In a similar way,  the collinear matrix element, $\langle 0 |C_{\ell_4}^f (x,0)|0\rangle_l$ can be written as
\begin{equation} \label{cl4}
\langle 0| C_{\ell_4}^f (x,0)| 0\rangle_l =  k_f\int \frac{d^4 p_4}{(2\pi)^3} e^{-i\omega_4 n_4\cdot x/2} 
\omega_4 J_{l}^f (p_4^2, M, \mu), 
\end{equation}
where the semi-inclusive antilepton jet function $J_{l}^f (p_4^2, M, \mu)$ is given as
\begin{align} \label{semijetbar}
J_{l}^f (p_4^2,  M, \mu)  &= \int \frac{d\overline{n}_4\cdot p_l}{\overline{n}_4 \cdot p_l}
\int \frac{d^2 \mathbf{p}_l^{\perp}}{\omega_4} \sum_X   \langle 0| 
 \overline{\ell}_{L4} (0) \delta (n_4\cdot p_4 +n_4\cdot \mathcal{P}) 
\delta^{(2)}  (p_{4\perp} + \mathcal{P}_{\perp}) |l X\rangle \nonumber \\
&\times   \langle l X|
\frac{\fms{\overline{n}}_4}{2}T^f [\delta (\omega_4 +\overline{n}_4\cdot \mathcal{P})
\ell_{L4} (0) ]   |0\rangle.
\end{align}
The fragmentation function or the FJF will be discussed later.

\subsubsection{The soft function}
The soft matrix elements  from eq.~\eqref{o1o2} are written as 
\begin{align} \label{intsoft}
&  \langle 0| \mathrm{Tr} \Bigl( 
T^c Y_2^{\dagger} T_J Y_1 (0) T^d Y_1^{\dagger} T_I Y_2 (x)\Bigr)
 \mathrm{Tr} \Bigl( T^e Y_3^{\dagger} T_J Y_4 (0) T^f Y_4^{\dagger} T_I Y_3 (x)\Bigr) |0\rangle
\nonumber \\
&=  \langle 0| \Bigl(
Y_1^{\dagger} T_I Y_2 (x) T^c \Bigr)^{ij} \Bigl( Y_4^{\dagger} T_I Y_3 (x) T^e\Bigr)^{kl} 
 \Bigl( Y_2^{\dagger} T_J Y_1 (0) T^d\Bigr)^{ji} \Bigl( Y_3^{\dagger} T_J Y_4 (0) T^f\Bigr)^{lk}
|0\rangle \nonumber \\
&= \int d\mathcal{T}_S \int d^4 p_s  e^{-i p_s \cdot x} \mathcal{S}^{cdef}_{IJ} 
(\mathcal{T}_S, M, \mu, p_s),
\end{align}
where $\mathcal{S}^{cdef}_{IJ} (\mathcal{T}_S, M, \mu, p_s)$ is given by 
\begin{align} 
\mathcal{S}^{cdef}_{IJ} (\mathcal{T}_S, M, \mu, p_s) &=  \langle 0| \Bigl(
Y_1^{\dagger} T_I Y_2 (0) T^c\Bigr)^{ij} \Bigl( Y_4^{\dagger} T_I Y_3 (0)  T^e\Bigr)^{kl} 
\delta^{(4)} (p_s +\mathcal{P})  \\ 
 & \times \delta \Bigl( \mathcal{T}_S - \sum_{X_s} \mathrm{min} (\{ n_i\cdot p_{X_s} \}) \Bigr)
\Bigl( Y_2^{\dagger} T_J Y_1 (0)  T^d\Bigr)^{ji} \Bigl( Y_3^{\dagger} T_J Y_4 (0)  T^f\Bigr)^{lk}
|0\rangle. \nonumber
\end{align}
Note that we have reshuffled the Wilson lines such that those with the coordinate $x$ are moved 
to the left.

The exponential factor $e^{-ip_s\cdot x}$ in eq.~\eqref{intsoft} disappears by appropriate reparameterization 
transformations~\cite{Ellis:2010rwa}, and
eq.~\eqref{intsoft} can be written as $\int d\mathcal{T}_S S^{cdef}_{IJ} (\mathcal{T}_S, M, \mu)$, where
the soft function for the 2-jettiness is defined as
\begin{align} \label{softdef}
S^{cdef}_{IJ} (\mathcal{T}_S, M, \mu)  &= \langle 0| \Bigl(
Y_1^{\dagger} T_I Y_2 T^c\Bigr)^{ij} \Bigl( Y_4^{\dagger} T_I Y_3 T^e\Bigr)^{kl} \nonumber \\ 
 & \times 
\delta \Bigl( \mathcal{T}_S - \sum_{X_s} \mathrm{min} (\{ n_i\cdot p_{X_s} \} )\Bigr)
\Bigl( Y_2^{\dagger} T_J Y_1 T^d\Bigr)^{ji} \Bigl( Y_3^{\dagger} T_J Y_4 T^f\Bigr)^{lk}
|0\rangle. 
\end{align} 
In this form, the virtual contribution comes from
the contraction of the soft Wilson lines to the left-hand side  or to
the right-hand side of the delta function. The real contribution can be obtained by
contracting the Wilson lines across the delta function.

\subsubsection{Factorized $N$-jettiness in \scone}
Combining all these components, the factorized cross section for the 2-jettiness, 
according to eq.~\eqref{observ}, can be written as
\begin{align} \label{taufac}
\frac{d\sigma}{d\mathcal{T}_2} &= \frac{8}{Q^2} \sum_{IJ}  \int \frac{dz_1}{z_1} \int \frac{dz_2}{z_2}
\int \frac{d^4 p_3}{(2\pi)^3}
\frac{d^4 p_4}{(2\pi)^3}  (2\pi)^4 \delta^{(4)} \Bigl(\frac{\omega_1 n_1}{2} 
+\frac{\omega_2 n_2}{2} - \frac{\omega_3 n_3}{2} -\frac{\omega_4 n_4}{2}\Bigr)
 \nonumber \\
&\times \Bigl(\omega_1 \omega_2 \omega_3 \omega_4 n_1 
\cdot n_4 n_2 \cdot n_3   D_I^*D_J\Bigr)   \int d\mathcal{T}_S
\delta \Bigl( \mathcal{T}_2 -\frac{t_1}{\omega_1} -\frac{t_2}{\omega_2} -\frac{p_3^2}{\omega_3} 
-\frac{p_4^2}{\omega_4} -\mathcal{T}_S \Bigr) \sum_{cdef} k_c k_d k_e k_f 
 \nonumber  \\
&\times \int dt_1 \, B_e^d (t_1, z_1, M, \mu) 
\int dt_2 \, B_{\bar{e}}^c (t_2, z_2,  M, \mu) J_{\mu}^e (p_3^2,  M, \mu)
J_{\bar{\mu} }^f (p_4^2, M, \mu)  S^{cdef}_{IJ} (\mathcal{T}_S, M, \mu) 
\nonumber \\
 &= \frac{8}{Q^2}   \int \frac{dz_1}{z_1} \int \frac{dz_2}{z_2}  \int  d\Phi (\{p_J\} )
(2\pi)^4 \delta^{(4)} \Bigl(\frac{\omega_1 n_1}{2} 
+\frac{\omega_2 n_2}{2} - \frac{\omega_3 n_3}{2} -\frac{\omega_4 n_4}{2}\Bigr)  \int d\mathcal{T}_S 
\nonumber \\
&\times  \sum_{cdef} k_c k_d k_e k_f  \int dt_1 \, B_e^d (t_1, z_1, M, \mu)  
\int dt_2 \, B_{\bar{e}}^c (t_2, z_2, M, \mu) J_{\mu}^e (p_3^2, M, \mu)
J_{\bar{\mu}}^f (p_4^2, M, \mu) \nonumber \\
&\times \sum_{IJ} H_{JI}S^{cdef}_{IJ} (\mathcal{T}_S,  M, \mu) 
\delta \Bigl( \mathcal{T}_2 -\frac{t_1}{\omega_1} -\frac{t_2}{\omega_2} -\frac{p_3^2}{\omega_3} 
-\frac{p_4^2}{\omega_4} -\mathcal{T}_S \Bigr).
\end{align} 
After integrating over the coordinate $x$, the exponential factors  
in eqs.~\eqref{cl1}, \eqref{cl2}, \eqref{cl3} and \eqref{cl4} yield the delta function,  
responsible for the momentum conservation. Note that the last delta function in eq.~\eqref{taufac} 
corresponds to $\delta (\mathcal{T}_2 - g(X))$ in eq.~\eqref{observ}.
The corresponding factorization formula for the $N$-jettiness in QCD in the 
framework of SCET is presented in
refs.~\cite{Stewart:2009yx,Stewart:2010tn}, and this result is an extension including the nonsinglet contributions.
 
The Mandelstam variables $s$, $t$, $u$  are given by $s =(p_1 + p_2)^2 = (p_3+p_4)^2$, 
$t=  (p_1 -p_3)^2 = (p_2 - p_4)^2$,  $u =  (p_1 -p_4)^2 = (p_2 - p_3)^2$, where $p_i$ are the partonic 
momenta. In terms of $\omega_i$, $u$ is given by
$u = \omega_1 \omega_4 n_1 \cdot n_4/2 = \omega_2 \omega_3 n_2 \cdot n_3/2.$
We set the hard coefficients $H_{JI}$ as
\begin{equation} \label{finhard}
H_{JI} = 4u^2 D_I^* D_J = \omega_1 \omega_2 \omega_3 \omega_4 n_1 \cdot n_4 n_2 \cdot n_3
D_I^* D_J \equiv C_I^* C_J,
\end{equation}
where the Wilson coefficients $C_I$ are defined as $C_I = 2u D_I$.
The phase space is denoted as $d\Phi (\{p_J\} )$, which is given by
\begin{equation}
d\Phi (\{p_J\} ) = \prod_J \frac{d^4 p_J}{(2\pi)^3}.
\end{equation}
At tree level, the jet function is proportional to $\delta (p_J^2)$, and when combined with $d^4 p_J /(2\pi)^3$,
it gives $\delta (p_J^2) d^4 p_J /(2\pi)^3 = d^3 \mathbf{p}_J/[2E_J (2\pi)^3]$, which is the phase space for the 
final-state particle $J$.

\subsection{$\mathrm{SCET_{II}}$: $\mathcal{T}^2 \sim M^2 \sim p_c^2 \ll Q\mathcal{T} \ll Q^2$}
In $\mathrm{SCET_{II}}$,  the soft momentum scales as $p_s^{\mu} \sim (\mathcal{T}, \mathcal{T}, \mathcal{T})$,
while the $n$-collinear momentum scales as $p_n^{\mu} \sim (Q, \mathcal{T}, \mathcal{T}^2/Q)$. Therefore 
the small component $p_n^+$ does not contribute to the jettiness, and the factorized form of the 2-jettiness
in Eq.~\eqref{taufac} should be changed. 
If we naively employed the PDFs $f_i^a$ and the fragmentation functions $D_i^a$ instead of
the beam functions and the jet functions, % without considering the matching relation considered above, 
there would be no contribution to the jettiness from these collinear 
functions~\cite{Lustermans:2019plv}.
Schematically, the 2-jettiness in \sctwo\ might be written as
\begin{equation} \label{fakefac}
\frac{d\sigma}{d\mathcal{T}_2} \sim f_e^d ( z_1, \mu) \otimes f_{\bar{e}}^c (z_2,\mu) 
D_{\mu}^e (z_3, \mu)   D_{\bar{\mu}}^f (z_4, \mu) 
\int d\mathcal{T}_S \sum_{IJ} H_{JI} S_{IJ}^{cdef} (\mathcal{T}_S, \mu) 
\delta (\mathcal{T}_2 -\mathcal{T}_S),
\end{equation}
where the phase space and the integration with respect to other variables are omitted.
The imminent problem in this formulation is that the sum of the anomalous dimensions does not cancel.
Here the soft anomalous dimension depends on the
jettiness $\mathcal{T}_S$. (This will be explicitly shown later.) But, if we use the factorization of the 
form in eq.~\eqref{fakefac}, there is no dependence of the anomalous dimensions 
on the jettiness in collinear functions. As a result, the 
total sum of the anomalous dimensions does not cancel.

\begin{figure}[t] 
\begin{center}
\includegraphics[height=7cm]{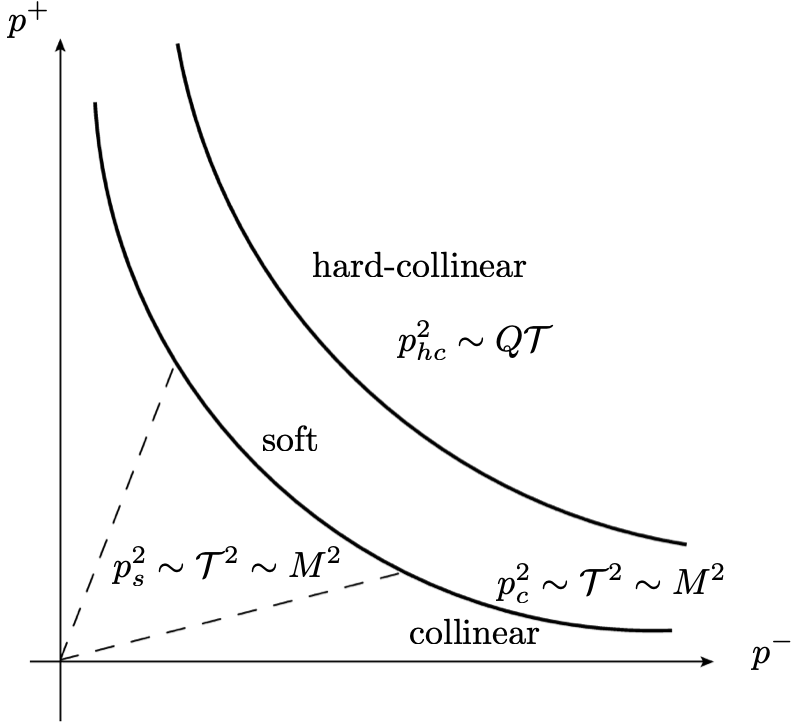} 
\end{center} \vspace{-0.6cm}
\caption{\label{hardcollinear}\baselineskip 3.0ex  In $\mathrm{SCET_{II}}$, the collinear and soft modes lie on
the same mass shell $p_c^2 \sim p_s ^2 \sim M^2 \sim \mathcal{T}^2$.  The hard-collinear modes with
$p_{hc}^2 \sim Q\mathcal{T}$ are integrated out to obtain $\mathrm{SCET_{II}}$ from $\mathrm{SCET_I}$.  } 
\end{figure}

Therefore, care must be taken in obtaining $\mathrm{SCET_{II}}$ from $\mathrm{SCET_I}$.
In fact, the collinear modes in \scone, scaling as $p_c^{\mu} \sim (Q, \sqrt{Q\mathcal{T}}, \mathcal{T})$, 
are integrated out to obtain \sctwo.  We call the collinear modes in \scone\ as the hard-collinear 
modes in \sctwo. In fig.~\ref{hardcollinear}, the hyperbolas for the collinear and soft modes and 
for the hard-collinear modes are shown.  By integrating out the hard-collinear modes, we obtain 
the corresponding matching coefficients between \scone\ and \sctwo.

The relation between the beam function and the PDF can be obtained by the operator product expansion
(OPE) of the relevant operators.  Let us consider the operators  
\begin{equation}
\mathcal{O}_{\ell}^a (t, \omega, M, \mu)=\overline{\ell}_{L} (0)\frac{\fms{\overline{n}}}{2}
T^a \delta ( t +\omega  n \cdot \mathcal{P} )   
[\delta (\omega -\overline{n} \cdot \mathcal{P}) \ell_{L} (0) ],
\end{equation}
of which the matrix elements yield the beam function in eq.~\eqref{beamdef}, and
\begin{align}
\mathcal{Q}_{\ell}^a (\omega, M, \mu) &= \overline{\ell}_{L} (0)\frac{\fms{\overline{n}}}{2}
T^a  [\delta (\omega -\overline{n} \cdot \mathcal{P}) \ell_{L} (0) ], \nonumber \\
\mathcal{Q}_{\bar{\ell}}^a (\omega, M, \mu) &= \mathrm{Tr}\, \Bigl( \frac{\fms{\overline{n}}}{2} T^a 
\ell_L (0) [\delta (\omega -\overline{n} \cdot \mathcal{P})  \overline{\ell}_L (0)] \Bigr), \nonumber \\
\mathcal{Q}_W (\omega, M, \mu) &= -\omega \delta_{bc} \mathcal{B}_{n\perp\mu}^b (0)
 [\delta (\omega -\overline{n} \cdot \mathcal{P})  \mathcal{B}_{n\perp}^{\mu c} (0)],
\end{align}
of which the matrix elements yield the PDFs\footnote{For SU($N$) with $N>2$, 
there should be an additional nonsiglet gauge boson operator $\mathcal{O}_W^a$, where $\delta_{ab}$
is replaced by the structure constant $d_{abc}$. It is also true in eq.~\eqref{fragop} below.}.
 [See eq.~\eqref{defpdf} below.] 

Using the OPE in the limit $M^2/t \rightarrow 0$, we can expand the operators 
$\mathcal{O}_{\ell}^a$ in terms of a sum of the operators $\mathcal{Q}_{\ell}^a$~\cite{Stewart:2010qs} 
\begin{equation} \label{oper}
\mathcal{O}_i^a (t, \omega, M, \mu) =  \sum_{j, b} 
\int \frac{d\omega'}{\omega'} \mathcal{I}_{ij}^{ab} \Bigl(t, \frac{\omega}{\omega'} ,\mu\Bigr) 
\mathcal{Q}_j^b (\omega', M, \mu) + O  ( M^2/t ).
\end{equation}
 Eq.~\eqref{oper} shows the
matching relation between the operators $\mathcal{O}_{\ell}^a$ in \scone, and the operators  
$\mathcal{Q}_{\ell}^b$ in \sctwo, where  $\mathcal{I}^{ab}_{ij}$ are the
corresponding Wilson coefficients. 
When we take the electron matrix element of eq.~\eqref{oper},  
we obtain the OPE for the beam functions, with $z = \omega/\overline{n}\cdot p$, 
\begin{equation}
B_i^a (t, z, M, \mu) = \sum_{j,b} \int_z^1 \frac{dz'}{z'} \mathcal{I}_{ij}^{ab} (t, z/z', \mu) 
f_j^b (z', M, \mu) + O(M^2/t).
\end{equation}
 
We can establish the relation between the semi-inclusive jet functions in \scone, and the fragmentation
functions in \sctwo\ in the same way. Let us consider the operators in \scone
\begin{equation}
O_{\ell}^a (p^2, M, \mu)  = \mathrm{Tr} \Bigl( \ell_L (0) 
\delta (p^2 + \omega n\cdot \mathcal{P})
\delta^{(2)} (\mathcal{P}_{\perp}) [ \delta (\omega + \overline{n}\cdot \mathcal{P}) \overline{\ell}_L (0)] 
\frac{\fms{\overline{n}}}{2} T^a \Bigr), 
\end{equation}
which yield the semi-inclusive jet functions, and the operators in \sctwo 
\begin{align} \label{fragop}
Q_{\ell}^a (M, \mu) &= \mathrm{Tr} \Bigl( \ell_L (0) 
\delta^{(2)} (\mathcal{P}_{\perp}) [ \delta (\omega + \overline{n}\cdot \mathcal{P}) \overline{\ell}_L (0)] 
\frac{\fms{\overline{n}}}{2} T^a \Bigr), \nonumber \\
Q_{\bar{\ell}}^a (M, \mu) &= \overline{\ell}_L (0) \frac{\fms{\overline{n}}}{2} T^a
\delta^{(2)} (\mathcal{P}_{\perp}) 
[ \delta (\omega + \overline{n}\cdot \mathcal{P})  \ell_L (0)],  \nonumber \\
Q_W (M, \mu) &=-\omega \delta_{bc} \mathcal{B}_{n\perp\mu}^b (0)
\delta^{(2)} (\mathcal{P}_{\perp}) 
 [\delta (\omega +\overline{n} \cdot \mathcal{P})  \mathcal{B}_{n\perp}^{\mu c} (0)],
\end{align}
where we will choose the frame such that the transverse momentum is zero in taking the matrix elements.
We employ the OPE in the limit $M^2/p^2 \rightarrow 0$ to expand the operators $O_{\ell}^a$ in 
terms of the operators $Q_{\ell}^a$ as
\begin{equation} \label{opej}
O_i^a (p^2,   M, \mu) =  \sum _{j,b}  \int \frac{d\omega'}{\omega'} \mathcal{J}_{ij}^{ab}
(p^2, \omega/\omega', \mu) Q_j^b (\omega', M, \mu)+ O(M^2/p^2).
\end{equation}
This equation shows the matching relation between $O_i^a$ in \scone, and  
$Q_j^b$ in \sctwo, where  $\mathcal{J}_{ij}^{ab}$ are the corresponding
matching coefficients.
In order to obtain the relation between the semi-inclusive jet function in \scone, and the fragmentation
function in \sctwo, we take the vacuum expectation values, but with the state $i$ included in the 
intermediate states.  The result is written as
\begin{equation}
J_{i} ^a (p^2, \mu) = \sum_{j,b}\int_0^1 dz \int_z^1 \frac{dz'}{z'} \mathcal{J}_{ij}^{ab} (p^2, z/z', \mu) 
D_j^b (z', \mu) + O (M^2/p^2).
\end{equation}
 
In addition, the soft Wilson lines are obtained by integrating out the offshell modes 
when the soft gauge particles are emitted from the collinear source, which are given by
\begin{equation}
S_i =  \sum_{\mathrm{perm.}} \exp\Bigl[-g \frac{n_i\cdot A_s}{n_i\cdot \mathcal{P}} \Bigr],
\end{equation}
where $A_s$ is the soft gauge field. As a result, the soft Wilson line $Y_i$ in \scone\ is replaced 
by  $S_i$, which is the rescaled version in \sctwo.  
 
With these ingredients, the cross section for the 2-jettiness in \sctwo\ is written as
\begin{align}\label{sctwofac}
 \frac{d\sigma}{d\mathcal{T}_2} &=
 \frac{8}{Q^2}   \int \frac{dz_1 dz_2}{z_1 z_2} \int  d\Phi (\{p_J\} )
(2\pi)^4 \delta^{(4)} \Bigl(\frac{\omega_1 n_1}{2} 
+\frac{\omega_2 n_2}{2} - \frac{\omega_3 n_3}{2} -\frac{\omega_4 n_4}{2}\Bigr) 
 \sum_{cdef} k_c k_d k_e k_f  \nonumber \\
&\times \int d\mathcal{T}_S  \sum_{ij, ab}  \int dt_1  \int_{z_1}^1 \frac{dz_1^{\prime}}{z_1^{\prime}} 
\mathcal{I}_{ei}^{da} \Bigl(t_1, \frac{z_1}{z_1^{\prime}}, \mu\Bigr) f_i^a (z_1^{\prime}, \mu)    \int dt_2
\frac{dz_2^{\prime}}{z_2^{\prime}} 
\mathcal{I}_{\bar{e}j}^{cb} \Bigl(t_2, \frac{z_2}{z_2^{\prime}}, \mu\Bigr) f_j^b (z_2^{\prime}, \mu) 
\nonumber \\
&\times \sum_{kl, pq} \int dz_3 \int_{z_3}^1 \frac{dz_3^{\prime}}{z_3^{\prime}} 
\mathcal{J}_{\mu k}^{ep} \Bigl(p_3^2, \frac{z_3}{z_3^{\prime}}, \mu\Bigr) D_k^p (z_3^{\prime}, \mu)  
 \int dz_4 \int_{z_4}^1 \frac{dz_4^{\prime}}{z_4^{\prime}} 
\mathcal{J}_{\bar{\mu}l}^{fq} \Bigl(p_4^2, \frac{z_4}{z_4^{\prime}}, \mu\Bigr) D_l^q (z_4^{\prime}, \mu) \nonumber \\
&\times \sum_{IJ} H_{JI}S^{cdef}_{IJ} (\mathcal{T}_S, \mu) 
\delta \Bigl( \mathcal{T}_2 -\frac{t_1}{\omega_1} -\frac{t_2}{\omega_2} -\frac{p_3^2}{\omega_3} 
-\frac{p_4^2}{\omega_4} -\mathcal{T}_S \Bigr).
\end{align} 
In addition to the contribution of the soft function to the 2-jettiness, note that there are contributions 
from the hard-collinear contributions. The expression in eq.~\eqref{sctwofac} also conforms to the consistent
RG behavior of the 2-jettiness. That is, the sum of the anomalous dimensions of all the factorized parts should be
zero so that the 2-jettiness is independent of the factorization scale.  It cancels only when we use eq.~\eqref{sctwofac}.

We will present the matching coefficients $\mathcal{I}_{ij}^{ab}$ and $\mathcal{J}_{ij}^{ab}$ 
explicitly, but a practical way to calculate the collinear parts in \sctwo\ is to compute the beam functions 
and the jet functions using the power counting
in \scone, and perform the soft zero-bin subtraction. That is, we compute the combination of 
the matching coefficients and the PDF or the fragmentation functions together in \sctwo, which are equivalent 
to the computation of the beam functions and the FJF in \scone. 
 
The factorization for the $N$-jettiness is established both in \scone\ and \sctwo. However, there is an important
caveat that Glauber exchange between spectator partons may violate factorization when the weak charges of the final
states are specified~\cite {Baumgart:2018ntv}. The possible breakdown of the factorization may start at order $\alpha^4$ 
of the magnitude $\sim \alpha^4 \ln^4 (M^2 /Q^2)$, 
and it is due to the fact that the group-theory factors for the exchange of two Glauber gauge bosons in different configurations
across the unitarity cuts
are different and the overall effects do not cancel. This should be considered seriously in ascertaining the factorization
in electroweak interaction, but it is beyond the scope of this paper, and will not be considered here. 
 
\section{Treatment of rapidity divergence\label{rapidity}} 
In SCET, the rapidity divergence shows up because the collinear and the soft  modes reside in disparate 
phase spaces.  When these modes have the same invariant mass, they are distinguished by their rapidities.  
The rapidity divergence appears without regard to the UV and IR
divergences, hence it has to be regulated independently. As mentioned in section~\ref{intro}, there are
various methods to regulate the rapidity divergence~\cite{collins_2011, Idilbi:2007ff,Idilbi:2007yi, Becher:2011dz,
Chiu:2011qc,Chiu:2012ir, Li:2016axz, Ebert:2018gsn}.
In ref.~\cite{Chay:2020jzn}, one
of the authors has constructed consistent rapidity regulators both for the collinear and
the soft sectors, and we use this prescription here.

The essential idea is to attach a regulator of the form
$(\nu/\overline{n}\cdot k)^{\eta}$ for the $n$-collinear field, where the rapidity divergence arises. 
And the rapidity regulator in the soft sector should have the same form as that of the collinear rapidity regulator
because we track the same source of the radiation, which causes the rapidity divergence, 
as in the collinear sector. However, it
can be written in such a way to conform to the expression of the soft Wilson line. As an example, 
let us specify the rapidity regulator for the collinear current 
$\overline{\xi}_{n_1} W_{n_1}S_{n_1}^{\dagger}\Gamma
S_{n_2}W_{n_2}^{\dagger} \xi_{n_2}$ with $n_1\cdot n_2 \sim \mathcal{O} (1)$, which is not necessarily
back-to-back. The collinear and soft Wilson lines $W_{n_i}$ and $S_{n_i}$ are inserted to make 
the current collinear and soft gauge 
invariant. For the collinear Wilson line $W_{n_1}$ and the soft Wilson line $S_{n_2}$,
the modified Wilson lines with the rapidity regulator are given as
\begin{align} \label{rareg} 
W_{n_1} &= \sum_{\mathrm{perm.}} \exp \Bigl[
-\frac{g}{\overline{n}_1\cdot \mathcal{P}} \Bigl(\frac{\nu}{|\overline{n}_1\cdot
\mathcal{P}|}\Bigr)^{\eta} \overline{n}_1 \cdot A_{n_1}\Bigr], \nonumber \\ S_{n_2} 
&=\sum_{\mathrm{perm.}} \exp\Bigl[- \frac{g}{n_2\cdot \mathcal{P}}
\Bigl(\frac{\nu}{|n_2\cdot \mc{P}|} \frac{n_1\cdot n_2}{2}\Bigr)^{\eta} n_2\cdot A_s
\Bigr], 
\end{align} 
where $\mathcal{P}$ is the operator extracting the momentum. 
The remaining Wilson lines $W_{n_2}$ and $S_{n_1}$ can be obtained by switching $n_1$ and $n_2$. 
The point in selecting the rapidity regulator is to trace the same emitted gauge bosons both in the collinear
and the soft sectors, which are eikonalized to produce the Wilson lines. Note that the
rapidity divergences from $W_{n_1}$ and $S_{n_2}$ have the same origin because the collinear and soft
gauge bosons are emitted from the $n_2$-collinear quark for both of the Wilson lines. 
For the soft momentum $k$, in the limit $\overline{n}_1\cdot k\rightarrow \infty$ 
where the rapidity divergence occurs in the soft sector,  it becomes $k^{\mu} \approx
(\overline{n}_1 \cdot k) n_1^{\mu}/2$ and the soft rapidity regulator approaches
\begin{equation} 
\Bigl(\frac{\nu}{n_2\cdot k} \frac{n_1\cdot n_2}{2}\Bigr)^{\eta}
\xrightarrow[\bar{n}_1\cdot k \to \infty]{} \Bigl(\frac{\nu}{\overline{n}_1\cdot k}\Bigr)^{\eta},
\end{equation} 
which has the same form as the collinear rapidity regulator for
$W_{n_1}$.  Another pair possessing the same source of rapidity divergence is
$W_{n_2}^{\dagger}$ and $S_{n_1}^{\dagger}$.
Tracking the same source of the emission of gauge bosons in the soft sector gives the correct
directional dependence in the soft anomalous dimensions, in the sense that they cancel the 
total anomalous dimensions when combined with other factorized 
parts~\cite{Bertolini:2017efs}.

The source of the rapidity divergence can be understood as follows: 
For the soft modes with small rapidities, they cannot recognize the region with large rapidity, 
in which the collinear modes reside. But these  collinear modes
are obtained by traversing the boundary from the soft sector to the collinear sector.
Technically, with the momentum $k^{\mu}  = (k^-, k_{\perp}, k^+)$, 
the rapidity divergence arises when $k^+$ or $k^-$ approaches infinity while $\mathbf{k}_{\perp}^2$ is fixed.
Therefore we modify the region with large rapidity such that the rapidity divergence can be extracted. 

On the other hand, for the $n$-collinear modes, $k^-$ cannot approach infinity in the real contribution because 
it cannot exceed the large scale $Q$. Therefore, in our choice of the rapidity regulators,
there is no rapidity divergence in the naive collinear contribution,
though there appears the divergence associated with the region $k^- \rightarrow 0$. 
However, the true collinear contribution is obtained by performing the zero-bin 
subtraction~\cite{Chiu:2011qc,Chiu:2012ir} in which the collinear contribution in the soft limit is removed to
avoid double counting.  The zero-bin subtraction can be regarded as the matching between the collinear
part with large rapidity and the soft part with small rapidity when the soft part is obtained by integrating out
the region with large rapidity.  

The divergence in the collinear part as $k^- \rightarrow 0$ with fixed $\mathbf{k}_{\perp}^2$ is cancelled 
by the zero-bin subtraction in analogy to the cancellation of the IR divergence in matching.  
Note that the rapidity divergence from the collinear sector
has the opposite sign compared to the rapidity divergence in the soft sector. It is the reason why
the rapidity divergences cancel when the collinear and the soft sectors
are combined. It is consistent with the fact that the full QCD does not have any rapidity divergence 
since there is no such kinematic
constraint, separating the collinear and the soft modes.  However, the structure
of rapidity divergence and its evolution in each sector sheds light on the intricate
nature of the theory.

In electroweak interaction, in which the weak nonsinglets can appear in each factorized part,
the behavior of the rapidity divergence is strikingly different from QCD. In
collinear quantities such as the jet function, and the FJF, etc., the
rapidity divergence cancels for the gauge singlets in each function. Technically, this happens due to the
fact that the real and virtual contributions to the rapidity divergence are equal, but with the opposite sign. 
For gauge nonsinglets, the group theory factors are
different for real and virtual contributions, hence producing nontrivial rapidity
divergence in each sector. We will present the resummed result at next-to-leading logarithmic (NLL) 
order for the 2-jettiness in $e^- e^+\rightarrow \mu^- \ \mathrm{jet}\ \mu^+\ \mathrm{jet}+X$, 
evolving both under the renormalization
scale and the rapidity scale.  The cancellation of the rapidity divergence when all the
contributions are added becomes more sophisticated.  But the non-cancellation of the rapidity divergence
in each sector for the gauge singlets induces the additional evolution with respect to the rapidity scale 
in the process of resummation.

\section{Collinear functions\label{collinear}} 
\subsection{Beam function and PDF\label{incol}} 
The beam functions for QCD are defined in 
refs.~\cite{Stewart:2009yx,Stewart:2010qs}, and they describe the initial-state radiations from 
the incoming particles. We can extend them to those for the weak
interaction. The singlet and nonsinglet beam functions are defined as the matrix elements
with a target electron $e$ in our case, which are given as [See eq.~\eqref{beamdev}.]
\begin{equation} \label{beams}
B_e^a (t, z = \omega/P^-, M, \mu) =
\langle e (P)| \theta (\omega) \overline{\ell}_n (0) \delta (t+ \omega
n\cdot \mathcal{P}) \frac{\fms{\overline{n}}}{2} T^a\Bigl[ \delta (\omega
-\overline{n}\cdot \mathcal{P} ) \ell_n (0)\Bigr]|e(P)\rangle, 
\end{equation} 
where $T^0 = 1$ and $T^a= t^a$ are the weak generators.  The beam functions at tree level are given as
\begin{equation}
B^{a(0)}_e (t, z, M, \mu) = \delta (t) \delta (1-z) \mathrm{Tr} (T^a P_e),
\end{equation}
where  $P_e$ is the projection operator in the weak charge space to project out the electron  $e$,
because the beam function is initiated by the incoming electron. 
Otherwise, the nonsinglet beam function vanishes because the contributions from
the electron and the electron neutrino cancel. 
In QCD, since the initial state consists of a color singlet, say, a proton, the color average 
is performed in the beam function. However it is not true in this case. If we consider the 
imaginary situation in which the color can be measured in QCD, we should define the beam
function with fixed color charges. The beam functions for antileptons can be defined accordingly as
\begin{equation}  
B_{\bar{e}}^a (t, z,  M, \mu) = \mathrm{Tr} \,  \langle e^+ | \theta(\omega)\ell_{L} (0) \delta (t + \omega n 
\cdot \mathcal{P}) [\delta (\omega -\overline{n} \cdot \mathcal{P}) \overline{\ell}_{L} (0) 
\frac{\fms{\overline{n}}}{2} T^a] |e^+\rangle.
\end{equation}

\begin{figure}[b] 
\begin{center}
\includegraphics[height=4.2cm]{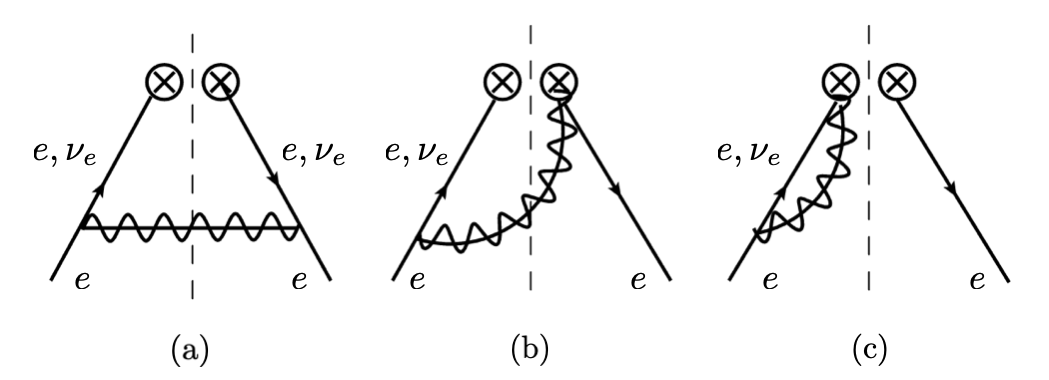} 
\end{center} \vspace{-0.6cm}
\caption{\label{beampdf}\baselineskip 3.0ex Feynman diagrams for the beam function and the
PDF at one loop. The mirror images of (b) and (c) are omitted. The wavy lines with solid lines 
denote collinear gauge bosons. The dashed lines denote the final cuts.
For the beam function, the virtuality $t$ and the longitudinal momentum fraction $z$ are measured,
while only the momentum fraction $z$ is measured for the PDF.}
\end{figure}

The Feynman diagrams for the beam function 
at order $\alpha$ are shown in figure~\ref{beampdf}. We choose the reference frame in 
which the transverse momentum of the incoming particle is
zero. Our computational method is different from what was performed in
ref.~\cite{Stewart:2010qs}. Here all the massless fermions are on the mass shell with the nonzero
gauge boson mass $M$.  In our computation, we express the
beam functions and the PDFs in terms of the variables $t$ and $z =
\omega/p^-$, while we put $t^{\prime} = -\omega p^+= -z p^+ p^- =0$. Nonzero $t'$ plays 
the role of an IR regulator in ref.~\cite{Stewart:2010qs}.  
In our calculation, the massless fermions are on the mass shell and 
there is no IR divergence because of the physical nonzero gauge boson mass $M$.

The radiative correction of the singlet beam function at
order $\alpha$ is different in finite terms from the result in ref.~\cite{Stewart:2010qs},
but the matching coefficients between the beam functions and the PDFs turn out to be the same. 
However, what is new here is that we include the nonsinglet
contributions which show distinct behavior, compared to the singlet contributions. 
In \scone, since $t \gg M^2$, we take the limit of small $M$ in the final result. In \sctwo,
the hard-collinear contribution from the matching coefficients $\mathcal{I}$ contains
$t$, and we also take the small $M$ limit.
 
The contribution of figure~\ref{beampdf} (a) apart from the group theory factor is given by 
\begin{align} \label{mabeam} 
M_a &= 2\pi g^2 \mums \int \dd{\ell} \frac{2 (2-2\eps) p^- \bm{\ell}_{\perp}^2}{(p-\ell)^2} 
\delta ( t-\omega \ell^+) \delta (\omega -p^- +\ell^-)   \nonumber \\
&\times \delta (\ell^2 -M^2) \theta (\ell^-) \theta (p^- -\ell^-)
\nonumber \\
&=\frac{\alpha}{2\pi} \theta\Bigl( (1-z) t-zM^2\Bigr) \theta (z) \theta
(1-z) \frac{(1-z)t -zM^2}{(t-zM^2)^2}\nonumber \\
&\longrightarrow (1-z) \theta (z) \theta (1-z) \Bigl[ \Bigl( -1 + \ln \frac{(1-z)
\mu^2}{z^2 M^2}\Bigr) \delta (t) + \frac{1}{\mu^2} \mathcal{L}_0
\Bigl(\frac{t}{\mu^2}\Bigr) \Bigl]. 
\end{align} 
The last result is obtained by taking the small $M$ limit. The detailed calculation taking this limit 
is presented in appendix~\ref{beamzero}.
There is no zero-bin contribution for $M_a$ at leading order. 

The naive collinear contribution in figure~\ref{beampdf} (b), 
with the momentum $\ell$ of the gauge boson, yields 
\begin{align} \label{mbbeam} 
\tilde{M}_b &= -4\pi g^2  \mums \theta (t) \int \dd{\ell} \frac{p^- (p^- -\ell^- )}{\ell^- (p-\ell)^2}
\Bigl( \frac{\nu}{\ell^-}\Bigr)^{\eta} \delta ( t-\omega \ell^+) \delta (\omega -p^- +\ell^-)
\nonumber \\
&\times \delta (\ell^2 -M^2) \theta (\ell^-) \theta (p^- -\ell^-) \nonumber \\
&= \frac{\alpha}{2\pi} \dms \Bigl( \frac{\nu}{p^-}\Bigr)^{\eta} \theta (z) \theta (1-z) \theta (t)
\Bigl( \frac{(1-z)t}{z} -M^2\Bigr)^{-\eps} \frac{z}{(1-z)^{1+\eta}} \frac{1}{t-zM^2}
\nonumber \\
& =\frac{\alpha}{2\pi} \frac{z}{1-z} \frac{1}{t-zM^2} \theta \Bigl( (1-z)t -z M^2\Bigr) \theta(z) \theta(1-z)\nonumber \\
&\longrightarrow \frac{\alpha}{2\pi} \Bigl[ \delta (t) \delta (1-z) \frac{1}{2} \ln^2 \frac{\mu^2}{M^2}+
\delta (t) z\Bigl(\ln \frac{\mu^2}{z^2 M^2} \mathcal{L}_0 (1-z) + \mathcal{L}_1 (1-z)
\Bigr) \nonumber \\ 
& + \delta (1-z) \Bigl( \frac{1}{\mu^2} \mathcal{L}_0 \Bigl(
\frac{t}{\mu^2}\Bigr) \ln \frac{\mu^2}{M^2} +\frac{1}{\mu^2} \mathcal{L}_1 \Bigl(
\frac{t}{\mu^2}\Bigr) \Bigr) + \frac{z}{\mu^2} \mathcal{L}_0 \Bigl( \frac{t}{\mu^2}\Bigr)
\mathcal{L}_0 (1-z)\Bigr]. 
\end{align} 
We put $\eps=\eta =0$ because there is neither UV nor rapidity divergence.

The zero-bin contribution  from figure~\ref{beampdf} (b), in which the rapidity regulator in eq.~\eqref{rareg} is implemented, is  given by 
\begin{align} 
M_b^{\varnothing} &= 
-4\pi g^2 \mums \theta (t) \int\dd{\ell} \frac{(p^-)^2}{\ell^- (p -\ell)^2} 
\Bigl( \frac{\nu}{\ell^-}\Bigr)^{\eta} \delta ( t-\omega \ell^+) \delta (\omega -p^-)
 \nonumber \\
&\times \delta (\ell^2 -M^2) \theta(\ell^-) \theta (p^-) \nonumber \\
&= \frac{\alpha}{2\pi} \dms \delta (1-z) \frac{\theta (t)}{t}  \int_{\omega M^2/t}^{\infty} 
\frac{d\ell^-}{\ell^-}  \Bigl( \frac{\nu}{\ell^-}\Bigr)^{\eta}  \Bigl( \frac{t\ell^-}{\omega} -M^2\Bigr)^{-\eps}
\nonumber  \\
&= \frac{\alpha}{2\pi}  \Bigl( \frac{\mu^2 e^{\gamma_{\mathrm{E}}}}{M^2}\Bigr)^{\eps}
\Bigl( \frac{\nu \mu^2}{\omega M^2} \Bigr)^{\eta} \frac{\Gamma (\eps+\eta)}{\Gamma(1+\eta)}
\Bigl[ \frac{1}{\eta} \delta (t) +\frac{1}{\mu^2} \mathcal{L}_0 \Bigl(\frac{t}{\mu^2}\Bigr) \Bigr]
\nonumber \\
&=\frac{\alpha}{2\pi} \Bigl\{ \delta (t) \Bigl[
\Bigl(\frac{1}{\eta} + \ln \frac{\nu}{\omega} \Bigr) \Bigl( \frac{1}{\eps} +\ln
\frac{\mu^2}{M^2}\Bigr) -\frac{1}{\eps^2} +\frac{1}{2} \ln^2 \frac{\mu^2}{M^2}
+\frac{\pi^2}{12} \Bigr]\nonumber \\ 
& +\frac{1}{\mu^2} \mathcal{L}_0 
\Bigl(\frac{t}{\mu^2}\Bigr) \Bigl( \frac{1}{\eps} + \ln \frac{\mu^2}{M^2}
\Bigr)\Bigr\}\delta (1-z),
\end{align} 
where the following relation is used.
\begin{equation}
\frac{(\mu^2)^{-\eta}}{t^{1-\eta}} = \frac{1}{\eta} \delta (t) +\frac{1}{\mu^2} \mathcal{L}_0 \Bigl(\frac{t}{\mu^2}\Bigr)
+\mathcal{O}(\eta).
\end{equation}
The distributions $\mathcal{L}_n (z)$ are listed in appendix~\ref{sijf}~\cite{Ligeti:2008ac}. 
 
 By performing the contour integral in the complex $\ell^+$-plane, the naive collinear contribution 
of figure~\ref{beampdf} (c) is given as 
\begin{align} 
\tilde{M}_c &= 2ig^2 \mums \delta (t) \delta (1-z) 
\int \dd{l} \frac{p^- -\ell^-}{\ell^- (\ell^2 -M^2+i0) \Bigl((p-\ell)^2 +i0\Bigr)} 
\Bigl( \frac{\nu}{\ell^-}\Bigr)^{\eta}
\nonumber \\
&= -\frac{\alpha}{2\pi}  \Bigl( \frac{\mu^2 e^{\gamma_{\mathrm{E}}}}{M^2}\Bigr)^{\eps} 
\Gamma (\eps) \Bigl( \frac{\nu}{\omega}\Bigr)^{\eta} 
\delta (t) \delta (1-z) \int_0^1 dx \frac{(1-x)^{1-\eps}}{x^{1+\eta}}, 
\end{align}
where $x = \ell^-/p^-$. And the zero-bin contribution is given as
\begin{align}
M_c^{\varnothing} &= 2ig^2 \mums \delta (t) \delta (1-z) 
\int \dd{l} \frac{p^-}{\ell^- (\ell^2 -M^2+i0) (-p^-\ell^+ + i0)} \Bigl( \frac{\nu}{\ell^-}\Bigr)^{\eta}
\nonumber \\
&= -\frac{\alpha}{2\pi}  \Bigl( \frac{\mu^2 e^{\gamma_{\mathrm{E}}}}{M^2}\Bigr)^{\eps} 
\Gamma (\eps) \Bigl( \frac{\nu}{\omega}\Bigr)^{\eta} \delta (t) \delta (1-z)\int_0^{\infty}  \frac{dx}{x^{1+\eta}}. 
\end{align}
The net contribution with the zero-bin subtraction is given as
\begin{align}
M_c &= \tilde{M}_c -M_c^{\varnothing} \nonumber \\
&= -\frac{\alpha}{2\pi}  \Bigl( \frac{\mu^2 e^{\gamma_{\mathrm{E}}}}{M^2}\Bigr)^{\eps} 
\Gamma (\eps) \Bigl( \frac{\nu}{\omega}\Bigr)^{\eta} \delta (t) \delta (1-z)
\Bigl( \int_0^1 dx \frac{(1-x)^{1-\eps} -1}{x^{1+\eta}}
-\int_1^{\infty}  \frac{dx}{x^{1+\eta}}\Bigr)
\nonumber \\
&=\frac{\alpha}{2\pi} \delta (t) \delta (1-z) \Bigl[ \Bigl( \frac{1}{\eps} +\ln
\frac{\mu^2}{M^2}\Bigr) \Bigl( \frac{1}{\eta} +\ln \frac{\nu}{\omega} +1\Bigr) +
1-\frac{\pi^2}{6} \Bigr],
\end{align} 
where the zero-bin contribution is split such that there is no divergence near $x=0$. 
Note that the rapidity divergence occurs at
large $\ell^-$ region due to the zero-bin subtraction which cancels the divergence at
small $\ell^-$ from the naive collinear contribution. The result is consistent with
ref.~\cite{Chay:2020jzn}. The zero-bin contributions can be computed for the jet functions, FJF, etc. in the same
spirit. The wavefunction renormalization $Z^{(1)}$ and 
the residue $R^{(1)}$ at one loop are given by 
\begin{equation} \label{wfren}
 Z^{(1)} +R^{(1)} = \frac{\alpha}{2\pi} \Bigl(
-\frac{1}{2\eps} -\frac{1}{2} \ln \frac{\mu^2}{M^2} +\frac{1}{4}\Bigr). 
\end{equation}

We express the singlet and nonsinglet beam functions $B^0$ and $B^a$ in terms of $B_s$ and $B_n$ by extracting and separating
the group theory factors as\footnote{We will also separate the group theory factors and express the singlet and 
nonsinglet functions in a similar way for the PDFs 
[eq.~\eqref{pdf}], the matching coefficients between the beam functions and the PDFs [eq.~\eqref{iij}], the jet functions 
[eq.~\eqref{jetex}], the FJFs [eq.~\eqref{fjfex}], and the matching coefficients between the fragmentation functions and the FJFs
[eq.~\eqref{jij}].
}
\begin{equation} 
B^0_e (t,z,  M, \mu) = B_s (t, z, M, \mu) \mathrm{Tr} (T^0 P_e), \ 
B^a_e (t, z, M, \mu) = B_n (t, z, M, \mu) \mathrm{Tr} (T^a P_e).
\end{equation}

The bare beam functions $B_s$ and $B_n$ are given at NLO as
\begin{align}
B_s^{(1)} (t, z, M, \mu) &= C_F \Bigl(M_a +2(\tilde{M}_b
-M_b^{\varnothing}) + 2 M_c +(Z^{(1)}+R^{(1)}) \delta (t) \delta (1-z) \Bigr)  \\ 
&=\frac{\alpha C_F}{2\pi} \Bigl\{  \delta (t) \delta (1-z) \Bigl( \frac{2}{\eps^2} +\frac{3}{2\eps}  
+\frac{9}{4} -\frac{\pi^2}{2}\Bigr) \nonumber \\
&+\delta (t) \Bigl[ P_{\ell \ell}(z) \ln \frac{\mu^2}{z^2 M^2}  + (1+z^2) \mathcal{L}_1 (1-z) -(1-z) \theta (z)
\theta (1-z) \Bigr]\nonumber \\
& + \delta (1-z) \Bigl[ -\frac{2}{\eps}  \frac{1}{\mu^2} \mathcal{L}_0
\Bigl(\frac{t}{\mu^2}\Bigr) +\frac{2}{\mu^2} \mathcal{L}_1
\Bigl(\frac{t}{\mu^2}\Bigr) \Bigr]  + (1+z^2) \mathcal{L}_0 (1-z) 
\frac{1}{\mu^2} \mathcal{L}_0 \Bigl( \frac{t}{\mu^2}\Bigr)\Bigr\}, \nonumber \\
B_n^{(1)} (t, z, M, \mu) &=\Bigl[\Bigl(C_F -\frac{C_A}{2}\Bigr)  
\Bigl(M_a +2(\tilde{M}_b -M_b^{\varnothing}) \Bigr)  + C_F
\Bigl(2M_c +(Z^{(1)}+R^{(1)}) \delta (t) \delta (1-z) \Bigr) \Bigr]\nonumber \\ 
&= B_{s} (t, z, M, \mu) \nonumber \\
& -\frac{\alpha C_A}{4\pi} \Bigl\{  -2 \delta (t) \delta (1-z) \Bigl[ \Bigl(\frac{1}{\eta} +\ln \frac{\nu}{\omega}\Bigr)\Bigl(
\frac{1}{\eps} +\ln \frac{\mu^2}{M^2}\Bigr) -\frac{1}{\eps^2} +\frac{\pi^2}{12} \Bigr] \nonumber \\ 
&+ \delta (t) \Bigl[ (1+z^2) \mathcal{L}_0 (1-z) \ln \frac{\mu^2}{z^2 M^2} + (1+z^2) \mathcal{L}_1 (1-z)
-(1-z) \theta (z) \theta (1-z)\Bigr] \nonumber \\
&+\delta (1-z) \Bigl[-\frac{2}{\eps} \frac{1}{\mu^2} \mathcal{L}_0
\Bigl(\frac{t}{\mu^2}\Bigr)  +\frac{2}{\mu^2} \mathcal{L}_1 \Bigl(\frac{t}{\mu^2}\Bigr) \Bigr]
+ (1+z^2) \mathcal{L}_0 (1-z) \frac{1}{\mu^2} \mathcal{L}_0
\Bigl(\frac{t}{\mu^2}\Bigr) \Bigr\}. \nonumber
\end{align} 
Here the splitting function $P_{\ell \ell }(z)$ for $\ell \rightarrow \ell W$ is the same as the quark
splitting function $P_{qq} (z)$, and is given by 
\begin{equation} 
P_{\ell\ell} (z) = P_{qq} (z) = \mathcal{L}_0 (1-z) (1+z^2) +\frac{3}{2}\delta (1-z) = \Bigl[ \theta (1-z)
\frac{1+z^2}{1-z}\Bigr]_+. 
\end{equation} 
Note that the group theory factors for 
the real emission ($M_a$, $M_b$) and for the virtual
correction ($M_c$, $Z^{(1)}$ and $R^{(1)}$) are the same for the singlet, while they are
different for the nonsinglet. Due to this fact, the rapidity divergence cancels in the singlet,
while it does not in the nonsinglet.
 
The $\mu$-anomalous dimensions $\gamma^{\mu}_B$ and the $\nu$-anomalous dimensions
$\gamma^{\nu}_B$ of the beam functions  are given as 
 \begin{align} \label{beamanom}
&\gamma_{Bs}^{\mu} =-2C_F \Gamma_c 
 \frac{1}{\mu^2} \mathcal{L}_0
\Bigl(\frac{t}{\mu^2}\Bigr)-2\gamma_{\ell} \delta (t), \ \gamma_{Bs}^{\nu}=0 \nonumber \\ 
&\gamma_{Bn}^{\mu}
= \gamma_{B_s}^{\mu} +C_A \Gamma_c \Bigl[ \delta (t) \ln \frac{\nu}{\omega} +
\frac{1}{\mu^2} \mathcal{L}_0 \Bigl(\frac{t}{\mu^2}\Bigr) \Bigr], \ \gamma^{\nu}_{Bn} =
C_A \Gamma_c\delta (t) \ln \frac{\mu}{M},
\end{align} 
where $\gamma_{\ell}^{(1)} =-3C_F /(4\pi)$ at NLO.

The 2-jettiness in eqs.~\eqref{taufac} and \eqref{sctwofac} is expressed in terms 
of the convolution of the collinear and the soft 
functions, but it is convenient to express it in terms of the Laplace transform because 
the 2-jettiness is expressed in terms of the products. 
For the beam function, we make a Laplace transform with respect
to the jettiness $k=t/\omega$, which is written as
\begin{equation}
\tilde{B}_i \Bigl( \ln \frac{\omega Q_L}{\mu^2} ,z, M, \mu \Bigr) = \int_0^{\infty} dk \, e^{-s k} 
B_i (\omega k, z, M, \mu), \ s  = \frac{1}{e^{\gamma_{\mathrm{E}}}  Q_L}.
\end{equation}
When we make a Laplace transform of the various collinear functions, each part should contain the 
same scale, and we choose 
$s =1/(Q_L e^{\gamma_{\mathrm{E}}})$. Here $Q_L$ is an arbitrary scale involved in the Laplace transform.
Various distributions appearing in 
eq.~\eqref{beamanom} are expressed as regular functions in the Laplace transforms. 
(See appendix~\ref{laplace}.) As we will show later,
the evolution of the jettiness is independent of the factorization scale $\mu_F$, as well as $Q_L$. 

The anomalous dimensions of the Laplace-transformed beam functions are given as 
\begin{align} 
\tilde{\gamma}_{Bs}^{\mu} &= 2  C_F\Gamma_{c} \ln \frac{\mu^2}{\omega Q_L} -2\gamma_{\ell},
\  \tilde{\gamma}_{Bs}^{\nu} =0,    \nonumber \\ 
\tilde{\gamma}_{Bn}^{\mu} &= \tilde{\gamma}_{Bs}^{\mu} - C_A \Gamma_{c } \ln
\frac{\mu^2}{\nu Q_L},   \
\tilde{\gamma}_{Bn}^{\nu} = C_A \Gamma_c \ln \frac{\mu}{M}.  \label{beamanon}
\end{align}
To be precise, these anomalous dimensions are those of the beam functions in \scone, but in \sctwo, 
they are regarded as the anomalous dimensions of the combination of the matching coefficients and the PDFs.
Here  $\Gamma_c(\alpha)$ is the cusp anomalous dimension~\cite{Korchemsky:1987wg,Korchemskaya:1992je}, 
which can be expanded as
\begin{equation} 
\Gamma_c (\alpha) = \frac{\alpha}{4\pi}
\Gamma_c^0 + \Bigl( \frac{\alpha}{4\pi}\Bigr)^2 \Gamma_c^1 +\cdots, 
\end{equation} 
with
\begin{equation} 
\Gamma_c^0 =4, \ \Gamma_c^1 = \Bigl( \frac{268}{9}-\frac{4}{3}\pi^2
\Bigr) C_A -\frac{40 n_f}{9}. 
\end{equation} 
To NLL accuracy, the cusp anomalous dimension to two loops is needed.
 
 The PDFs are defined in terms of the matrix elements with a target $e$  as
\begin{equation} \label{defpdf}
 f^a_e (z =\omega/P^-, M, \mu) = \langle e (P)|\theta (\omega) \overline{\ell}_n (0)
\frac{\fms{\overline{n}}}{2} T^a\Bigl[ \delta (\omega -\overline{n}\cdot \mathcal{P} )
\ell_n (0)\Bigr]|e(P)\rangle,
\end{equation} 
and it is normalized at tree level as
\begin{equation}
f_e^{a(0)} (z) =\delta (1-z) \mathrm{Tr} (T^a P_e).
\end{equation}
 
The Feynman diagrams for the PDFs at one loop are shown in figure~\ref{beampdf}. 
Note that the Feynman diagrams are the same as those for the beam functions, but the
measured quantities are different. Including the zero-bin subtractions, the matrix elements 
apart from the group theory factors are given as 
\begin{align}  \label{pdfmat}
M_a &= \frac{\alpha}{2\pi} (1-z)
\Bigl( \frac{1}{\eps} -2 +\ln \frac{\mu^2}{z M^2}\Bigr), \nonumber \\ 
M_b &= \frac{\alpha}{2\pi}
\Bigl( \frac{1}{\eps} +\ln \frac{\mu^2}{z M^2}\Bigr) \Bigl[ -\delta (1-z)
\Bigl(\frac{1}{\eta} + \ln \frac{\nu}{\omega} \Bigr) +z\mathcal{L}_0 (z)\Bigr], \nonumber
\\ M_c &= \frac{\alpha}{2\pi} \delta (1-z) \Bigl[\Bigl(\frac{1}{\eps} + \ln
\frac{\mu^2}{M^2} \Bigr) \Bigl( \frac{1}{\eta} + \ln \frac{\nu}{\omega} +1\Bigr) +
1-\frac{\pi^2}{6}\Bigr].  
\end{align} 

The singlet and nonsinglet PDFs are expressed in terms of $f_{\ell s}$ and $f_{\ell n}$ as
\begin{equation} \label{pdf}
f_{\ell}^0 (z, M, \mu) = f_{\ell s} (z, M, \mu)  \mathrm{Tr} (T^0 P_{\ell}),  
\ f_{\ell}^a (z, M, \mu) = f_{\ell  n} (z, M, \mu)  \mathrm{Tr} (T^a P_{\ell}),
\end{equation}
where $f_{\ell s}$ and $f_{\ell n}$ at NLO are given as
\begin{align} 
f_{\ell s}^{(1)} (z, M, \mu) &= C_F \Bigl(M_a + 2M_b + 2M_c +(Z^{(1)}+R^{(1)}) \delta(1-z)\Bigr) \nonumber \\ 
&=\frac{\alpha C_F}{2\pi} \Bigl[ \Bigl(\frac{1}{\eps} + \ln \frac{\mu^2}{zM^2}\Bigr)
P_{\ell\ell}(z) +\Bigl( \frac{9}{4}-\frac{\pi^2}{3}\Bigr) \delta (1-z) -2 (1-z) \theta (1-z)\Bigr], \nonumber \\
 f_{\ell n}^{(1)} (z, M, \mu) &= \Bigl(C_F -\frac{C_A}{2}\Bigr) (M_a + 2 M_b) +C_F \Bigl(2 M_c
+(Z^{(1)}+R^{(1)}) \delta(1-z)\Bigr) \nonumber \\
&= f_{\ell s}(z, M, \mu) + \frac{\alpha C_A}{2\pi}\Bigl[ \delta
(1-z) \Bigl( \frac{1}{\eps} + \ln \frac{\mu^2}{M^2}\Bigr) \Bigl(\frac{1}{\eta} +\ln
\frac{\nu}{\omega}\Bigr)+(1-z) \theta (1-z) \nonumber \\ 
& - \frac{1}{2}\Bigl(\frac{1}{\eps} +\ln
\frac{\mu^2}{zM^2} \Bigr) (1+z^2)\mathcal{L}_0 (1-z) \Bigr].
\end{align} 
Here the rapidity divergence shows up in the
nonsinglet PDF for the same reason as in the beam functions. It coincides with the result
in ref.~\cite{Manohar:2018kfx}. The $\mu$- and $\nu$-anomalous dimensions of the PDFs are
given as 
\begin{align}  \label{pdfgam}
\gamma_{fs}^{\mu} &= C_F \Gamma_c  P_{\ell\ell} (z), \
\gamma_{fs}^{\nu} =0 \\ \gamma_{fn}^{\mu} &= \gamma_{fs}^{\mu} +C_A \Gamma_c 
\Bigl( \delta (1-z) \ln \frac{\nu}{\omega} -\frac{1}{2} (1+z^2) \mathcal{L}_0
(1-z)\Bigr), \ \gamma_{fn}^{\nu} = C_A \Gamma_c  \ln \frac{\mu}{M}. \nonumber 
\end{align}

It is noteworthy to compare the distinction between QCD and the weak interaction. The results
for the singlets correspond to QCD and there is only a single logarithm in the singlet PDF, and it satisfies the
usual DGLAP equation
\begin{equation}
\frac{d f_{\ell s}(z, M, \mu)}{d\ln \mu} = \frac{\alpha}{\pi} \int \frac{dz'}{z'}  \sum_{i=\ell, W} P_{\ell i} 
\Bigl(\frac{z}{z'}\Bigr)  f_{i s} (z', M, \mu ),
\end{equation}
where $P_{\ell W} (z)$ is the analog of the splitting function $P_{qg} (z)$ in QCD. 
On the other hand, the nonsinglet 
PDF shows the double logarithms due to the mismatch of the real and virtual contributions, along with
the rapidity divergence. Due to the double logarithms, the nonsinglet PDF satisfies more complicated
RG equations. It happens to all the collinear functions and the soft function, which necessitates the introduction
of \sctwo.

The beam functions are related to the PDFs as  
\begin{equation} \label{beammat}
B_i^a (t, z, M, \mu) =\sum_{j,b} \int_z^1 \frac{d\xi}{\xi} \mathcal{I}_{ij}^{ab} \Bigl( t, \frac{z}{\xi}, \mu\Bigr) f_j^b
(\xi, M, \mu), 
\end{equation} 
where $\mathcal{I}_{ij}^{ab}$ are the matching coefficients, which describe the collinear initial-state radiation
and can be computed perturbatively.  Here $i$, $j$ are the indices for particle species, and $a$, $b$ 
are the weak indices.  The only nonzero matching coefficients are those which are diagonal in weak-charge space,  
from which the singlet and nonsinglet matching coefficients 
$\mathcal{I}_{\ell\ell}^s$ and $\mathcal{I}_{\ell\ell}^{n}$  are defined as
\begin{equation} \label{iij}
\mathcal{I}_{\ell\ell}^s (t,z,\mu)=\mathcal{I}_{\ell\ell}^{00} (t,z,\mu) , \ \ \  
\mathcal{I}_{\ell\ell}^n(t,z,\mu)=\mathcal{I}_{\ell\ell}^{aa} (t,z,\mu),  \ \ \ \mathcal{I}_{\ell\ell}^{a0} (t,z,\mu)=0. 
\end{equation}
 
The matching coefficients $\mathcal{I}_{\ell\ell}^s$ for the singlet and $\mathcal{I}_{\ell\ell}^n$ for the nonsinglet at NLO are given as 
\begin{align} \label{ill}
&\mathcal{I}_{\ell\ell}^{s(1)} (t,z,\mu) = B_{\ell s}^{(1)} (t,z, M, \mu) -f_{\ell s}^{(1)} (z, M, \mu) \delta (t) \nonumber \\ 
&=\frac{\alpha C_F}{2\pi}\Bigl\{ -\frac{\pi^2}{6} \delta (t) \delta (1-z)  
+\delta (t) \Bigl[  (1+z^2) \mathcal{L}_1 (1-z)   
+\theta (1-z) \Bigl( 1-z - \frac{1+z^2}{1-z} \ln z\Bigr)\Bigr]\nonumber \\
&+\Bigl(P_{\ell\ell} (z) -\frac{3}{2}\delta (1-z) \Bigr) \frac{1}{\mu^2} \mathcal{L}_0
\Bigl( \frac{t}{\mu^2}\Bigr) + \frac{2}{\mu^2} \mathcal{L}_1 \Bigl( \frac{t}{\mu^2}\Bigr)
\delta (1-z) \Bigr\}, \nonumber \\
&\mathcal{I}_{\ell\ell}^{n(1)} (t,z,\mu)  =B_{\ell n}^{(1)} (t,z, M, \mu) 
-f_{\ell n}^{(1)} (z, M, \mu) \delta (t) = \frac{C_F-C_A/2}{C_F} \mathcal{I}_{\ell \ell s}^{(1)}.
\end{align} 
The finite terms in the beam functions and PDFs are different compared to
the result in ref.~\cite{Stewart:2010qs} due to the presence of the gauge boson mass $M$, but
the matching coefficient $\mathcal{I}_{\ell\ell}^{s(1)}$ is the same.  Note that there is no dependence
on $M$ in the matching coefficients, because they should be independent of the low-energy physics.
The matching coefficient $\mathcal{I}_{\ell\ell}^{n(1)}$
for the nonsinglet is new, but interestingly enough, it is proportional to $\mathcal{I}_{\ell\ell}^{s(1)}$ for the singlet.

\subsection{Semi-inclusive jet functions\label{outcol}} 
\begin{figure}[b] 
\vspace{-0.2cm} 
\begin{center}
\includegraphics[height=4cm]{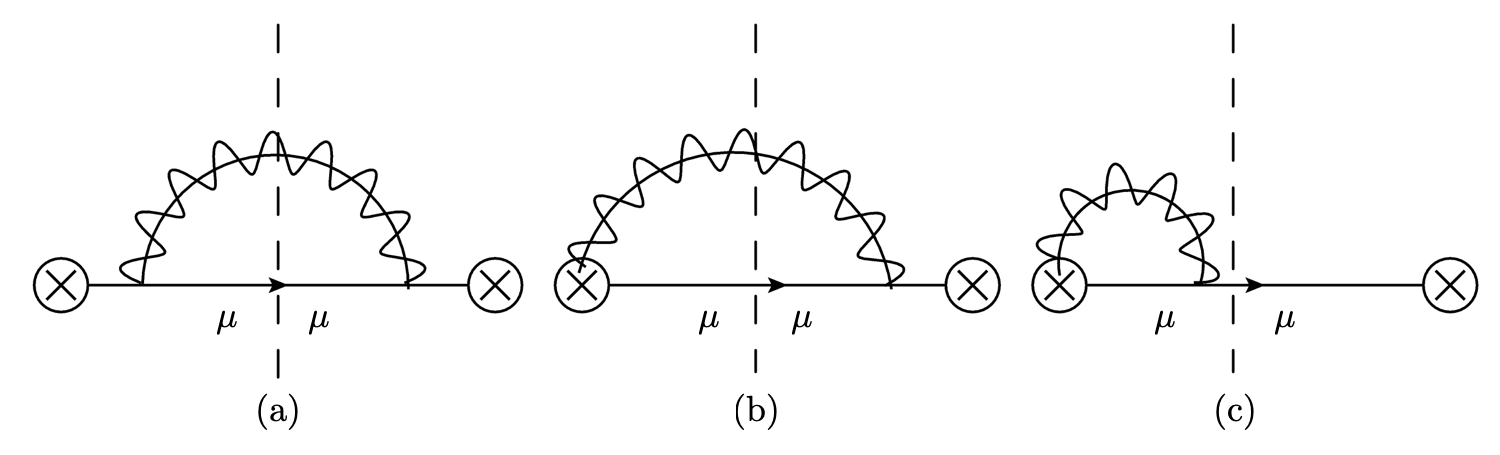} 
\end{center} \vspace{-0.6cm}
\caption{\label{jetfunction}  Feynman diagrams for the semi-inclusive jet function 
at one loop. The dotted lines denote the cut. Mirror images for (b) and (c) are omitted.
For the muon semi-inclusive jet functions, the fermions cut by the dotted lines are muons.} 
\end{figure}

The semi-inclusive jet functions are defined in eq.~\eqref{semijet}. The Feynman diagrams 
at order $\alpha$ are shown in Fig.~\ref{jetfunction}. 
As in computing the beam functions, the final result is obtained by taking the limit of small $M$. 
 All the contributions including the zero-bin contributions without the group theory factors 
 at NLO are given as (See appendix~\ref{sijf}.)
\begin{align} \label{jetmat}
M_a &= \frac{\alpha}{2\pi} \frac{\theta(p^2 -M^2)}{p^2}  \Bigl[\frac{1}{2}   -\frac{M^2}{p^2} +\frac{1}{2} 
\Bigl( \frac{M^2}{p^2}\Bigr)^2 \Bigr]   \longrightarrow  
\frac{\alpha}{2\pi} \Bigl[ \delta (p^2) \Bigl(-\frac{3}{4} +\frac{1}{2} \ln \frac{\mu^2}{M^2} 
\Bigr) +\frac{1}{2\mu^2} \mathcal{L}_0 \Bigl(\frac{p^2}{\mu^2} \Bigr)
\Bigr], \nonumber \\
\tilde{M}_b &= \frac{\alpha}{2\pi}
 \theta(p^2 -M^2)  \frac{1}{p^2}  
\Bigl[ -1 + \frac{M^2}{p^2} - \ln \frac{M^2}{p^2} \Bigr] \nonumber \\
&\longrightarrow\frac{\alpha}{2\pi} \Bigl[ \delta (p^2) \Bigl( 1 - \ln \frac{\mu^2}{M^2} 
+\frac{1}{2} \ln^2 \frac{\mu^2}{M^2} \Bigr) -\frac{1}{\mu^2} \mathcal{L}_0 \Bigl( \frac{p^2}{\mu^2}\Bigr)
\Bigl( 1 - \ln \frac{\mu^2}{M^2}\Bigr) +\frac{1}{\mu^2} \mathcal{L}_1 \Bigl( \frac{p^2}{\mu^2}\Bigr) \Bigr], 
\nonumber  \\
M_b^{\varnothing} &= \frac{\alpha}{2\pi} \Bigl\{ \delta (p^2) \Bigl[ \Bigl(
\frac{1}{\eta} +\ln \frac{\nu}{\omega}\Bigr) \Bigl( \frac{1}{\eps} +\ln
\frac{\mu^2}{M^2}\Bigr) -\frac{1}{\eps^2} +\frac{1}{2} \ln^2 \frac{\mu^2}{M^2}
+\frac{\pi^2}{12}\Bigr] \nonumber \\ 
& + \Bigl( \frac{1}{\eps}+\ln \frac{\mu^2}{M^2}\Bigr)
\frac{1}{\mu^2} \mathcal{L}_0
\Bigl(\frac{p^2}{\mu^2}\Bigr)  \Bigr\}, \nonumber \\
M_c &= \frac{\alpha}{2\pi} \delta (p^2) \Bigl[ \Bigl( \frac{1}{\eta} +\ln
\frac{\nu}{\omega}\Bigr) \Bigl( \frac{1}{\eps} +\ln \frac{\mu^2}{M^2}\Bigr)
+\frac{1}{\eps} +\ln \frac{\mu^2}{M^2} + 1-\frac{\pi^2}{6}\Bigr],
\end{align}
and the wavefunction renormalization and the residue are given by eq.~\eqref{wfren}.
We put $p^2 = \omega p^+$, where $\omega = \overline{n}\cdot p$, and $p^+$ is the jettiness from the jet. 

The bare singlet and nonsinglet semi-inclusive jet functions $J_l^0$ and $J_l^a$, 
which include the lepton $l$ in the final state, are given as
\begin{equation} \label{jetex}
J_l^0 (p^2 , M, \mu) =  J_s (p^2, M, \mu) \mathrm{Tr} (P_l T^0), \ J_l^a (p^2 , M, \mu) = 
 J_n (p^2, M, \mu) \mathrm{Tr} (P_l T^a),
\end{equation}
where $P_l$ is the projection operator to the lepton $l$ ($l= \mu, \nu_{\mu}$).
With the appropriate group theory factors, the bare singlet and the nonsinglet jet functions at NLO
are given as 
\begin{align} 
J_s^{(1)} (p^2, M, \mu) &= C_F\Bigl(M_a +
2(\tilde{M}_b -M_b^{\varnothing}) +2 M_c + (Z^{(1)} +R^{(1)})\delta (p^2) \Bigr) \\
&=\frac{\alpha C_F}{2\pi} \Bigl[ \delta (p^2) \Bigl( \frac{2}{\eps^2} +\frac{3}{2\eps}
+\frac{7}{2} -\frac{\pi^2}{2}\Bigr) -\Bigl( \frac{2}{\eps}
+\frac{3}{2}\Bigr)\frac{1}{\mu^2} \mathcal{L}_0 \Bigl(\frac{p^2}{\mu^2}\Bigr)
+\frac{2}{\mu^2} \mathcal{L}_1 \Bigl(\frac{p^2}{\mu^2}\Bigr) \Bigr], \nonumber \\
J_n^{(1)} (p^2, M, \mu) &= \Bigl( C_F -\frac{C_A}{2}\Bigr) 
\Bigl(M_a + 2(\tilde{M}_b - M_b^{\varnothing} )\Bigr)
+C_F \Bigl(2 M_c +  (Z^{(1)} +R^{(1)})\delta (p^2) \Bigr) \nonumber \\ %
&= J_s (p^2, M, \mu) + \frac{\alpha
C_A}{2\pi}\Bigl\{ \delta (p^2) \Bigl[\frac{1}{\eta}\Bigl( \frac{1}{\eps} +\ln
\frac{\mu^2}{M^2}\Bigr) -\frac{1}{\eps^2} +\frac{1}{\eps} \ln \frac{\nu}{\omega} + \ln
\frac{\mu^2}{M^2} \ln \frac{\nu}{\omega} \nonumber \\ 
&+\frac{3}{4} \ln \frac{\mu^2}{M^2}
-\frac{5}{8} +\frac{\pi^2}{12}\Bigr] + \Bigl( \frac{1}{\eps} +\frac{3}{4}\Bigr)
\frac{1}{\mu^2} \mathcal{L}_0 \Bigl(\frac{p^2}{\mu^2}\Bigr) -\frac{1}{\mu^2} \mathcal{L}_1
\Bigl(\frac{p^2}{\mu^2}\Bigr) \Bigr\}.  \nonumber 
\end{align} 
Here again the nonsinglet jet function develops the rapidity divergence. 
 
 The anomalous dimensions are given as 
\begin{align} \label{anojet}
\gamma_{Js}^{\mu} &= -2C_F \Gamma_c\frac{1}{\mu^2} \mathcal{L}_0 \Bigl(\frac{p^2}{\mu^2}\Bigr)
-2\gamma_{\ell} \delta (p^2), \ \gamma_{Js}^{\nu} = 0,  \nonumber \\
\gamma_{Jn}^{\mu} &=
\gamma_{Js}^{\mu} +C_A \Gamma_c \Bigl[ \delta (p^2) \ln \frac{\nu}{\omega}
+\frac{1}{\mu^2} \mathcal{L}_0 \Bigl(\frac{p^2}{\mu^2}\Bigr) \Bigr],  
\ \gamma_{Jn}^{\nu} = C_A \Gamma_c \delta (p^2) \ln \frac{\mu}{M}.
\end{align} 
The Laplace transforms of the anomalous dimensions with $s=1/(Q_L e^{\gamma_{\mathrm{E}}})$
conjugate to the jettiness $p^+$ are given by 
\begin{align} \label{jetanol}
\tilde{\gamma}_{Js}^{\mu} &= 2C_F \Gamma_c \ln \frac{\mu^2}{\omega Q_L} - 2\gamma_{\ell}, \
\tilde{\gamma}_{Js}^{\nu} = 0, \nonumber \\ 
\tilde{\gamma}_{Jn}^{\mu} &= \tilde{\gamma}_{Js}^{\mu} -C_A \Gamma_c \ln
\frac{\mu^2}{\nu Q_L}, \ 
\tilde{\gamma}_{Jn}^{\nu} = C_A\Gamma_c  \ln \frac{\mu}{M},
\end{align}
which are the same as the anomalous dimensions of the beam functions in eq.~\eqref{beamanon}.

\subsection{Fragmentation functions and fragmenting jet functions \label{ffandfjf}} 
The collinear parts pertaining to the final states are described either by the fragmentation functions
or by the FJFs. The fragmentation function is used when a single
particle is observed with no properties of the jet to be probed. The FJF 
describes the fragmentation of a parton $i$ to another parton $j$ within a
jet originating from $i$ with the measurement of the momentum fraction and the invariant
mass of the jet. 
 
The fragmentation function from $\ell$ to the lepton $l$ is extended from the definition in 
QCD~\cite{Procura:2009vm, Jain:2011xz} to
\begin{equation} \label{ffdef}
D_{l}^{a} (z, M, \mu) =  \int \frac{d^2 p_{l}^{\perp}}{z} \sum_X \mathrm{Tr} \,
\langle 0|T^a \frac{\fms{\overline{n}}}{2} [\delta(\omega + \overline{n}\cdot \mathcal{P})
\delta^{(2)} (\mathcal{P}_{\perp})\ell_L (0) ]|lX\rangle \langle lX|\overline{\ell}_L (0)
|0\rangle,
\end{equation}
where $z =p_l^-/p_{\ell}^- = p_l^-/\omega$ is the fraction of the largest lightcone components 
of the observed lepton $l$ originating from $\ell$. 
At tree level, the fragmentation functions are normalized as
\begin{equation}
D_{l}^{a (0)} (z, M, \mu) = \delta (1-z) \mathrm{Tr} (P_l T^a).
\end{equation}
The matrix elements for the fragmentation functions are the same as those for the PDFs in eq.~\eqref{pdfmat}, 
and we will not present them here. 

The FJF is defined as 
\begin{align} \label{fjfdef}
\mathcal{G}_{l}^{a} (p^2, z, M, \mu ) &=  \frac{2(2\pi)^3}{\omega z} \int d^2 \mathbf{p}_{l\perp}
\sum_X   \\
&\times \mathrm{Tr}\,  \langle 0|T^a \frac{\fms{\overline{n}}}{2} \ell_L (0) |lX\rangle  \langle lX|
\delta (\omega+\overline{n}\cdot \mathcal{P}) \delta^{(2)} (\mathcal{P}_{\perp}) 
\delta (p^+ +n\cdot\mathcal{P}) 
\overline{\ell}_L (0)|0\rangle, \nonumber
\end{align} 
where $p^2 = \omega p^+$ is the invariant mass of the collinear jet. The small component $p^+$ is 
the jettiness from the fragmented lepton. The FJF is normalized at tree level as
\begin{equation}
\mathcal{G}_{l}^{a (0)} (p^2, z, M, \mu ) = 2 (2\pi)^3 \delta (p^2) \delta (1-z) \mathrm{Tr}(P_l T^a).
\end{equation}

\begin{figure}[b] 
\begin{center}
\includegraphics[width=15cm]{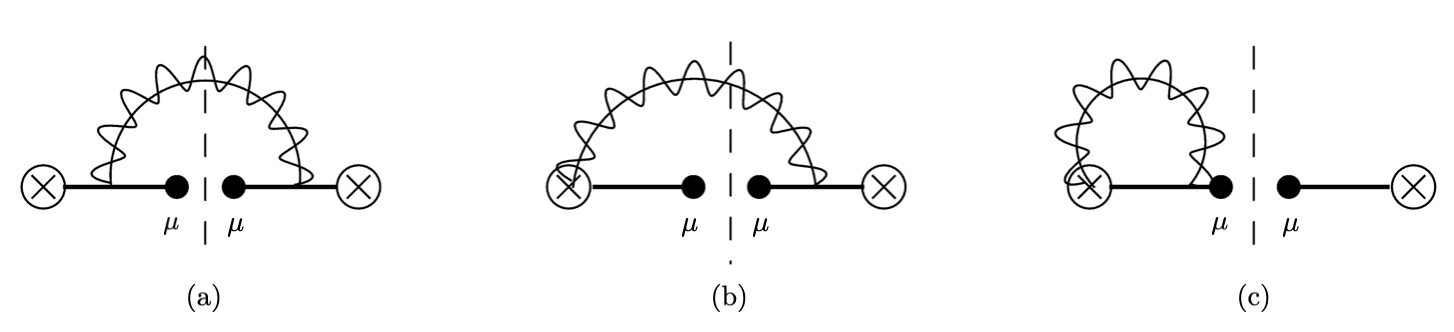} 
\end{center} \vspace{-0.6cm}
\caption{\label{fragfig}  Feynman diagrams for the fragmentation functions and the FJFs 
at one loop. The dotted lines denote the cut. The mirror images of (b) and (c) are omitted. The dots represent
the observed particles in the final state. Note that $\omega$ is measured for the fragmentation function,
while $\omega$ and $p^+$ are measured for the FJF.} 
\end{figure}

The relations between the FJF, the semi-inclusive jet functions, and the fragmentation functions can be found
by comparing eqs.~\eqref{semijet}, \eqref{ffdef} and \eqref{fjfdef}.  The FJF probes the differential distributions
with respect to the invariant mass $p^2$ and the momentum fraction $z$.  If we integrate the FJF over 
all the possible invariant masses, it yields the fragmentation functions which shows the distributions of the 
momentum fraction of the observed lepton. On the other hand, if we integrate the FJF over the momentum fraction
$z$, it yields the semi-inclusive jet function for the lepton $l$ in the final state. Therefore the relations can
be expressed as
\begin{equation}
D_{l}^{a} (z, M, \mu) = \int_0^{\infty} dp^2 \frac{\mathcal{G}_{l}^{a} 
(p^2, z, M, \mu)}{2 (2\pi)^3},  \ 
J_l^a (p^2, M, \mu) = \int_0^1 dz \frac{\mathcal{G}_{l}^{a} (p^2, z, M, \mu)}{2 (2\pi)^3}, \label{jetsemi}
\end{equation}
and it is confirmed at order $\alpha$. 
Our definition of the semi-inclusive (singlet) jet function is different from that in ref.~\cite{Kang:2016mcy}.
As explained by the authors in ref.~\cite{Kang:2016mcy}, their semi-inclusive jet function is similar to the 
fragmentation function. To be exact, it is the fragmentation function, in which the final-state hadron forms a jet.
Our semi-inclusive jet function describes the jet mass distribution with the lepton $l$ (or the leptonic
jet which includes $l$) in the final state.

In ref.~\cite{Jain:2011xz}, integrating the FJF with an additional factor $z$ yields the inclusive jet function 
$J_i (p^2, \mu)$. The additional factor of $z$ is due to the symmetrization of the final states when all the 
hadrons in the final state are summed over.  However, we refer to the semi-inclusive jet function
for the lepton $l$ specified in the final state in eq.~\eqref{jetsemi}. If we sum over all the final leptons to yield
the inclusive jet functions, the additional factor of $z$ should be included.

The Feynman diagrams for the fragmentation functions and the FJFs are shown in figure~\ref{fragfig}. 
Though we present the same Feynman diagrams for both functions, note that the observed quantities are different.
The matrix elements for the FJF without the group theory factors are given at NLO as  
\begin{align}  \label{fjfmat}
M_a &= \frac{\alpha}{2\pi} \theta(z)
\theta (1-z) \theta \Bigl( p^2 - \frac{M^2}{1-z}\Bigr) \frac{1}{p^2} \Bigl( 1-z -\frac{M^2}{p^2}\Bigr)
  \nonumber \\ 
&\longrightarrow \frac{\alpha}{2\pi} \theta(z) \theta (1-z) (1-z) \Bigl[ 
\delta (p^2) \Bigl( \ln \frac{(1-z) \mu^2}{M^2} -1\Bigr) +\frac{1}{\mu^2} \mathcal{L}_0 
\Bigl( \frac{p^2}{\mu^2}\Bigr) 
 \Bigr], \nonumber \\
 \tilde{M}_b
&= \frac{\alpha}{2\pi} \theta(z)
\theta (1-z) \theta \Bigl( p^2 - \frac{M^2}{1-z}\Bigr) \frac{1}{p^2} \frac{z}{1-z}, \nonumber \\
&\longrightarrow \frac{\alpha}{2\pi} \Bigl[ \delta (p^2) \delta (1-z) \frac{1}{2}\ln^2 \frac{\mu^2}{M^2} +
\delta (p^2) \Bigl( z \ln \frac{\mu^2}{M^2} \mathcal{L}_0 (1-z) +z \mathcal{L}_1 (1-z)
\Bigr) \nonumber \\ 
& + \delta (1-z) \Bigl( \frac{1}{\mu^2} \mathcal{L}_0 \Bigl(
\frac{p^2}{\mu^2}\Bigr) \ln \frac{\mu^2}{M^2} +\frac{1}{\mu^2} \mathcal{L}_1 \Bigl(
\frac{p^2}{\mu^2}\Bigr) \Bigr) + \frac{z}{\mu^2} \mathcal{L}_0 \Bigl( \frac{p^2}{\mu^2}\Bigr)
\mathcal{L}_0 (1-z)\Bigr], \nonumber \\
M_b^{\varnothing} &= \frac{\alpha}{2\pi} \delta
(1-z)\Bigl\{ \Bigl( \frac{1}{\eps} +\ln \frac{\mu^2}{M^2}\Bigr)\Bigl[\delta (p^2) \Bigl(
\frac{1}{\eta} + \ln \frac{\nu}{\omega}\Bigr) +\frac{1}{\mu^2} \mathcal{L}_0
\Bigl(\frac{p^2}{\mu^2}\Bigr)\Bigr]  \nonumber \\
&+\delta(p^2) \Bigl(-\frac{1}{\eps^2} +\frac{1}{2} \ln^2 \frac{\mu^2}{M^2}
+\frac{\pi^2}{12} \Bigr)\Bigr\}\nonumber \\ 
M_c &= \frac{\alpha}{2\pi} \delta (p^2) \delta (1-z)
\Bigl[ \Bigl( \frac{1}{\eps} +\ln \frac{\mu^2}{M^2}\Bigr) \Bigl( \frac{1}{\eta} +\ln
\frac{\nu}{\omega} +1\Bigr) +1-\frac{\pi^2}{6}\Bigr].
\end{align}
Here we also express the matrix elements in the limit of small mass $M$. 
The detailed derivation of taking the limit is presented in appendix~\ref{fjfsmall}.

The singlet and nonsinglet FJFs are written as
\begin{align} \label{fjfex}
\mathcal{G}_{l}^{0} (p^2, z, M, \mu ) &=  
\mathcal{G}_{l s}  (p^2, z, M, \mu ) \mathrm{Tr} (P_l T^0), \nonumber \\
 \mathcal{G}_{l}^{a} (p^2, z, M, \mu ) 
&=  \mathcal{G}_{ln} (p^2, z, M, \mu)\mathrm{Tr} (P_l T^a).
\end{align}
The bare FJFs  for the singlet and the nonsinglet at NLO  are given as
\begin{align} 
\frac{\mathcal{G}_{l s}^{(1)} (p^2, z, M, \mu)}{2(2\pi)^3} 
&= C_F \Bigl(M_a +2(\tilde{M}_b -M_b^{\varnothing} + M_c) +(Z^{(1)}+R^{(1)}) 
\delta(p^2) \delta (1-z)\Bigr)  \label{gsing0}     \\ 
&= \frac{\alpha C_F}{2\pi} \Bigl\{  
\delta (p^2) \delta (1-z) \Bigl( \frac{2}{\eps^2} +\frac{3}{2\eps}   +\frac{9}{4} -\frac{\pi^2}{2} \Bigr) \nonumber \\ 
&+\delta (p^2) \Bigl[ P_{\ell\ell} (z) \ln \frac{\mu^2}{M^2} + (1+z^2) \mathcal{L}_1 (1-z) 
-(1-z) \theta (z) \theta (1-z) \Bigr]\nonumber \\
&+\delta (1-z)  \Bigl[ -\frac{2}{\eps} \frac{1}{\mu^2} \mathcal{L}_0 \Bigl(\frac{p^2}{\mu^2}\Bigr)   
+\frac{2}{\mu^2} \mathcal{L}_1 \Bigl(\frac{p^2}{\mu^2}\Bigr) \Bigr] 
+(1+z^2) \mathcal{L}_0 (1-z) 
\frac{1}{\mu^2} \mathcal{L}_0 \Bigl(\frac{p^2}{\mu^2}\Bigr)\Bigr\},   \nonumber \\
\frac{\mathcal{G}_{l n}^{(1)} (p^2, z, M, \mu)}{2(2\pi)^3} 
&= (C_F -C_A/2) \Bigl(M_a + 2(\tilde{M}_b -M_b^{\varnothing}) \Bigr)     \\
&+ C_F \Bigl(2M_c + (Z^{(1)}+R^{(1)}) \delta (p^2) \delta (1-z)\Bigr)   \nonumber \\
&= \frac{\mathcal{G}_{l s}^{(1)}  (p^2, z, M, \mu) }{2(2\pi)^3} \nonumber \\
&- \frac{\alpha C_A}{4\pi} \Bigl\{ 
\delta (p^2) \delta (1-z) \Bigl[  \frac{2}{\eps^2} -2\Bigl(\frac{1}{\eps} 
+\ln \frac{\mu^2}{M^2}\Bigr) \Bigl( \frac{1}{\eta} 
+ \ln \frac{\nu}{\omega} \Bigr)  -\frac{\pi^2}{6}\Bigr] \nonumber  \\
& + \delta (p^2) \Bigl[ (1+z^2) \mathcal{L}_0 (1-z) \ln \frac{\mu^2}{M^2} + (1+z^2) \mathcal{L}_1 (1-z) 
-(1-z) \theta (z) \theta (1-z)\Bigr] \nonumber \\
&+ \delta (1-z) \Bigl[-\frac{2}{\eps} \frac{1}{\mu^2} \mathcal{L}_0 \Bigl(\frac{p^2}{\mu^2}\Bigr) 
 +  \frac{2}{\mu^2} \mathcal{L}_1 \Bigl(\frac{p^2}{\mu^2}\Bigr)\Bigr] + (1+z^2) \mathcal{L}_0 (1-z) 
\frac{1}{\mu^2} \mathcal{L}_0 \Bigl(\frac{p^2}{\mu^2}\Bigr) \Bigr\}.   \nonumber
\end{align} 
 The singlet FJF in eq.~\eqref{gsing0} is the same as the result in ref.~\cite{Jain:2011xz} though the individual 
contributions are different. However, the result for the nonsinglet FJFs is new, and note that the dependence 
on the rapidity scale remains in the nonsinglet FJFs, while there is none for the singlet FJFs.

The anomalous dimensions are given as 
\begin{align}
\gamma_{Fs}^{\mu} &=   -2C_F \Gamma_c
\frac{1}{\mu^2} \mathcal{L}_0 \Bigl(\frac{p^2}{\mu^2}\Bigr) -2\gamma_{\ell} \delta (p^2), 
\ \gamma_{Fs}^{\nu} =0,
\nonumber \\ 
\gamma_{Fn}^{\mu}&= \gamma_{Fs}^{\mu} + C_A  \Gamma_c \Bigl[\delta
(p^2) \ln \frac{\nu}{\omega} + \frac{1}{\mu^2} \mathcal{L}_0 \Bigl(\frac{p^2}{\mu^2}\Bigr)
\Bigr], \ \gamma_{Fn}^{\nu} = C_A \Gamma_c \delta (p^2) \ln \frac{\mu}{M}, 
\end{align}
and their Laplace transforms are given by 
\begin{align} 
\tilde{\gamma}_{Fs}^{\mu} &=
2C_F\Gamma_c \ln \frac{\mu^2}{\omega Q_L} -2 \gamma_{\ell}, \ \tilde{\gamma}_{Fs}^{\nu} =0,
\nonumber \\ \tilde{\gamma}_{Fn}^{\mu} &= \tilde{\gamma}_{Fs}^{\mu} -C_A \Gamma_c
 \ln \frac{\mu^2}{\nu Q_L}, \ \tilde{\gamma}_{Fn}^{\nu} = C_A \Gamma_c \ln \frac{\mu}{M}. 
\end{align}
Because there is a relation between the semi-inclusive jet function and the FJF in eq.~\eqref{jetsemi},
the anomalous dimensions of the jet functions and the FJFs are the same. 
[See eqs.~\eqref{anojet} and \eqref{jetanol}.]

The matching between the fragmentation function and the FJF are written as
\begin{equation}
\mathcal{G}_i^{a} (p^2, z, M, \mu) = \sum_j \int_z^1 \frac{dz'}{z'} 
\mathcal{J}_{ij}^{ab} \Bigl( p^2, \frac{z}{z'}, \mu\Bigr) D_j^{b} (z', M, \mu)
\end{equation}
at leading power, where $\mathcal{J}_{ij}^{ab}$ are the matching coefficients. At tree-level, it is given by
\begin{equation}
\mathcal{J}_{ij}^{ab(0)} (p^2, z,\mu) = 2 (2\pi)^3 \delta_{ij} \delta^{ab}\delta (p^2) \delta (1-z).
\end{equation}
If we write
\begin{equation} \label{jij}
\mathcal{J}_{\ell\ell}^{00} (p^2, z,\mu) = \mathcal{J}_{\ell\ell}^s (p^2, z,\mu), \ 
\mathcal{J}_{\ell\ell}^{ab} (p^2, z,\mu)= \delta^{ab} 
\mathcal{J}_{\ell\ell}^n (p^2, z,\mu), \ \mathcal{J}_{\ell\ell}^{a0} (p^2, z,\mu)=0,
\end{equation}
the singlet matching coefficient $\mathcal{J}_{\ell\ell}^s$ and the nonsinglet matching coefficient
$\mathcal{J}_{\ell\ell}^n$ at order $\alpha$ are given as
\begin{align}
\frac{\mathcal{J}_{\ell\ell}^{s(1)}(p^2, z,\mu)}{2 (2\pi)^3} 
&= \frac{\alpha C_F}{2\pi} \Bigl\{ -\frac{\pi^2}{6} \delta (p^2) \delta (1-z) \nonumber \\
&+\delta (p^2) \Bigl( P_{\ell\ell} (z) \ln z + (1+z^2) \mathcal{L}_1 (1-z) + \theta(1-z) (1-z)\Bigr) 
\nonumber \\
&+\delta (1-z) \frac{2}{\mu^2} \mathcal{L}_1 \Bigl(\frac{p^2}{\mu^2}\Bigr) + (1+z^2) \mathcal{L}_0 (1-z)
\frac{1}{\mu^2} \mathcal{L}_0 \Bigl(\frac{p^2}{\mu^2}\Bigr) \Bigr\}, \nonumber \\
\frac{\mathcal{J}_{\ell\ell}^{n(1)}(p^2, z,\mu)}{2 (2\pi)^3} 
&= \frac{C_F -C_A/2}{C_F} \frac{\mathcal{J}_{\ell\ell}^{s(1)}(p^2, z,\mu)}{2 (2\pi)^3}.
\end{align}
The matching coefficients for the singlet $\mathcal{J}_{\ell\ell}^{s(1)}$ are the same
as those in QCD~\cite{Jain:2011xz}. 
As in the case of the matching coefficients $\mathcal{I}_{\ell\ell}^{s,n}$ for the beam functions and the PDFs,
the new nonsinglet matching coefficient is proportional to the singlet matching coefficient.  
These matching coefficients do not depend on the gauge boson mass $M$, as in the matching 
coefficients in eq.~\eqref{ill}.

\section{Hard function\label{hardfun}}
The hard functions are represented by a matrix in the basis of the singlet and nonsinglet 
operators $O_I$, so are the soft functions. 
The hard functions  are defined in eq.~\eqref{finhard} as $H_{JI} = 4u^2 D_I^* D_J =C_I^* C_J$, where $D_I$
is the Wilson coefficient of the operators $O_I$ for the process $\ell_e \overline{\ell}_e 
\rightarrow \ell_{\mu} \overline{\ell}_{\mu}$. Here $\ell_e$ and $\ell_{\mu}$ are electron and muon 
doublets respectively because the relevant hard processes involve lepton doublets at higher orders. 
The hard coefficients are determined by matching the results of the full theory onto the effective theory. 

We can utilize the results of the hard functions in previous literature. 
The detailed form of the hard function at one loop for $2\rightarrow 2$ partonic
processes can be found in ref.~\cite{Kelley:2010fn} for QCD, and in ref.~\cite{Chiu:2009mg}
for electroweak interactions. 
Since we deal with the left-handed
fields, all the right-handed contributions are put to zero. 
 
The Wilson coefficients $C_{ILL}$ to order $\alpha$ are given by~\cite{Kelley:2010fn,Chiu:2009mg}
\begin{align}
C_{1LL} (u,s,t) &=  2g^2\frac{u}{s} \Bigl\{ 1 + \frac{\alpha}{4\pi} \Bigl[ -2C_F L(s)^2 +X_1 (u,s,t) 
L(s) +Y  \nonumber \\ 
&+\Bigl( \frac{C_A}{2} - 2C_F \Bigr) Z(u, s, t)\Bigr] \Bigr\}, \nonumber \\
C_{2LL} (u, s, t) &= 2g^2 \frac{u}{s} \frac{\alpha}{4\pi} \Bigl[ X_2 (u, s, t) L(s) -\frac{C_F}{2C_A} Z(u,s,t) \Bigr],
\end{align}
where
\begin{align}
X_1 (u, s, t) &= 6C_F -\beta_0 + 8C_F [L(u) - L(t)] -2 C_A [ 2 L(u) -L(s) - L(t)], \nonumber \\
X_2 (u, s, t) &= \frac{2C_F}{C_A} [L(u) -L(t)], \nonumber \\
Z (u, s, t) &= \frac{s}{u} \Bigl( \frac{s+2t}{u} [L(t) - L(s)]^2 + 2[L(t) -L(s)] + \frac{s+2t}{u} \pi^2\Bigr),  \nonumber \\
Y &= C_A \Bigl( \frac{10}{3} +\pi^2 \Bigr) +C_F \Bigl( \frac{\pi^2}{3} -16\Bigr) + \frac{5}{3} \beta_0.
\end{align}
The function $L(x)$ as a function of the Mandelstam variables is given by 
\begin{equation} 
L(t) = \ln \frac{-t}{\mu^2}, \ L(u) = \ln \frac{-u}{\mu^2}, \ L(s) = \ln \frac{s}{\mu^2} -i\pi. 
\end{equation} 

The RG equation for $C_I$ can be written as
\begin{equation}
\frac{d}{d\ln \mu} \begin{pmatrix} C_1 \\ C_2 \end{pmatrix} = \bm{\Gamma}_H  
\begin{pmatrix} C_1 \\ C_2 \end{pmatrix},
\end{equation}
where $\bm{\Gamma}_H$ is the anomalous dimension matrix for the Wilson coefficients. 
Then the RG equation for the hard function $\mathbf{H}$ is given as
\begin{equation}
\frac{d}{d\ln \mu} \mathbf{H} = \mathbf{\Gamma}_H \mathbf{H} + 
\mathbf{H} \mathbf{\Gamma}_H^{\dagger}. 
\end{equation}
 The anomalous dimension matrix is given by~\cite{Kelley:2010fn,Chiu:2009mg}
\begin{equation} \label{gamh} 
\mathbf{\Gamma}_H (u, s, t)= \Bigl[ 2C_F \Gamma_c (\alpha ) L(s)
+4\gamma_{\ell}-\frac{\beta(\alpha)}{\alpha}\Bigr] \mathbf{1} + \Gamma_c (\alpha)
\mathbf{M}, 
\end{equation} 
where the $\beta/\alpha$ term compensates
the $g^2 (\mu)$ dependence in the leading Wilson coefficients. Here the beta function
$\beta$ is given by 
\begin{equation} 
\beta (\alpha) =\frac{d\alpha}{d\ln \mu} =  -2\alpha \sum_{k=0} \beta_k \Bigl(\frac{\alpha}{4\pi}\Bigr)^{k+1}, 
\ \ \beta_0 = \frac{11}{3} C_A -\frac{2}{3} n_f. 
\end{equation}

The matrix $\mathbf{M}$ can be written as 
\begin{equation} 
\mathbf{M} = -\sum_{i<j}
\mathbf{T}_i \cdot \mathbf{T}_j \Bigl[L(s_{ij}) -L(s)\Bigr], 
\end{equation} 
where $s=s_{12}=s_{34}$, $t=s_{13} =s_{24}$ and $u=s_{14} = s_{23}$.
By explicitly computing the color factors  $\mathbf{T}_i \cdot \mathbf{T}_j$~\cite{Chiu:2009mg}, 
the matrix $\mathbf{M}$ is written as 
\begin{align} \label{mij} 
\mathbf{M}† (u, s, t) &= -\sum_{i<j} \mathbf{T}_i\cdot \mathbf{T}_j \Bigl[ L(s_{ij}) - L(s)\Bigr]   \\ 
 &=
\begin{pmatrix} 
\displaystyle \Bigl(2C_F - C_A/2\Bigr) \ln \frac{n_{13}
n_{24}}{n_{14} n_{23}} -\frac{C_A}{2} \ln \frac{n_{12} n_{34}}{n_{14} n_{23}} &&
\displaystyle \ln \frac{n_{13} n_{24}}{n_{14} n_{23}} \\ \displaystyle -C_F \Bigl(C_F
-C_A/2\Bigr) \ln \frac{n_{13} n_{24}}{n_{14} n_{23}} & &0 
\end{pmatrix} 
+ i\pi \begin{pmatrix}
\displaystyle  C_A & 0 \\
0& 0
\end{pmatrix}, \nonumber
\end{align} 
with $n_{ij} = n_i\cdot n_j/2$. 
    
\section {Soft function\label{softs}} 

The soft functions for the $N$-jettiness or more general jet observables have been discussed 
in refs.~\cite{Jouttenus:2011wh, Bertolini:2017efs}. The authors have considered 
the differential jettiness, that is, the individual jettiness in the $N$ jets. [See eq.~\eqref{indijet}.] 
Here we consider the total jettiness which corresponds to the sum of the individual jettiness. 
The soft function for the jettiness is defined in eq.~\eqref{softdef}.
Since the calculation was performed in massless cases in these references, they set
the virtual contribution to zero since they consist of scaleless integrals. In our scheme with the nonzero 
gauge boson mass, there is nonzero virtual contribution, and we present the results here.

\subsection{Hemisphere soft function}
\begin{figure}[t] 
 \begin{center}
\includegraphics[height=5.8cm]{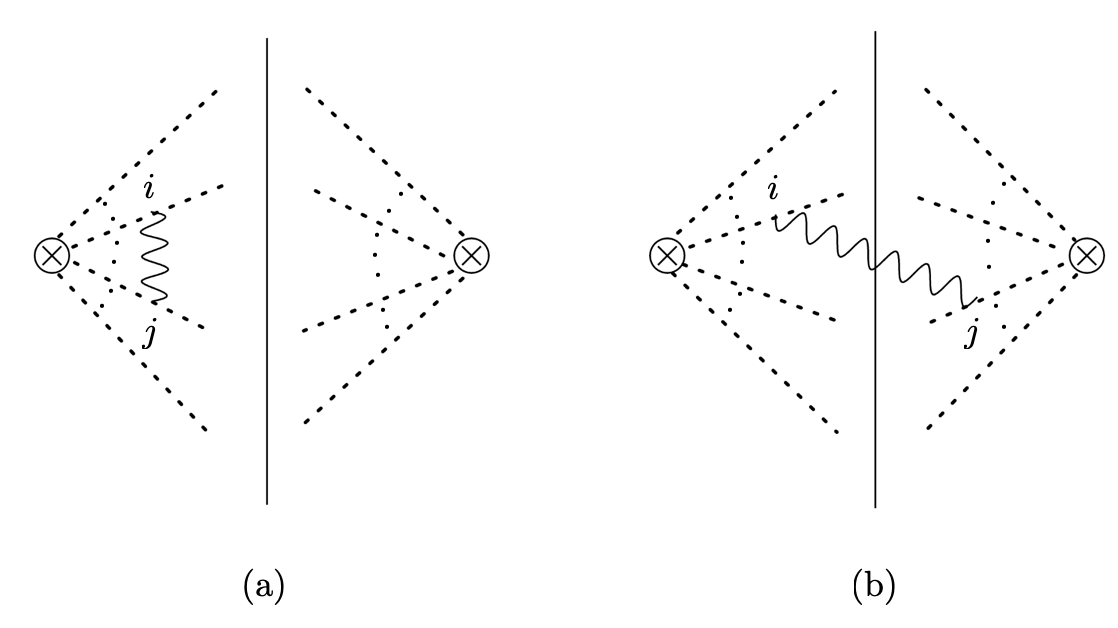} 
\end{center} \vspace{-0.6cm}
\caption{\label{softwilson}\baselineskip 3.0ex  Schematic diagrams for the emission of a soft gauge
boson in $S_{ij}$. 
The vertical lines are the final-state cut. Diagram (a)
denotes the virtual contribution, and diagram (b) denotes the real contribution.  
The gauge bosons attached to the same $i$ do not contribute. In order to obtain the soft function, all the possible
pairs of contractions with different $i$ and $j$ should be included.} 
\end{figure}

The diagrams for the emission of  a gauge boson between the soft Wilson lines $S_i$ and $S_j^{\dagger}$
($Y_i$ and $Y_j^{\dagger}$ in \scone) are shown in figure~\ref{softwilson}, where $S_i$ is the soft 
Wilson line in the $n_i$ direction. 
Because $n_i^2 =0$, the emission from the soft Wilson lines with $i=j$ vanishes.  
Figure~\ref{softwilson} (a)  [figure~\ref{softwilson} (b)] denotes the
virtual contribution (the real contribution). In computing
the total soft function, we include all the possible combinations of $i$ and $j$ with the appropriate group theory
factors.  The following calculations are based on the contractions of a gauge boson from the soft Wilson lines 
$S_i$ and $S_j^{\dagger}$.  
The additional minus signs when the contractions are performed between
$S_i$ and $S_j$ or $S_i^{\dagger}$ and $S_j^{\dagger}$ are included in the group theory factors. 

The real contribution is decomposed into the hemisphere and the non-hemisphere parts.
Only the hemisphere part involves the divergence, and the non-hemisphere parts are finite~\cite{Jouttenus:2011wh,Bertolini:2017efs}. 
Our case is relevant to ref.~\cite{Jouttenus:2011wh}, which corresponds to the case $\beta_i =2$ in ref.~\cite{Bertolini:2017efs}. We concentrate on the hemisphere soft function to extract the anomalous 
dimensions of the $N$-jettiness soft function. The virtual contribution is not affected by the process of
choosing the hemisphere functions in the real contribution.

The virtual contribution in figure~\ref{softwilson} (a) aside from the group theory factor is written as 
\begin{equation} \label{svirt}
S_{ij,\mathrm{hemi}}^V = -2\pi g^2 \mums \delta (k) \int \dd{q} \delta (q^2 -M^2)
\frac{2n_{ij}}{q_i q_j} R(q_i, q_j),
\end{equation}
where $q_i = n_i\cdot q$ and $q_j = n_j\cdot q$, and $R(q_i, q_j)$ is the rapidity regulator, which 
can be written at NLO as~\cite{Chay:2020jzn} 
\begin{equation} 
R(q_i, q_j) = \Bigl( \frac{\nu n_{ij}}{q_j}\Bigr)^{\eta} \theta (q_j-q_i) +
\Bigl( \frac{\nu n_{ij}}{q_i}\Bigr)^{\eta} \theta (q_i-q_j).
\end{equation}
Since the integrand in eq.~\eqref{svirt} is symmetric under $i\leftrightarrow j$, the contributions from both terms
in the rapidity regulator is the same. Here we pick up the first term in the rapidity regulator and multiply two 
to get the virtual contribution.
It is given as  
\begin{align} \label{svirt1}
S_{ij,\mathrm{hemi}}^V &= -8\pi g^2 \mums \delta (k) \int \dd{q} \delta (q^2 -M^2)
\frac{n_{ij}}{q_i q_j} \Bigl( \frac{\nu n_{ij}}{q_j} \Bigr)^{\eta} \theta (q_j -q_i)   \\
&= -\frac{\alpha}{\pi} \frac{(\mu^2 e^{\gamma_{\mathrm{E}}})^{\eps}}{\Gamma (1-\eps)}  \delta (k)
\int  dq_j   dq_i 
\Bigl( \frac{q_i q_j}{n_{ij}} - M^2 \Bigr)^{-\eps}
\frac{1}{q_i q_j} \Bigl( \frac{\nu n_{ij}}{q_j} \Bigr)^{\eta}  
\theta( q_i q_j > n_{ij} M^2)\theta (q_j -q_i)\nonumber \\
&= -\frac{\alpha}{\pi} \frac{(\mu^2 e^{\gamma_{\mathrm{E}}})^{\eps}}{\Gamma (1-\eps)}  \delta (k)
(\nu n_{ij})^{\eta} \int_{M^2}^{\infty} dy_1 \frac{(y_1 -M^2)^{-\eps}}{y_1^{1+\eta/2}} \int_1^{\infty} dy_2
\frac{1}{2y_2^{1+\eta/2}} \nonumber \\ 
&= -\frac{\alpha}{\pi} \delta (k)  e^{\gamma_{\mathrm{E}} \eps} \Bigl( \frac{\mu^2}{M^2}\Bigr)^{\eps} 
\Bigl( \frac{\nu \sqrt{n_{ij}}}{M}\Bigr)^{\eta} \frac{1}{\eta} 
\frac{\Gamma ( \eps + \eta/2)}{\Gamma (1+\eta/2)} \nonumber \\
&=\frac{\alpha}{2\pi} \delta (k)   \Bigl[ \frac{1}{\eps^2} -\frac{2}{\eta}
\Bigl( \frac{1}{\eps} +\ln \frac{\mu^2}{M^2}\Bigr)
-\frac{1}{\eps} \ln \frac{n_{ij} \nu^2}{\mu^2} 
+\frac{1}{2} \ln^2 \frac{\mu^2}{M^2} - \ln \frac{n_{ij} \nu^2}{M^2}\ln \frac{\mu^2}{M^2}
-\frac{\pi^2}{12}\Bigr].  \nonumber 
\end{align}
In the third line, we change variables $y_1 = q_i q_j/n_{ij}$, $y_2 = q_j/q_i$. Then $q_i$ and $q_j$ are written as
$q_i = \sqrt{n_{ij} y_1/y_2}$, $q_j = \sqrt{n_{ij} y_1 y_2}$ with $dq_i dq_j = n_{ij} dy_1 dy_2 /(2y_2)$,
with $y_1 >M^2$ and $y_2 >1$.

Here we focus on the real part. 
The imaginary part can be obtained by implementing the $i\epsilon$ prescription for the soft Wilson lines~\cite{Chay:2004zn}. 
The result can be summarized as follows: The 
logarithmic term can be expressed in the form $\ln (\sigma_{ij} n_{ij} -i\eps)$, where $\sigma_{ij} =-1$
when $i$ and $j$  are both incoming or outgoing, and $\sigma_{ij}=1$ when one is incoming and the other
is outgoing. When $\sigma_{ij} =-1$, the imaginary part is induced.

The real soft contribution without the group theory factor at order $\alpha$ is given as
\begin{equation}
S^{R(1)}_{ij} = -2\pi g^2 \mums  \int \dd{q} \delta (q^2 -M^2) 
\frac{2n_{ij}}{q_i q_j} R(q_i, q_j) F(k, \{ q_i\}), 
\end{equation}
where the function $F$ constrains the phase space on the emission of a single gauge boson, 
from which the hemisphere soft function is to be extracted. 

For 2-jettiness, we consider four independent labels $i$, $j$, $l$, $m$  and the constraint function $F$ is given by
\begin{align} \label{softjetti}
F(k, \{ q_i \} ) &=  \theta(q_j - q_i) \Bigl[ \delta (k-q_i) \theta(q_l - q_i) \theta (q_m -q_i) + 
\delta (k-q_l ) \theta (q_i - q_l)
\theta (q_m -q_l)  \nonumber \\
& +\delta (k-q_m) \theta (q_i - q_m) \theta (q_l - q_m) \Bigr] +(i \leftrightarrow j) \nonumber  \\
 &= \theta (q_j - q_i) \Bigl[ \delta (k-q_i) + \theta (q_i - q_m) \theta (q_l - q_m) 
\Bigl( \delta (k-q_m) -\delta (k-q_i)\Bigr) \nonumber \\
&+ \theta (q_i -q_l) \theta (q_m - q_l) \Bigl( \delta (k-q_l) -\delta (k-q_i)\Bigr) \Bigr] 
+ (i \leftrightarrow j) \nonumber \\
&\equiv F_{ij, \mathrm{hemi}} (k, \{q_i \}) + F_{ij, ml} (k, \{q_i \}) 
+  F_{ij, lm} (k, \{q_i \}) + (i \leftrightarrow j).
\end{align}
(See ref.~\cite{Jouttenus:2011wh} for 1-jettiness.) In obtaining the second relation, the theta functions in the 
first term is replaced by 
\begin{equation}
\theta(q_l - q_i) \theta (q_m -q_i) =\Bigl( 1-\theta(q_i - q_l) \Bigr) \Bigl( 1- \theta (q_i -q_m) \Bigr).  
\end{equation}

The hemisphere measurement function for the full hemisphere $q_j > q_i$ is given by
\begin{equation}
F_{ij, \mathrm{hemi}} (k, \{q_i \}) = \theta(q_j - q_i)  \delta (k-q_i). 
\end{equation}
\begin{figure}[t] 
\begin{center}
\includegraphics[width = 15.5cm]{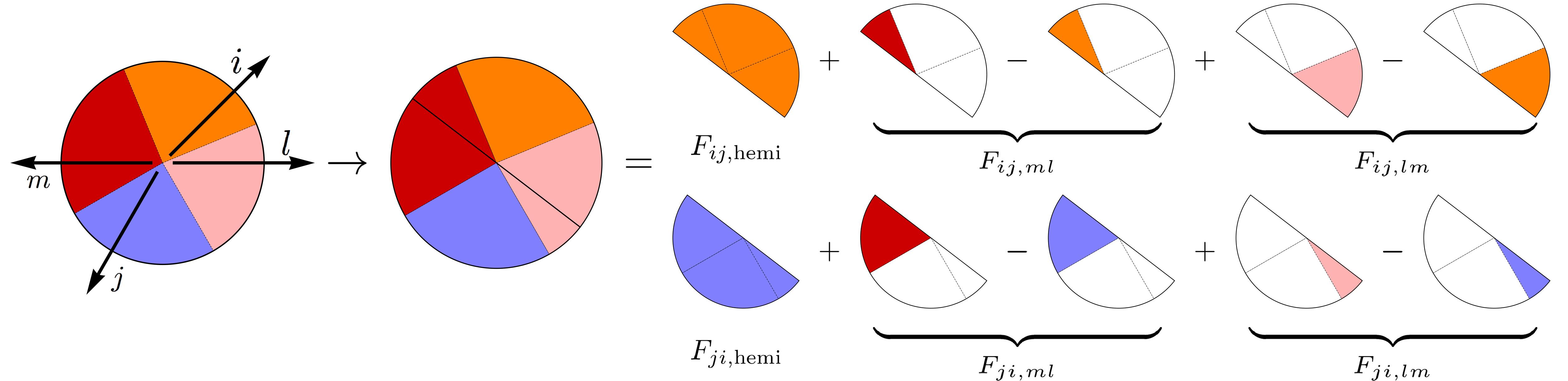} 
\end{center} \vspace{-0.6cm}
\caption{\label{hemisphere}\baselineskip 3.0ex  The real soft contribution is decomposed into the
hemisphere soft functions $F_{ij, \mathrm{hemi}}$ and $F_{ji, \mathrm{hemi}}$ and 
the non-hemisphere soft functions. The hemisphere functions contain all the divergences, 
while the non-hemisphere functions are finite. Here we focus
on the hemisphere soft functions.} 
\end{figure}
And the non-hemisphere functions are given by
\begin{align}
F_{ij, ml} (k, \{q_i \}) &= \theta (q_j - q_i)\theta (q_i - q_m) \theta (q_l - q_m) 
\Bigl( \delta (k-q_m) -\delta (k-q_i)\Bigr),  \nonumber \\
F_{ij, lm} (k, \{q_i \}) &= \theta (q_j - q_i) \theta (q_i -q_l) \theta (q_m - q_l) 
\Bigl( \delta (k-q_l) -\delta (k-q_i)\Bigr),
\end{align}
which are the non-hemisphere measurement function for region $m$ and $l$ respectively.
Note that the constraint function $F$ is constructed for the 
gauge boson emitted from the soft Wilson lines $S_i$ and $S_j^{\dagger}$ ($Y_i$ and $Y_j^{\dagger}$). 
The hemisphere function for the $i$ and $j$ jet directions contains the collinear
and the soft divergences. The contribution to the $l$ and $m$ directions only contains the soft divergence, 
which is subtracted from the corresponding region of the hemisphere parts.  
Here we focus on the hemisphere soft functions, from which the anomalous dimensions are obtained.
The decomposition of the soft real contribution into the hemisphere functions and the non-hemisphere functions
are schematically shown in figure~\ref{hemisphere}. 
As can be seen in the figure, the soft divergence in the 
non-hemisphere parts is cancelled by the corresponding subtraction from the hemisphere parts.

\begin{figure}[b] 
\begin{center}
\includegraphics[height=6cm]{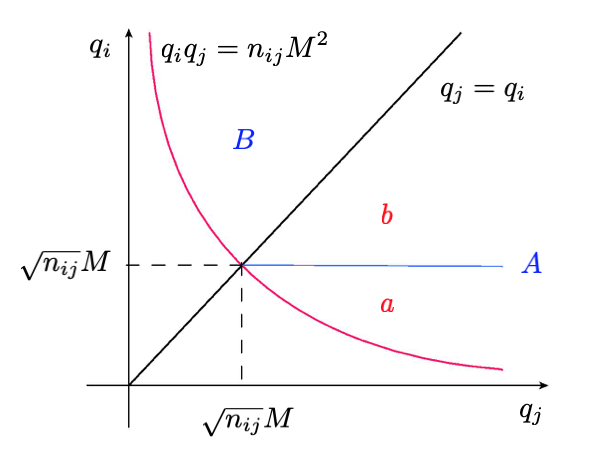} 
\end{center} \vspace{-0.6cm}
\caption{\label{softphase}\baselineskip 3.0ex Phase space for a real, soft gauge boson emission.
Region $A$ [$F_{ij,\mathrm{hemi}} (k, \{ q_i \})$] corresponds to $q_i q_j >n_{ij} M^2$ and $q_j> q_i$. 
Region $B$ [$F_{ji,\mathrm{hemi}} (k, \{ q_i \})$] corresponds
to $q_i q_j > n_{ij} M^2$ and $q_i >q_j$. Region $A$ is subdivided into the region $a$ with 
$k < \sqrt{n_{ij}} M$ and the region $b$ with $k > \sqrt{n_{ij}} M$.}
\end{figure}

The phase space for the real emission is shown in figure~\ref{softphase}, and we compute the real
contribution in the phase space $A$, which corresponds to the hemisphere constraint 
$F_{ij,\mathrm{hemi}} (k, \{ q_i \})$. 
The real contribution from the region $A$ is given by 
\begin{equation} 
S_{ij,\mathrm{hemi}}^{RA} =
-\frac{\alpha}{2\pi} \frac{(\mu^2 e^{\gamma_{\mathrm{E}}} )^{\eps}}{\Gamma(1-\eps)} (\nu
n_{ij} )^{\eta} \int dq_j \frac{1}{kq_j^{1+\eta}} \Bigl( \frac{kq_j}{n_{ij}} -M^2
\Bigr)^{-\eps} \theta (q_j -k) \theta (kq_j -n_{ij} M^2). 
\end{equation} 
We divide the phase space into the region $a$ with $k< \sqrt{n_{ij}} M$ and the region $b$ with $k>
\sqrt{n_{ij}} M$, and the integral in the region $a$ is given as 
\begin{align} 
I_a&= -\frac{\alpha}{2\pi} \frac{(\mu^2 e^{\gamma_{\mathrm{E}}} )^{\eps}}{\Gamma(1-\eps)}
\frac{(\nu n_{ij} )^{\eta}}{k} \int_{n_{ij} M^2/k}^{\infty} dq_j \frac{1}{q_j^{1+\eta}}
\Bigl( \frac{kq_j}{n_{ij}} -M^2 \Bigr)^{-\eps} \theta (k< \sqrt{n_{ij}} M) \nonumber  \\
&= -\frac{\alpha}{2\pi} e^{\gamma_{\mathrm{E}} \eps} \Bigl( \frac{\mu^2}{M^2} \Bigr)^{\eps}
\Bigl( \frac{\mu}{\nu}\Bigr)^{\eta} \Bigl( \frac{\nu^2}{M^2}\Bigr)^{\eta}
\frac{\Gamma(\eps+\eta)}{\Gamma(1+\eta)} \frac{1}{\mu}\Bigl( \frac{\mu}{k}\Bigr)^{1-\eta}
\nonumber  \\
&= -\frac{\alpha}{2\pi} \Bigl\{ \delta (k) \Bigl[ \frac{1}{\eta}\Bigl(
\frac{1}{\eps} +\ln \frac{\mu^2}{M^2}\Bigr) -\frac{1}{\eps^2} +\frac{1}{\eps} \ln
\frac{\nu}{\mu} +\ln \frac{\nu}{M} \ln \frac{\mu^2}{M^2} +\frac{\pi^2}{12}\Bigr] \nonumber
\\ 
&+ \Bigl( \frac{1}{\eps} +\ln \frac{\mu^2}{M^2}\Bigr) \frac{1}{\mu} \mathcal{L}_0
\Bigl( \frac{k}{\mu}\Bigr) \theta (k< \sqrt{n_{ij}} M)\Bigr\}, 
\end{align} 
where the following relation is used. 
\begin{equation} 
\frac{1}{\mu} \Bigl(
\frac{\mu}{k}\Bigr)^{1-\eta} = \frac{1}{\eta}\delta (k) + \frac{1}{\mu} \mathcal{L}_0
\Bigl( \frac{k}{\mu}\Bigr) +\mathcal{O}(\eta) . 
\end{equation}
The integral in the region $b$ is written as 
\begin{align} 
I_b &=-\frac{\alpha}{2\pi} \frac{(\mu^2 e^{\gamma_{\mathrm{E}}} )^{\eps}}{\Gamma(1-\eps)} 
(\nu n_{ij} )^{\eta} \frac{n_{ij}}{k^{1+\eps}} \int_k^{\infty} dq_j \frac{1}{q_j^{1+\eta}} \Bigl(
q_j - \frac{n_{ij} M^2}{k}\Bigr)^{-\eps} \theta (k > \sqrt{n_{ij}} M) \nonumber \\ 
&= -\frac{\alpha}{2\pi} \frac{1}{k} \Bigl( \frac{1}{\eps} + \ln \frac{n_{ij} \mu^2}{k^2} \Bigr)
\theta (k > \sqrt{n_{ij}} M). 
\end{align} 
Note that $I_a + I_b$ is remains the same when $i$ and $j$ are switched. 
Therefore the real contribution is twice 
$S_{ij,\mathrm{hemi}}^{RA}$ because the integration in the region $B$ is the same. The real
hemisphere contribution is given by
\begin{align} \label{realsoft}
S_{ij,\mathrm{hemi}}^{R} &=
-\frac{\alpha}{2\pi} \Bigl\{ \delta (k) \Bigl[ \frac{2}{\eta}\Bigl( \frac{1}{\eps} +\ln
\frac{\mu^2}{M^2}\Bigr) -\frac{2}{\eps^2} +\frac{1}{\eps} \ln \frac{\nu^2}{\mu^2} 
+\ln \frac{\nu^2}{M^2} \ln \frac{\mu^2}{M^2} +\frac{\pi^2}{6}\Bigr]  \\ 
&+ \Bigl(
\frac{1}{\eps} +\ln \frac{\mu^2}{M^2}\Bigr) \frac{2}{\mu} \mathcal{L}_0 \Bigl(
\frac{k}{\mu}\Bigr) \theta (k< \sqrt{n_{ij}} M) + \frac{2}{k} \Bigl( \frac{1}{\eps} + \ln
\frac{n_{ij} \mu^2}{k^2} \Bigr) \theta (k > \sqrt{n_{ij}} M)\Bigr\}. \nonumber
\end{align}

With the virtual contribution in eq.~\eqref{svirt1} and the real contribution in eq.~\eqref{realsoft}, the 
hemisphere soft function  can be written as
\begin{equation}
\mathbf{S}_{\mathrm{hemi}} (a_1, a_2, a_3, a_4) = \sum_{i\neq j} \Bigl[ \mathbf{S}_{ij}^V (a_1, a_2, a_3, a_4) 
S_{ij, \mathrm{hemi}}^V + \mathbf{S}_{ij}^R (a_1, a_2, a_3, a_4) S_{ij, \mathrm{hemi}}^R \Bigr].
\end{equation}
Here we represent the color structure of the soft function in the form 
 $\mathbf{S} (a_1, a_2, a_3, a_4)$,  in which the indices $a_i$ represent 
the presence of the nonsinglets from the originating $i$-th collinear particle  ($i=1,2$  
for the incoming particles, and $i=3, 4$ for the 
outgoing particles  in our convention). For example, the soft color matrix  
with all the singlets is given 
by $\mathbf{S}(0,0,0,0)$ and the soft color matrix with the 
nonsinglet contributions from 1 and 3 is denoted as $\mathbf{S} (1,0,1,0)$, etc.. 
For the soft color matrix at tree level, see appendix~\ref{stree}. 
With these color matrices, the real soft function for the 
nonsinglets 1 and 3 is written as $\sum_{i \neq  j} \mathbf{S}^{R}_{ij} (1, 0, 1, 0)  S_{ij}^R$, 
where $S_{ij}^R$ is the soft correction, which is given by eq.~\eqref{realsoft}.
Using this convention, the hemisphere soft function from $\mathbf{S}_{\mathrm{hemi}} (1,0,1,0)$ 
incorporates $S_{IJ}^{c0e0}$ by putting the generators $T^c$ and $T^e$ for the first and the third indices 
respectively, and $T^0$ in the second and the fourth indices.
 
 The virtual and real contributions to the $\mu$ and $\nu$ soft anomalous dimensions 
without the group theory factors are given as
\begin{align}
\frac{d S^V_{ij, \mathrm{hemi}}}{d\ln \mu} &= -\frac{\alpha}{\pi}\delta (k) \ln \frac{n_{ij} \nu^2}{\mu^2}, \ 
\frac{d S^R_{ij, \mathrm{hemi}}}{d\ln \mu} = -\frac{\alpha}{\pi} \Bigl[ \delta (k) \ln \frac{\nu^2}{\mu^2} 
+\frac{2}{\mu} \mathcal{L}_0 \Bigl(\frac{k}{\mu}\Bigr)\Bigr], \
\nonumber \\
\frac{d S^V_{ij, \mathrm{hemi}}}{d\ln \nu} &= -\frac{\alpha}{\pi} \delta (k) \ln \frac{\mu^2}{M^2}, \ \
\frac{d S^R_{ij, \mathrm{hemi}}}{d\ln \nu} =  -\frac{\alpha}{\pi} \delta (k) \ln \frac{\mu^2}{M^2}.
\end{align}
The derivatives of the Laplace-transformed soft parts are given by
\begin{align} \label{lasoft}
\frac{d \tilde{S}^V_{ij, \mathrm{hemi}}}{d\ln \mu}  &= -\frac{\alpha}{\pi} \ln \frac{n_{ij} \nu^2}{\mu^2}, \ 
\frac{d \tilde{S}^R_{ij, \mathrm{hemi}}}{d\ln \mu}  = -\frac{\alpha}{\pi}     \ln \frac{\nu^2 Q_L^2}{\mu^4}, \
  \\
\frac{d \tilde{S}^V_{ij, \mathrm{hemi}}}{d\ln \nu} &= -\frac{\alpha}{\pi}  \ln \frac{\mu^2}{M^2}, \ \
\frac{d \tilde{S}^R_{ij, \mathrm{hemi}}}{d\ln \nu} =  -\frac{\alpha}{\pi} \ln \frac{\mu^2}{M^2}.
\end{align}

It is noteworthy to compare eq.~\eqref{lasoft} with the results in ref.~\cite{Manohar:2018kfx}, 
in which the soft anomalous dimensions are  given in inclusive cross sections. In ref.~\cite{Manohar:2018kfx}, 
the virtual and the real contributions at order $\alpha$ are the same except the sign, and the 
$\mu$-anomalous dimensions depend on $n_{ij}$. 
However, in the case of the jettiness in which we give constraints on 
the phase space in the real emissions, the contribution to the anomalous dimension 
from the virtual corrections is the same as in ref.~\cite{Manohar:2018kfx}, 
but the contribution from the real emission is different, especially it is independent of $n_{ij}$. 

Due to this difference, the $N$-jettiness soft function with four nonsinglets does not have mixing in contrast to
the inclusive soft function, in which the mixing is induced at one loop. The virtual corrections do not cause 
mixing, while the mixing is cancelled in the sum of the real contributions because the real soft anomalous 
dimensions in  eq.~\eqref{lasoft} is independent of $n_{ij}$.  The proof that there is no mixing 
in the $N$-jettiness soft function with four nonsinglets at one loop is presented in detail in 
appendix~\ref{snomix}.

\subsection{Soft anomalous dimensions} 
The soft functions are written as matrices in the operator basis. The Laplace transform of the soft function
is given as
\begin{equation}
\tilde{\mathbf{S}} \Bigl(\ln \frac{Q_L}{\mu}, M, \mu\Bigr) = \int_0^{\infty} dk e^{-sk} \, \mathbf{S}(k, M, \mu), 
\ s = \frac{1}{e^{\gamma_{\mathrm{E}}}  Q_L}.
\end{equation}
The RG equation with respect to the renormalization scale $\mu$ for the soft function in 
Laplace transform is written as
\begin{equation} \label{rgsoft}
\frac{d}{d\ln \mu}  \tilde{\mathbf{S}} =   \tilde{\bm{\Gamma}}_S^{\mu\dagger}  \tilde{\mathbf{S}}
 +  \tilde{\mathbf{S}}   \tilde{\bm{\Gamma}}_S^{\mu},
\end{equation}
where $\tilde{\bm{\Gamma}}_S^{\mu}$ is the $\mu$-anomalous dimension matrix.
At NLO, eq.~\eqref{rgsoft} is written as
\begin{equation} \label{rgsoft1}
\frac{d}{d\ln \mu} \tilde{\mathbf{S}}^{(1)} =  \tilde{\bm{\Gamma}}_S^{\mu\dagger (1)} 
\tilde{\mathbf{S}}^{(0)} + \tilde{\mathbf{S}}^{(0)} \tilde{\bm{\Gamma}}_S^{\mu (1)},
\end{equation}
where $\mathbf{S}^{(1)}$ is the renormalized soft function. 
The soft anomalous dimensions can be extracted from the requirement that the sum of the anomalous 
dimensions from all the factorized parts should cancel. 
The soft anomalous dimension
at one loop is given by
\begin{equation}
\tilde{\bm{\Gamma}}_S^{\mu(1)} = -\Bigl(\tilde{\bm{\Gamma}}_H^{(1)}  +\frac{1}{2} 
(\tilde{\gamma}_{B_1}^{\mu (1)} +\tilde{\gamma}_{B_2}^{\mu (1)}
+\tilde{\gamma}_{J_3}^{\mu (1)} +\tilde{\gamma}_{J_4}^{\mu (1)}) 
\otimes \mathbf{1}\Bigr),
\end{equation}
where $\tilde{\gamma}_C^{\mu}$ is the anomalous dimensions of the beam functions or the jet functions.
The $\mu$-anomalous dimensions of the soft function with $k$ nonsinglets ($k=0, 2, 3, 4$) 
are  given as
\begin{equation} \label{gamsmu}
 (\tilde{\bm{\Gamma}}_S^{\mu}  )_k^{(1)} = -C_F \Gamma_c (\alpha) 
\Bigl(\ln \frac{\mu^4 n_{12} n_{34}}{Q_L^4} -2i\pi\Bigr)\times
\mathbf{1} - \Gamma_c(\alpha) \mathbf{M} +\frac{C_A\Gamma_c}{2} k\ln
\frac{\mu^2}{\nu Q_L} \times \mathbf{1},
\end{equation} 
where $\mathbf{M}$ is the matrix in eq.~\eqref{mij} appearing in the hard function. As in the hard function,
the imaginary part in the identity matrix does not contribute to the evolution.

The RG equation with respect to the rapidity scale $\nu$ is written as
\begin{equation}
\frac{d}{d\ln \nu} \tilde{\mathbf{S}} = \tilde{\Gamma}_S^{\nu} \tilde{\mathbf{S}}.
\end{equation}
After color algebra~\cite{Sjodahl:2012nk}, the $\nu$-soft anomalous dimensions with $k$ 
nonsinglets  ($k \neq 1$) are given as 
\begin{equation}  \label{gamsnu}
(\tilde{\Gamma}^{\nu}_S)_k^{(1)} 
= -\frac{C_A\Gamma_c}{2} k\ln \frac{\mu^2}{M^2}.
\end{equation}
Note that there is no contribution from those with one nonsinglet $k=1$ due to charge conservation.

\section{Renormalization group evolution\label{rge}} 
In order to study the RG evolution, it is convenient to
make a Laplace transform of the $N$-jettiness which involves the convolution of the
jet, beam and the soft functions in the factorization formulae, eqs.~\eqref{taufac} and \eqref{sctwofac}.
The convolution becomes the product of the factorized parts with their Laplace transforms. 
Here  we choose $s =1/[Q_L\exp(\gamma_{\mathrm{E}})]$, which is the conjugate variable to the jettiness. 
Then we can take inverse Laplace transform and express the evolutions accordingly~\cite{Becher:2006mr}.
  
In order to resum large logarithms,
the RG evolutions of the factorized functions start from their own characteristic scales to the common factorization scale 
$\mu_F$ and the rapidity scale $\nu_F$. 
The characteristic scales are the scales which minimize the logarithms in the factorized functions, 
and they are give by
\begin{equation}
\mu_H  \sim \omega_i,  \ \mu_{Ci} \sim \sqrt{\omega_i M}, \ \mu_S \sim M, \ 
\nu_C \sim  \omega_i, \  \nu_S \sim M,  
\end{equation}
where $\omega_i$ ($i=1, \cdots, 4$) are the largest collinear components. 
The characteristic collinear scales (hard-collinear scales in \sctwo) $\mu_{Ci}$ apply both to the
singlets and the nonsinglets of the collinear functions. The collinear rapidity scale 
$\nu_C$ belongs to the nonsinglets. The soft scale $\mu_S$
applies to both the singlets and the nonsinglets, while the scale $\nu_S$ belongs to the nonsinglet soft functions. 

\subsection{Hard function} 

The anomalous dimension of the hard function is given by eq.~\eqref{gamh}.
The evolution of the hard function from the high-energy scale $\mu_H$ to the factorization scale
$\mu_F$ is written as 
\begin{equation} 
\mathbf{H}(\mu_F) =\Pi_H (\mu_F, \mu_H) \bm{\Pi}_H
(\mu_F, \mu_H) \mathbf{H} (\mu_H) \bm{\Pi}_H^{\dagger} (\mu_F, \mu_H), 
\end{equation} 
where $\Pi_H (\mu_F, \mu_H)$ is the evolution from the identity matrix of the anomalous
dimension, and $\bm{\Pi}_H (\mu_F, \mu_H)$ is obtained by the exponentiating the matrix
$\mathbf{M}$. They can be written as 
\begin{equation} 
\Pi_H (\mu_F,\mu_H) = \exp \Bigl[ -8
C_F S(\mu_F, \mu_H) -4 C_F a_{\Gamma} (\mu_F,\mu_H) \ln \frac{\mu_H^2}{s} +
8a_{\gamma_{\ell}} (\mu_F, \mu_H) \Bigr], 
\end{equation} 
where 
\begin{align} 
&\int_{\mu_H}^{\mu_F} \frac{d\mu}{\mu} \Gamma_c (\alpha) \ln \frac{\mu^2}{s} 
 \equiv 2S_{\Gamma} (\mu_F, \mu_H) +a_{\Gamma} (\mu_F, \mu_H) \ln
\frac{\mu_H^2}{s}, 
\end{align} 
and $S_{\Gamma}$ and $a_{f}$ are given as 
\begin{equation} 
S_{\Gamma} (\mu,\mu_i) =
\int_{\alpha(\mu_i)}^{\alpha (\mu)} d\alpha \frac{\Gamma_c (\alpha)}{\beta (\alpha)} \int_{\alpha
(\mu_i)}^{\alpha} \frac{d\alpha^{\prime}}{\beta(\alpha^{\prime})}, \ \
a_f (\mu, \mu_i) = 
\int_{\alpha (\mu_i)}^{\alpha (\mu)} \frac{d\alpha}{\beta(\alpha)} f(\alpha), 
\end{equation} 
for an arbitrary function $f(\alpha)$.  The explicit results of $S_{\Gamma}$, $a_{\Gamma}$ and
$a_{\gamma_\ell}$ are presented at next-to-next-to-leading-logarithmic accuracy 
in ref.~\cite{Stewart:2010qs}. And $\bm{\Pi}_H$ is obtained by exponentiating $\mathbf{M}$ as 
\begin{equation} 
\bm{\Pi}_H (\mu_F, \mu_H) =\exp \Bigl[ a_{\Gamma} (\mu_F,\mu_H) \mathbf{M}\Bigr]. 
\end{equation}

\subsection{Collinear functions} 
We present the evolution of the beam functions as a representative of the collinear functions. 
Since the semi-inclusive jet functions and the FJFs have the same  anomalous dimensions as the 
beam functions, their evolutions can be described in a similar way.

The RG equation with respect to $\mu$ for the singlet beam function is given by
\begin{equation}
\frac{d}{d\ln \mu} \tilde{B}_s (\mu) = \tilde{\gamma}_{Bs}^{\mu}\tilde{B}_s (\mu),
\end{equation}
where the $\mu$-anomalous dimension $\tilde{\gamma}_{Bs}^{\mu}$
for the singlet is given by eq.~\eqref{beamanon}.
The evolution from the collinear scale $\mu_C$ to the factorization scale $\mu_F$ is given by 
\begin{equation} 
\tilde{B}_s (\mu_F) =U_{Bs} (\mu_F, \mu_C) \tilde{B}_s (\mu_C), 
\end{equation} 
where the evolution kernel $U_{Bs}$ is given by
\begin{equation} \label{pib}
U_{Bs} (\mu_F, \mu_C) =\exp\Bigl[ 4 C_F S_{\Gamma} (\mu_F,\mu_C) 
-2 C_F a_{\Gamma} (\mu_F, \mu_C) \ln
\frac{\omega Q_L}{\mu_C^2} -2 a_{\gamma_{\ell}} (\mu_F, \mu_C)\Bigr]. 
\end{equation}

For the nonsinglet, the $\mu$ and $\nu$ RG equations are given by
\begin{equation}
\frac{d}{d\ln \mu} \tilde{B}_n (\mu, \nu) = \tilde{\gamma}_{Bn}^{\mu} \tilde{B}_n (\mu, \nu),  \ 
\frac{d}{d\ln \nu} \tilde{B}_n (\mu, \nu) = \tilde{\gamma}_{Bn}^{\nu} \tilde{B}_n (\mu, \nu),
\end{equation}
where the $\mu$-anomalous dimensions $\tilde{\gamma}_{Bn}^{\mu}$ and the $\nu$-anomalous
dimensions $\tilde{\gamma}_{Bn}^{\nu}$ for the nonsinglet are given by eq.~\eqref{beamanon}.
The order of the evolutions with respect to $\mu$ and $\nu$ is irrelevant, and here we evolve 
the beam function with respect to $\nu$ first, and then with respect to $\mu$.
Since the $\nu$-anomalous dimension contains a large logarithm with $\mu \sim \sqrt{\omega Q_L}\gg M$, 
we resum this large logarithm 
first in expressing the evolution with respect to $\nu$~\cite{Chiu:2012ir}.
Then the evolution is written as
\begin{equation}
\tilde{B}_n (\mu_F, \nu_F) = U_{Bn} (\mu_F, \mu_C;\nu_F) V_{Bn} (\nu_F, \nu_C; \mu_C) 
\tilde{B}_n (\mu_C, \nu_C),
\end{equation}
where the $\mu$-evolution kernel $U_{Bn}$ and the $\nu$-evolution kernel $V_{Bn}$ are given by
\begin{align} 
V_{B_n} (\nu_F, \nu_C;  \mu_C) &= \exp \Bigl[ C_A a_{\Gamma} (\mu_C, M) 
\ln \frac{\nu_F}{\nu_C}\Bigr],  \\
U_{Bn} (\mu_F, \mu_C;\nu_F) &=   U_{Bs} (\mu_F, \mu_C) 
\exp \Bigl[ - 2C_A S_{\Gamma}(\mu_F, \mu_C) + C_A a_{\Gamma} 
(\mu_F, \mu_C) \ln \frac{\nu_F Q_L}{\mu_C^2} \Bigr]. \nonumber 
\end{align} 
 
The anomalous dimensions of the semi-inclusive jet functions in eq.~\eqref{jetanol} are the same as those
of the beam functions. Therefore the evolution of the semi-inclusive jet functions is the same as the FJF. However,
the largest lightcone component in each $i$-th collinear direction is denoted by $\omega_i$, which determines
the characteristic collinear scale in each direction. Let us denote the collinear functions as $C_i (\mu_F, \nu_F)$,
where it corresponds to the beam functions for $i=1, 2$ and to the semi-inclusive jet functions or the 
FJFs for $i=3, 4$. Then the evolution of the collinear functions are written as ($i= 1, 2, 3, 4$)
\begin{align}
\tilde{C}_{is} (\mu_F) &=U_{is} (\mu_F, \mu_{Ci}) \tilde{C}_{is} (\mu_{Ci}), \nonumber \\
\tilde{C}_{in} (\mu_F, \nu_F) &= U_{in} (\mu_F, \mu_{Ci};\nu_F) V_{in} (\nu_F, \nu_C; \mu_{Ci}) 
\tilde{C}_{in} (\mu_{Ci}, \nu_C),
\end{align}
where the evolution kernels are given as
\begin{align}
U_{is} (\mu_F, \mu_{Ci}) &=\exp\Bigl[ 4 C_F S_{\Gamma} (\mu_F,\mu_{Ci}) 
-2 C_F a_{\Gamma} (\mu_F, \mu_{Ci}) \ln
\frac{\omega_i Q_L}{\mu_{Ci}^2} -2 a_{\gamma_{\ell}} (\mu_F, \mu_{Ci})\Bigr], \nonumber \\
U_{in} (\mu_F, \mu_{Ci};\nu_F) &=   U_{is} (\mu_F, \mu_{Ci}) 
\exp \Bigl[ - 2C_A S_{\Gamma}(\mu_F, \mu_{Ci}) + C_A a_{\Gamma} 
(\mu_F, \mu_{Ci}) \ln \frac{\nu_F Q_L}{\mu_{Ci}^2} \Bigr], \nonumber \\
V_{in} (\nu_F, \nu_C; \mu_{Ci}) &= \exp \Bigl[ C_A a_{\Gamma} (\mu_{Ci},M) 
\ln \frac{\nu_F}{\nu_C}\Bigr].
\end{align}
 
\subsection{Soft function} 
The $\mu$ and $\nu$ anomalous dimensions for the soft function are given by 
eqs.~\eqref{gamsmu} and \eqref{gamsnu} respectively.
The  evolution of the soft function can be proceeded as in the case of the hard function and 
the soft evolution with the $k$ nonsinglets can be written as 
\begin{equation} 
\tilde{\mathbf{S}}_k (\mu_F, \nu_F) = \Pi_{Sk} (\mu_F, \mu_S; \nu_F, \nu_S) 
\bm{\Pi}_S^{\dagger} (\mu_F, \mu_S) \tilde{\mathbf{S}}_k (\mu_S, \nu_S) \bm{\Pi}_S (\mu_F, \mu_S), 
\end{equation} 
where the evolution kernel $\Pi_{Sk} (\mu_F, \mu_S; \nu_F, \nu_S)$ comes from the identity matrix, 
and the evolution kernel $\bm{\Pi}_S (\mu_F, \mu_S)$ from $\mathbf{M}$.

Here we also evolve with respect to $\nu$ first, and then $\mu$. 
The evolution kernel $\Pi_{Sk}$ can be written as 
\begin{equation}
\Pi_{Sk} (\mu_F, \mu_S; \nu_F, \nu_S) = U_{Sk} (\mu_F, \mu_S; \nu_F) V_{Sk} (\nu_F, \nu_S; \mu_S), 
\end{equation}
where
\begin{align}
V_{Sk} (\nu_F, \nu_S; \mu_S) &= \exp \Bigl[ -k C_A a_{\Gamma} (\mu_S, M) 
\ln \frac{\nu_F}{\nu_S}\Bigr],  \nonumber \\
U_{Sk} (\mu_F, \mu_S; \nu_F) &= \exp \Bigl[ -8 C_F S_{\Gamma} (\mu_F, \mu_S ) 
-4C_F a_{\Gamma} (\mu_F, \mu_S) \ln \frac{\mu_S^2}{Q_L^2} 
\nonumber \\
&-2C_F a_{\Gamma} (\mu_F, \mu_S) \ln (n_{12} n_{34}) 
+ 2k C_A S_{\Gamma} (\mu_F, \mu_S) \nonumber \\
&- kC_A a_{\Gamma} (\mu_F, \mu_S) \ln \frac{\nu_F Q_L}{\mu_S^2} \Bigr],
\end{align}
and $\bm{\Pi}_S (\mu_F, \mu_S)$ is given by
\begin{equation}
\bm{\Pi}_S (\mu_F, \mu_S) = \exp \Bigl[ -a_{\Gamma} (\mu_F, \mu_S) \mathbf{M}\Bigr].
\end{equation}
  
With the addition of nonsinglets, the anomalous dimension of the hard function is not affected, 
and the sum of all the other $\nu$-anomalous dimensions for any number of
nonsinglets is cancelled. 
\begin{align} 
&(\tilde{\Gamma}_S^{\nu})_k^{(1)} 
  + \Bigl(\tilde{\gamma}_{B}^{\nu} (\omega_1)
+\tilde{\gamma}_{B}^{\nu} (\omega_2) +\tilde{\gamma}_{J}^{\nu} (\omega_3)
+\tilde{\gamma}_{J}^{\nu} (\omega_4)\Bigr)_k =0,
\end{align} 
where it is understood that the soft function with the $k$ nonsinglets should be employed, 
if there are $k$ nonsinglets in the collinear parts.
 
 \section{Conclusion and outlook\label{conc}}
The analysis of the $N$-jettiness in high-energy electroweak processes is more sophisticated due to the presence 
of the nonsinglet contributions. It may be looked upon as a mere copy of QCD, but
the most notable distinction, compared with QCD, is that a lot of different channels 
involving nonsinglet contributions enter the expression of the $N$-jettiness. In QCD, only the projection
to the color singlets for the collinear and the soft functions survives. As a result, the definitions 
of the factorized collinear and soft functions  
should be extended to the nonsinglet contributions  to include all the possible channels. With these
additional ingredients, we have established the factorization theorem for the $N$ jettiness in weak interaction. 
We have chosen the simplest $SU(2)$ gauge interaction only to show the distinction of the participation
of the nonsinglets in the initial and the final states. The extension to the full Standard Model is 
nontrivial due to the additional particles, the gauge mixing and the effect of the electroweak symmetry 
breaking, but is necessary for phenomenology. It will be the next subject to pursue, following this development.

According to the different hierarchy of scales, different effective theories are employed. 
When $\mathcal{T}^2 \sim M^2 \ll p_c^2 \sim Q\mathcal{T}\ll Q^2$, \scone\ is appropriate, 
and both the collinear and the soft functions contribute to the jettiness. 
On the other hand, when $\mathcal{T}^2 \sim M^2 \sim p_c^2 \ll Q\mathcal{T} \ll Q$, 
we employ \sctwo.  In this case, the collinear functions are the PDF or the fragmentation functions, 
which do not contribute to the jettiness, and only the soft function
seems to contribute to the jettiness.  However, the effect of the hard-collinear modes 
by integrating out the \scone-collinear modes through the matching coefficients contributes to the jettiness.  
Though the detailed physics is different in \scone\ and in \sctwo, 
if we combine the contributions of the matching coefficients and the PDF or the fragmentation functions, and 
identify them as the beam functions or the FJFs,
the jettiness in both cases can be obtained by computing the beam functions and the FJFs in \scone.   
Taking account of all the intricacies, we have established the factorization of the 2-jettiness both 
in \scone\ and in \sctwo. In the computation of the jettiness, the gauge boson mass is regarded as small, 
and we take the small mass limit in our final results.
Note that there is no IR divergence due to the physical gauge boson mass $M$, however small that is. 

When the nonsinglets participate in the scattering,  the main distinction is the existence of the rapidity 
divergence in the collinear and soft functions. 
As in QCD, there is no rapidity divergence in the singlet contributions. However, the rapidity divergence
arises when the nonsinglets are involved in the factorized parts owing to the different group 
theory factors between the real and the virtual contributions. 
Of course, in the final expression for the $N$-jettiness when all the factorized parts 
are added, the rapidity divergence cancels for  any number of nonsinglets. For the effective theory 
to be consistent, it should hold true because the full theory is free of the rapidity divergence.  
However, the effect of rapidity 
divergence in each collinear and soft sectors plays an important role in resumming large logarithms. 
Due to the presence of the double RG evolutions with respect to the renormalization scale $\mu$ 
and the rapidity scale $\nu$ for the nonsinglet contributions,  
we have to solve the coupled RG equation to evolve
with respect to both of them, and the results have been presented here.  

As mentioned in section~\ref{jettifac}, it is important to study the possible violation of the factorization in weak interaction
due to the Glauber exchange between the spectator partons. It arises when the nonsinglets participate in 
the scattering because the group theory factors are not the same for different configurations of the Glauber gauge bosons across
the unitarity cuts. 
The presence of the rapidity divergence due to the nonsinglets appear from a similar source. Therefore it is critical to look into 
the Glauber exchange in considering the factorization. It is beyond the scope of this paper, and we will investigate this topic
in the future.
 
Despite the fact that we only employed the $SU(2)$ gauge interaction, we reiterate that 
this opens up a lot of possibilities
in the phenomenology of the high-energy lepton colliders. The first task will be to include all the interactions 
to delineate the Standard Model completely. It also involves the electroweak symmetry breaking and the 
effect of the masses of the heavy particles. Next, the additional ingredients in the factorization should be provided
to yield theoretical predictions. We have considered $e^- e^+ \rightarrow \mu^-\mu^+$, but other modes 
such as $e^- e^+ \rightarrow W^+ W^-$, and the Higgs production should also be included 
for the study of the phenomenology. These topics will be our next areas of research.

\appendix

\section{Laplace transforms of the distributions  \label{laplace}} 

It is convenient to consider the Laplace transform of the $N$-jettiness, in which the factorized
parts are written as the products of the hard, collinear and soft functions. 
After the individual parts are evolved, we can make an inverse
Laplace transformation~\cite{Becher:2006mr} to obtain the original $N$-jettiness.  
Another advantage is that the anomalous dimensions with the Laplace transforms are ordinary functions, while the 
original anomalous dimensions may contain distributions. 
Therefore the solution of the RG equation in the Laplace transform can be written in a more tangible form.

Let us begin with the Laplace transform of the soft function, which is given as
\begin{align} \label{lapsoft}
\tilde{S} \Bigl(\ln \frac{Q_L}{\mu}, M, \mu\Bigr) &= \int_0^{\infty} dk e^{-sk} S(k, M, \mu), 
\ s = \frac{1}{e^{\gamma_{\mathrm{E}}}  Q_L},  \nonumber \\
S (k, M, \mu) &= \frac{1}{2\pi i} \int_{c-i\infty}^{c+i\infty} ds e^{sk} \tilde{S} 
\Bigl( \ln \frac{1}{e^{\gamma_{\mathrm{E}}} s\mu},  M, \mu \Bigr).
\end{align}
where the contour is chosen to stay to the right of all discontinuities ($c>0$) in the inverse Laplace transform.
The scale $Q_L$ is a conjugate variable to $s$. 
 
 Note that the variable $s$ in the Laplace transform should be common to all 
the factorized parts including the collinear part, that is, it should be conjugate to the jettiness. 
 It is straightforward to express the Laplace transform of the collinear functions with the same form as in 
eq.~\eqref{lapsoft} by noting that $t = \omega k$ in the beam functions and $p^2  = \omega k$ 
in the jet functions, where $k$ represents the jettiness.
\begin{align} \label{lapcol2}
\tilde{B}_i \Bigl( \ln \frac{\omega Q_L}{\mu^2} ,z, M, \mu \Bigr) &= \int_0^{\infty} dk \, e^{-s k} 
B_i (\omega k, z, M, \mu), \nonumber \\
 B_i (\omega k, z, M, \mu) &= \frac{1}{2\pi i} \int_{c-i\infty}^{c+i\infty} ds 
\, e^{s k} \tilde{B}_i \Bigl( \ln \frac{1}{e^{\gamma_{\mathrm{E}}} s \mu }, M, \mu \Bigr), \nonumber \\
\tilde{J}_i  \Bigl(\ln \frac{\omega Q_L}{\mu^2} , M, \mu\Bigr) &= \int_0^{\infty} dk e^{-s k} J_i (\omega k, M, \mu) , 
 \nonumber \\
J_i (\omega k, M, \mu) &= \frac{1}{2\pi i} \int_{c-i\infty}^{c+i\infty} ds e^{sk} \tilde{J}_i 
\Bigl( \ln  \frac{1}{e^{\gamma_{\mathrm{E}}} s\mu }, M, \mu \Bigr),
\end{align}
with $s=1/(e^{\gamma_{\mathrm{E}}} Q_L)$.
 
In the collinear and the soft functions, the distributions arise from the expressions
$\mu^{2\eps}/(\omega k)^{1+\eps}$ and $\mu^{\eps}/k^{1+\eps}$ respectively.
They can
be expanded in powers of $\eps$, and can be written as 
\begin{align} \label{dis}
\frac{\mu^{2\eps}}{(\omega k)^{1+\eps}} &= -\frac{1}{\eps} \delta (\omega k) +
\frac{1}{\mu^2}\mathcal{L}_0 \Bigl( \frac{\omega k}{\mu^2}\Bigr) -\eps
\frac{1}{\mu^2}\mathcal{L}_1 \Bigl(\frac{\omega k}{\mu^2}\Bigr) +\cdots, \nonumber \\
\frac{\mu^{\eps}}{k^{1+\eps}} &= -\frac{1}{\eps} \delta (k) + \frac{1}{\mu}\mathcal{L}_0
\Bigl(\frac{k}{\mu}\Bigr) - \eps \frac{1}{\mu}\mathcal{L}_1 \Bigl(\frac{k}{\mu}\Bigr)+\cdots.
\end{align}

The Laplace transforms of these functions are given as
\begin{equation} \label{lapdis} 
\int_0^{\infty} e^{-sk} dk \frac{\mu^{2\eps}}{(\omega
k)^{1+\eps}} =\frac{1}{\omega} \Bigl(\frac{s\mu^2}{\omega}\Bigr)^{\eps} \Gamma(-\eps), \ \
\int_0^{\infty} e^{-sk} dk \frac{\mu^{\eps}}{k^{1+\eps}} = (s\mu)^{\eps} \Gamma(-\eps).
\end{equation} 
By expanding eq.~\eqref{lapdis} in powers of $\eps$ and by comparing them
with eq.~\eqref{dis}, we can obtain the Laplace transforms of the distributions.
If we denote the Laplace transform of the function $f(k)$ by $L[f(k)]$,  
the Laplace transforms of the first three terms for the collinear parts are given by
\begin{align} 
&L[\delta (\omega k) ] = \frac{1}{\omega}, \  \ 
L\Bigl[\frac{1}{\mu^2}\mathcal{L}_0 \Bigl( \frac{\omega k}{\mu^2}\Bigr)  \Bigr] 
=-\frac{1}{\omega} \ln \frac{s e^{\gamma_{\mathrm{E}}}\mu^2}{\omega}
=   - \frac{1}{\omega} \ln \frac{\mu^2}{\omega Q_L}, \nonumber \\  
&L\Bigl[\frac{1}{\mu^2}\mathcal{L}_1 \Bigl( \frac{\omega k}{\mu^2}\Bigr) \Bigr] = 
\frac{1}{\omega} \Bigl(\frac{1}{2} \ln^2 \frac{s e^{\gamma_{\mathrm{E}}}\mu^2}{\omega} +\frac{\pi^2}{12}\Bigr)
=\frac{1}{\omega} \Bigl(\frac{1}{2} \ln^2 \frac{\mu^2}{\omega Q_L} +\frac{\pi^2}{12}\Bigr). 
\end{align}
and the first three terms from the soft part are given as
\begin{align}
&L[\delta (k)] = 1, \ \ L\Bigl[\frac{1}{\mu} \mathcal{L}_0 \Bigl( \frac{k}{\mu}\Bigr) \Bigr] 
=  -\ln  (s e^{\gamma_{\mathrm{E}}}\mu) = -\ln \frac{\mu}{Q_L}, \nonumber \\
& L\Bigl[\frac{1}{\mu} \mathcal{L}_1 \Bigl( \frac{k}{\mu} \Bigr) \Bigr]
=\frac{1}{2} \ln^2  (s e^{\gamma_{\mathrm{E}}}\mu)  +\frac{\pi^2}{12}=\frac{1}{2} \ln^2 \frac{\mu}{Q_L} +\frac{\pi^2}{12}. 
\end{align}

\section{Beam functions and the matching coefficients for small $M$\label{beamzero}} 
The matrix element $M_a$ in eq.~\eqref{mabeam} is given by 
\begin{equation} 
M_a = \frac{\alpha}{2\pi}
\theta\Bigl( (1-z) t-zM^2\Bigr) \theta (z) \theta (1-z) \frac{(1-z)t -zM^2}{(t-zM^2)^2}.
\end{equation} 
It is regarded as a distribution both in $t$ and $z$ for small $M$.
When $M_a$ is integrated over $t$ to an arbitrary renormalization scale $\mu^2$, it is given by 
\begin{equation} 
\int_{zM^2/(1-z)}^{\mu^2} dt M_a \rightarrow
\frac{\alpha}{2\pi} (1-z) \Bigl( -1 + \ln \frac{(1-z) \mu^2}{z^2 M^2}\Bigr) \theta (z) \theta (1-z),
\end{equation} 
which is regarded as the coefficient of $\delta (t)$. Because $M_a$ in this
limit behaves like $\sim 1/t$, it can be written as 
\begin{equation} 
M_a = \frac{\alpha}{2\pi} (1-z) \theta (z) \theta (1-z) \Bigl[ \Bigl( -1 + \ln \frac{(1-z) \mu^2}{z^2 M^2}\Bigr) \delta (t) +
f(z,\mu) \frac{1}{\mu^2} \mathcal{L}_0 \Bigl(\frac{t}{\mu^2}\Bigr) \Bigl], 
\end{equation}
where the function $f(z,\mu)$ is to be determined.

Note that we have the identity 
\begin{equation} 
\frac{\mu^{2\eps}}{t^{1+\eps}}
=-\frac{1}{\eps} \delta (t) + \frac{1}{\mu^2} \mathcal{L}_0 \Bigl(\frac{t}{\mu^2}\Bigr)
-\eps \frac{1}{\mu^2} \mathcal{L}_1 \Bigl(\frac{t}{\mu^2}\Bigr) +\cdots. 
\end{equation}
Taking the logarithmic derivative with respect to $\mu^2$, we get 
\begin{align} 
\mu^2 \frac{d}{d\mu^2} \frac{\mu^{2\eps}}{t^{1+\eps}} = \eps \frac{\mu^{2\eps}}{t^{1+\eps}} &=
- \delta (t) + \eps\frac{1}{\mu^2} \mathcal{L}_0 \Bigl(\frac{t}{\mu^2}\Bigr) +\cdots
\nonumber \\ 
&= \mu^2 \frac{d}{d\mu^2} \frac{1}{\mu^2} \mathcal{L}_0
\Bigl(\frac{t}{\mu^2}\Bigr) -\eps \mu^2 \frac{d}{d\mu^2} \frac{1}{\mu^2} \mathcal{L}_1
\Bigl(\frac{t}{\mu^2}\Bigr) + \cdots. 
\end{align} Comparing the coefficients of the
powers of $\eps$, we obtain the result 
\begin{equation} 
\mu^2 \frac{d}{d\mu^2} \Bigl[
\frac{1}{\mu^2} \mathcal{L}_0 \Bigl(\frac{t}{\mu^2}\Bigr) \Bigl] = -\delta (t), \ \mu^2
\frac{d}{d\mu^2} \Bigl[\frac{1}{\mu^2} \mathcal{L}_1 \Bigl(\frac{t}{\mu^2}\Bigr) \Bigr]
=-\frac{1}{\mu^2} \mathcal{L}_0 \Bigl(\frac{t}{\mu^2}\Bigr) , \cdots. 
\end{equation}

In order to determine $f(z,\mu)$, we use the fact that $M_a$ in eq.~\eqref{mabeam} is
independent of $\mu^2$,  $\mu^2 d M_a/d\mu^2 =0$. If we compare
the coefficients of $\delta (t)$, we obtain $f(z,\mu) =1$. The final result is given by
\begin{equation} 
M_a = \frac{\alpha}{2\pi} (1-z) \theta (z) \theta (1-z) \Bigl[ \Bigl( -1 + \ln \frac{(1-z)
\mu^2}{z^2 M^2}\Bigr) \delta (t) + \frac{1}{\mu^2} \mathcal{L}_0
\Bigl(\frac{t}{\mu^2}\Bigr) \Bigl]. 
\end{equation}

The naive collinear contribution $M_b$ in eq.~\eqref{mbbeam} is given by 
\begin{equation}
\tilde{M}_b =\frac{\alpha}{2\pi} \frac{z}{1-z} \frac{1}{t-zM^2} \theta \Bigl( (1-z)t -z
M^2\Bigr) \theta(z) \theta(1-z). 
\end{equation} 
For small $M$, it is proportional to
$1/t(1-z)$, and should be regarded as distributions both in $z$ and $t$. We
first integrate over $t$, and it is given as 
\begin{equation} \label{tintmb}
\int_{zM^2/(1-z)}^{\mu^2} dt \tilde{M}_b \rightarrow 
\frac{\alpha}{2\pi} \theta (z) \theta (1-z) \frac{z}{1-z} \ln \frac{(1-z) \mu^2}{z^2 M^2},
\end{equation} 
which should be regarded as a coefficient of $\delta (t)$, but it is also a
distribution in $z$. Therefore the most general form can be written as 
\begin{align}
\tilde{M}_b &= \frac{\alpha}{2\pi} \Bigl\{ \delta (t) \Bigl[ A \delta (1-z) 
+z\mathcal{L}_0 (1-z) \ln \frac{\mu^2}{z^2 M^2} + z\mathcal{L}_1 (1-z) \Bigr] \nonumber \\
&+ g(z,\mu) \frac{1}{\mu^2} \mathcal{L}_0 \Bigl(\frac{t}{\mu^2}\Bigr) + h(z,\mu)
\frac{1}{\mu^2} \mathcal{L}_1 \Bigl(\frac{t}{\mu^2}\Bigr) \Bigr\}. 
\end{align} 
Because the integration of $\tilde{M}_b$ over $z$ yields a term proportional to $\ln (t/\mu^2)$, 
the distribution $\mathcal{L}_1 (t/\mu^2)$ is included here. Or we can include $\mathcal{L}_n$ with 
$n\geq 2$, but the coefficients of those functions turn out to be zero.

Eq.~\eqref{tintmb} is further integrated over $z$ to yield 
\begin{equation}
\int_0^{\mu^2/(\mu^2 +M^2)} dz  \frac{z}{1-z} 
\ln \frac{(1-z) \mu^2}{z^2 M^2} =  -1 +\frac{\pi^2}{3} 
-\ln \frac{\mu^2}{M^2} +\frac{1}{2}\ln^2 \frac{\mu^2}{M^2},
\end{equation} 
and
\begin{equation}
\int_0^{\mu^2/(\mu^2 +M^2)} dz \Bigl(  z\mathcal{L}_0 (1-z) 
\ln \frac{\mu^2}{z^2 M^2} + z\mathcal{L}_1 (1-z)  \Bigr) = -1 + \frac{\pi^2}{3} -\ln \frac{\mu^2}{M^2}.
\end{equation}
Therefore $A$ is determined to be $A = [\ln^2 (\mu^2/M^2)]/2$.
The unknown quantities $g(z,\mu)$ and $h(z,\mu)$ are determined by requiring that $\tilde{M}_b$ is 
independent of $\mu$, that is, $\mu^2 d \tilde{M}_b/d\mu^2 =0$. They are given by
\begin{equation}
f(z, \mu) = \delta (1-z) \ln \frac{\mu^2}{z^2 M^2} +z \mathcal{L}_0 (1-z), \ g(z,\mu) = \delta (1-z).
\end{equation}

The final result is given by 
\begin{align} 
\tilde{M}_b &=\frac{\alpha}{2\pi} \Bigl[ \delta (t) \delta (1-z) \frac{1}{2} \ln^2 \frac{\mu^2}{M^2}+
\delta (t) z\Bigl(\ln \frac{\mu^2}{z^2 M^2} \mathcal{L}_0 (1-z) + \mathcal{L}_1 (1-z)
\Bigr) \nonumber \\ & + \delta (1-z) \Bigl( \frac{1}{\mu^2} \mathcal{L}_0 \Bigl(
\frac{t}{\mu^2}\Bigr) \ln \frac{\mu^2}{M^2} +\frac{1}{\mu^2} \mathcal{L}_1 \Bigl(
\frac{t}{\mu^2}\Bigr) \Bigr) + \frac{z}{\mu^2} \mathcal{L}_0 \Bigl( \frac{t}{\mu^2}\Bigr)
\mathcal{L}_0 (1-z)\Bigr]. 
\end{align}

\section{Semi-inclusive jet functions and FJF for small $M$ in $\mathrm{SCET_I}$\label{jetzero}}
\subsection{Semi-inclusive jet functions\label{sijf}}
In taking the limit of small $M$ from eq.~\eqref{jetmat}, we use the definition
\begin{equation}\label{ldef}
\mathcal{L}_n (x) \equiv \Bigl[ \frac{\theta (x) \ln^n x}{x}\Bigr]_+ = \lim_{\beta\rightarrow 0} 
\Bigl[ \frac{\theta (x-\beta) \ln^n x}{x} +\delta (x-\beta) \frac{\ln^{n+1} \beta}{n+1}\Bigr],
\end{equation} 
and the following identities:
\begin{align} \label{lrel}
& \lim_{\beta\rightarrow 0} \Bigl[ \frac{\theta (x-\beta) \ln (x-\beta)}{x} 
+\frac{1}{2}\delta (x-\beta) \ln^2 \beta\Bigl] = \mathcal{L}_1 (x) -\frac{\pi^2}{6} \delta(x),  
\nonumber \\
&\lim_{\beta \rightarrow 0} \frac{\theta (x-\beta)\beta}{x^2} = \delta (x), \ 
\lim_{\beta \rightarrow 0} \frac{\theta(x-\beta) \beta^2}{x^3} = \frac{1}{2} \delta (x). 
\end{align}

Putting $x=p^2/\mu^2$ and $\beta = M^2 /\mu^2$, and using the relations eqs.~\eqref{ldef} and 
\eqref{lrel}, $M_a$ and $M_b$ in eq.~\eqref{jetmat} are written as
\begin{align}
M_a &= \frac{\alpha}{2\pi} \frac{1}{\mu^2} \Bigl(  \frac{1}{2x} -\frac{\beta}{x^2} 
+\frac{1}{2} \frac{\beta^2}{x^3}\Bigr)  \theta (x-\beta)  
= \frac{\alpha}{2\pi} \Bigl[ \delta (p^2) \Bigl(-\frac{3}{4} +\frac{1}{2} \ln \frac{\mu^2}{M^2} 
\Bigr) +\frac{1}{2\mu^2} \mathcal{L}_0 \Bigl(\frac{p^2}{\mu^2} \Bigr)
\Bigr],
\nonumber \\
M_b &= \frac{\alpha}{2\pi} \frac{1}{\mu^2} \Bigl( - \frac{1+\ln \beta}{x}
+\frac{\ln x}{x} +\frac{\beta}{x^2}\Bigr)  \theta (x-\beta)    \\
&=\frac{\alpha}{2\pi} \Bigl[ \delta (p^2) \Bigl( 1 - \ln \frac{\mu^2}{M^2} 
+\frac{1}{2} \ln^2 \frac{\mu^2}{M^2} \Bigr) -\frac{1}{\mu^2} \mathcal{L}_0 \Bigl( \frac{p^2}{\mu^2}\Bigr)
\Bigl( 1 - \ln \frac{\mu^2}{M^2}\Bigr) +\frac{1}{\mu^2} \mathcal{L}_1 \Bigl( \frac{p^2}{\mu^2}\Bigr) \Bigr].
\nonumber 
\end{align}
The contributions $M_b^{\varnothing}$ and $M_c$, and the wavefunction renormalization with
the residue remain the same.

 \subsection{Fragmenting jet functions\label{fjfsmall}}
Out of the matrix elements in eq.~\eqref{fjfmat} for the FJFs, we need to consider $M_a$ and $\tilde{M}_b$ in the limit 
of small $M$ and the rest remains the same in the limit.  Firstly, $M_a$ is given by   
\begin{align}
M_a &= \frac{\alpha}{2\pi} 
\theta(z) \theta (1-z) \theta \Bigl( p^2 - \frac{M^2}{1-z}\Bigr) \frac{1}{p^2} 
\Bigl( 1-z -\frac{M^2}{p^2}\Bigr) \nonumber \\
&= \frac{\alpha}{2\pi} \theta(z) \theta (1-z) \theta (z-\beta)\frac{1-z}{\mu^2} 
\Bigl( \frac{1}{z} - \frac{\beta}{z^2}\Bigr),
 \end{align}
where the dimensionless variables $z=p^2/\mu^2$ , $\beta =M^2/[(1-z) \mu^2]$ are introduced. 
Using the definitions of the distributions and their properties in eqs.~\eqref{ldef} and 
\eqref{lrel}, $M_a$ is written as
\begin{equation}
M_a = \frac{\alpha}{2\pi} \theta(z) \theta (1-z) (1-z) \Bigl[ 
\delta (p^2) \Bigl( \ln \frac{(1-z) \mu^2}{M^2} -1\Bigr) +\frac{1}{\mu^2} \mathcal{L}_0 
\Bigl( \frac{p^2}{\mu^2}\Bigr) 
 \Bigr]. 
\end{equation}
 
In order to compute the limit of small $M$ in
 \begin{equation} 
\tilde{M}_b = \frac{\alpha}{2\pi} \theta(z)
\theta (1-z) \theta \Bigl( p^2 - \frac{M^2}{1-z}\Bigr) \frac{1}{p^2} \frac{z}{1-z}, 
\end{equation}
we first expand  $1/p^2$ as distributions in the form 
\begin{equation}
\frac{1}{p^2} \frac{1}{1-z} \theta \Bigl( p^2 - \frac{M^2}{1-z}\Bigr)  =  \delta (p^2) h(z,\mu) +\frac{1}{\mu^2} 
\mathcal{L}_0  \Bigl( \frac{p^2}{\mu^2}\Bigr) g(z,\mu) + \frac{1}{\mu^2} \mathcal{L}_1  
\Bigl( \frac{p^2}{\mu^2}\Bigr)  f(z,\mu)+\cdots,
\end{equation}
where the functions $h(z,\mu)$, $g(z,\mu)$ and $f(z,\mu)$ are to be determined.
Following the procedure explained in appendix~\ref{beamzero}, 
$\tilde{M}_b$ is given as
\begin{align} 
\tilde{M}_b
&= \frac{\alpha}{2\pi} \Bigl[ \delta (p^2) \delta (1-z) \frac{1}{2}\ln^2 \frac{\mu^2}{M^2} +
\delta (p^2) \Bigl( z \ln \frac{\mu^2}{M^2} \mathcal{L}_0 (1-z) +z \mathcal{L}_1 (1-z)
\Bigr) \nonumber \\ 
& + \delta (1-z) \Bigl( \frac{1}{\mu^2} \mathcal{L}_0 \Bigl(
\frac{p^2}{\mu^2}\Bigr) \ln \frac{\mu^2}{M^2} +\frac{1}{\mu^2} \mathcal{L}_1 \Bigl(
\frac{p^2}{\mu^2}\Bigr) \Bigr) + \frac{z}{\mu^2} \mathcal{L}_0 \Bigl( \frac{p^2}{\mu^2}\Bigr)
\mathcal{L}_0 (1-z)\Bigr]. 
\end{align}

\section{Color structures of the soft functions\label{samat}}
\subsection{Tree-level color matrices for the soft functions\label{stree}}
 
We present the color factors for the tree-level soft functions with different number of nonsinglets. 
\begin{align} \label{s0000}
\mathbf{S}^{(0)} (0,0,0,0) &= \begin{pmatrix}
C_A C_F /2 & 0 \\
0& C_A^2
\end{pmatrix},    \\
\mathbf{S}^{(0)} (1,1,0,0) &=  \frac{1}{2}\begin{pmatrix}
\displaystyle C_F - \frac{C_A}{2}   & 0 \\
0& 2 C_A 
\end{pmatrix}(12),   \  \mathbf{S}^{(0)} (0,0,1,1) =  \frac{1}{2} \begin{pmatrix}
\displaystyle C_F - \frac{C_A}{2}   & 0 \\
0& 2C_A
\end{pmatrix}(34), \nonumber \\
\mathbf{S}^{(0)} (1,0,1,0) & =  \frac{1}{2} \begin{pmatrix}
\displaystyle 2C_F - \frac{C_A}{2}   & 1 \\
1& 0
\end{pmatrix}(13), \ \mathbf{S}^{(0)} (0,1,0,1) =  \frac{1}{2}\begin{pmatrix}
\displaystyle 2C_F - \frac{C_A}{2}    &1 \\
1& 0 
\end{pmatrix}(24),    \nonumber \\
\mathbf{S}^{(0)} (1,0,0,1) &=  \frac{1}{2}\begin{pmatrix}
 2C_F - C_A    & 1 \\
1& 0 
\end{pmatrix}(14),   \  \mathbf{S}^{(0)} (0,1,1,0) =  \begin{pmatrix}
\displaystyle C_F - \frac{C_A}{2}   &\displaystyle \frac{1}{2} \\
\displaystyle\frac{1}{2}& 0
\end{pmatrix}(23),  \nonumber \\
\mathbf{S}^{(0)} (1,1,1,0) &=  \frac{1}{2} \begin{pmatrix}
\displaystyle  C_F - \frac{C_A}{2}  & 0  \\
1   &0
\end{pmatrix} (123) + \frac{1}{2}
\begin{pmatrix}
\displaystyle C_F -\frac{C_A}{2}  & 1  \\
0  &0
\end{pmatrix} (132), \nonumber \\
\mathbf{S}^{(0)} (1,1,0,1) &=   \frac{1}{2} \begin{pmatrix}
\displaystyle C_F - \frac{C_A}{2}  & 0  \\
1   &0
\end{pmatrix} (124) +\frac{1}{2}
\begin{pmatrix}
\displaystyle C_F -\frac{C_A}{2}  & 1  \\
0  &0
\end{pmatrix} (142), \nonumber \\
\mathbf{S}^{(0)} (1,0,1,1) &=  \frac{1}{2}\begin{pmatrix}
\displaystyle  C_F - \frac{C_A}{2}  & 0  \\
1    &0
\end{pmatrix} (143) + \frac{1}{2}
\begin{pmatrix}
\displaystyle C_F -\frac{C_A}{2}  &1  \\
0  &0
\end{pmatrix} (134), \nonumber \\
\mathbf{S}^{(0)} (0,1,1,1) &=  \frac{1}{2} \begin{pmatrix}
\displaystyle C_F - \frac{C_A}{2}  & 0  \\
1    &0
\end{pmatrix} (243) + \frac{1}{2}
\begin{pmatrix}
\displaystyle C_F -\frac{C_A}{2}   &1  \\
0  &0
\end{pmatrix} (234). \nonumber 
\end{align}
Here $(a_1 a_2 \cdots a_n) = \mathrm{Tr}(t^{a_1} t^{a_2} \cdots t^{a_n})$, and 
the color indices are to be contracted with the corresponding
collinear nonsinglet parts. Note that the soft color matrices with 
a single nonsinglet are zero due to weak charge conservation.\footnote{In SU(2), 
all the color matrices with the odd number of nonsinglets become zero.}
\begin{equation}
\mathbf{S} (1,0,0,0) = \mathbf{S} (0,1,0,0) = \mathbf{S} (0,0,1,0) = \mathbf{S}(0,0,0,1)=0.
\end{equation}

The color structure with all the four nonsinglets is given by
\begin{align} \label{s1111}
\mathbf{S}^{(0)} (1,1,1,1) &=  \frac{1}{2} \begin{pmatrix}
\displaystyle  C_F - \frac{C_A}{2}  & 0  \\
1    &0
\end{pmatrix} (1243) + \frac{1}{2}
\begin{pmatrix}
\displaystyle C_F -\frac{C_A}{2}  & 1  \\
0  &0
\end{pmatrix} (1342) \nonumber \\
&+ \begin{pmatrix}
 \displaystyle (C_F - \frac{C_A}{2})^2   &  \displaystyle C_F - \frac{C_A}{2} \\
 \displaystyle C_F - \frac{C_A}{2}   & 1
\end{pmatrix} (12)(34) +
\begin{pmatrix}
\displaystyle\frac{1}{4}   & 0  \\
0  &0
\end{pmatrix} (13)(24).
\end{align}

\subsection{No mixing in the soft function at order $\alpha$\label{snomix}}
We show that there is no mixing in the $N$-jettiness soft function for the case with all the four nonsinglets. 
In ref.~\cite{Manohar:2018kfx},  there are three independent functions and
there is a mixing among these terms at order $\alpha$. We treat the soft function in the operator basis
as a $2\times 2$ matrix. Here when we consider the jettiness, there is also a mixing
in the real contributions, while there is no mixing in the virtual contributions.
But it turns out that the mixing is cancelled in the sum.
 
The soft color matrix $\mathbf{S}^{(0)} (1, 1, 1, 1)$ in eq.~\eqref{s1111} includes the color factors
$(1243)$, $(1342)$, $(13)(24)$ and $(12)(34)$. There is mixing because there appear different color factors
at one loop, which are not present at tree level. 
It is helpful to look at the soft part in eq.~\eqref{intsoft} with all the nonsinglets, which is given as
\begin{equation} 
\langle 0| \mathrm{Tr} \Bigl( t^c S_2^{\dagger} T_J S_1 (0) t^d S_1^{\dagger} T_I S_2 (x)\Bigr)
\cdot \mathrm{Tr} \Bigl( t^e S_3^{\dagger} T_J S_4 (0) t^f S_4^{\dagger} T_I S_3 (x)\Bigr) |0\rangle.
\end{equation}
The contractions of the fields at 0 or $x$ yield virtual contributions, 
while the contraction of the fields with 0 and $x$ yield real contributions. 
 
Each contraction of  $i$ and $j$ in the virtual contributions in figure~\ref{softwilson} (a) does not produce 
new color structures, while the real contributions
produce new color structures.  
The contractions of $i$ with $j$ in the real contributions at one loop in figure~\ref{softwilson} (b) 
produce the  following color structures:
\begin{align}
S^{R(1)}_{12+21} (1, 1, 1, 1)&=-\Bigl( C_F - \frac{C_A}{2}\Bigr) \mathbf{S}^{(0)} (1, 1, 1, 1) 
- \frac{1}{8} \begin{pmatrix}
  1&0 \\
0&0
\end{pmatrix} \Bigl[(1423)+ (1324) \Bigr], \nonumber \\
S^{R(1)}_{13+31} (1, 1, 1, 1)&= -\Bigl( C_F - \frac{C_A}{2}\Bigr) \mathbf{S}^{(0)} (1, 1, 1, 1) 
- \frac{1}{8N^2} \begin{pmatrix}
 1 &    -2N\\
  -2N& 4N^2
\end{pmatrix} \Bigl[ (1432) + (1234) \Bigr] , \nonumber \\
S^{R(1)}_{14+41} (1, 1, 1, 1)&= -\Bigl( C_F - \frac{C_A}{2}\Bigr) \mathbf{S}^{(0)} (1, 1, 1, 1)  
+\frac{1}{8} 
\begin{pmatrix}
1&0\\0&0
\end{pmatrix} \Bigl[ (1423) + (1324)\Bigr]\nonumber \\
&   + \frac{1}{8N^2} \begin{pmatrix}
 1 &    -2N\\
  -2N& 4N^2
\end{pmatrix}\Bigl[ (1432) + (1234)\Bigr], \nonumber \\
S^{R(1)}_{23+32} (1, 1, 1, 1)&=  -\Bigl( C_F - \frac{C_A}{2}\Bigr) \mathbf{S}^{(0)} (1, 1, 1, 1) 
+\frac{1}{8} 
\begin{pmatrix}
1&0\\0&0
\end{pmatrix} \Bigl[ (1423)+(1324) \Bigr]\nonumber \\
&   +\frac{1}{8N^2} \begin{pmatrix}
 1 &    -2N\\
  -2N& 4N^2
\end{pmatrix}\Bigl[ (1432) + (1234)\Bigr],\nonumber \\
S^{R(1)}_{24+42} (1, 1, 1, 1)&= -\Bigl( C_F - \frac{C_A}{2}\Bigr) \mathbf{S}^{(0)} (1, 1, 1, 1) 
-\frac{1}{8N^2} \begin{pmatrix}
 1 &    -2N\\
  -2N& 4N^2
\end{pmatrix}\Bigl[ (1432) + (1234)\Bigr] , \nonumber \\
S^{R(1)}_{34+43} (1, 1, 1, 1)&= -\Bigl( C_F - \frac{C_A}{2}\Bigr) \mathbf{S}^{(0)} (1, 1, 1, 1)
-\frac{1}{8} 
\begin{pmatrix}
1&0\\0&0
\end{pmatrix}\Bigl[ (1423) + (1324)\Bigr],
\end{align}
where the extra terms in each contraction represents the new color structures which are not present in 
$\mathbf{S}^{(0)} (1, 1, 1, 1)$.  The first term $S^{R(1)}_{12+21} (1, 1, 1, 1)$, for example, represents the sum of the
contributions $i, j = 1, 2$ and $2, 1$, where the two contributions have different color factors $(1423)$ and 
$(1324)$.

When summed over all the contractions $(ij)$, these additional contributions 
cancel in $d \mathbf{S}/d\ln \mu$ from eq.~\eqref{rgsoft1},  
because $dS_{ij,\mathrm{hemi}}^R/ d\ln \mu$ are independent of $n_{ij}$. 
Though $dS_{ij,\mathrm{hemi}}^V/ d\ln \mu$ depends on $n_{ij}$, 
there is no mixing in the color structure for the virtual contributions.
As a consequence, there is no mixing in the soft anomalous dimensions at one loop. 
On the other hand, in inclusive cross sections in which both  
$dS_{ij}^R/ d\ln \mu$ and $dS_{ij}^V/ d\ln \mu$
depend on $n_{ij}$~\cite{Manohar:2018kfx}, there appears mixing in the 
contributions with four nonsinglets.

\acknowledgments{
This work is supported by Basic Science Research Program through the National 
Research Foundation of Korea (NRF) funded by 
the Ministry of Education (Grant No. NRF-2019R1F1A1060396). }

\bibliographystyle{JHEP1}
\bibliography{jettiness}
\end{document}